\documentclass[11pt,a4paper]{article}

\usepackage[utf8]{inputenc}
\usepackage[T1]{fontenc}
\usepackage{lmodern}
\usepackage[a4paper,margin=22mm]{geometry}
\usepackage{amsmath,amssymb,mathtools,bm}
\usepackage{graphicx}
\usepackage[numbers,sort&compress]{natbib}
\usepackage{microtype}
\usepackage{placeins}
\usepackage{caption}
\usepackage{hyperref}
\usepackage{xurl}

\graphicspath{{./figures/main/}{./figures/supporting_information/}}
\hypersetup{hidelinks,pdfauthor={Xi-Hao Chen et al.},pdftitle={A Non-Hermitian Biorthogonal Encoding Paradigm for Physical-Layer Secure Computational Imaging}}
\allowdisplaybreaks
\title{\textbf{A Non-Hermitian Biorthogonal Encoding Paradigm for\\Physical-Layer Secure Computational Imaging}}
\author{Xi-Hao Chen, Kan-Xu Jia, En-Rui Zhang, Yi-Zhu Zhang,\\
Xin-Peng Wei, Bu-Ran Yu, Qian-Qian Bao, and Shao-Ying Meng\\[0.7em]
\small Key Laboratory of Optoelectronic Devices and Detection Technology, School of Physics,\\
\small Liaoning University, Shenyang 110036, China\\[0.4em]
\small Corresponding author: Shao-Ying Meng (\href{mailto:mengshaoying@163.com}{mengshaoying@163.com})}
\date{}

\begin{document}
\maketitle

\begin{abstract}
The conventional paradigm of computational imaging, rooted in Hermitian systems, is fundamentally constrained by rigid orthogonal basis transformations, which bottleneck the balance between reconstruction fidelity, computational load, and physical-layer security. In this work, we propose a generalized secure computational imaging framework based on non-Hermitian biorthogonal symmetry breaking. By mapping spatial information into a biorthogonal operator space, we establish an asymmetric sensing architecture governed by distinct left-basis $\langle\phi_{m}\vert{}$ and right-basis $\vert{}\psi_{n}\rangle$ modes, satisfying the biorthogonality relation $\langle\phi_{m}\vert{}\psi_{n}\rangle = \delta_{mn}$. Within this manifold, precise tuning of the non-Hermitian parameter $\gamma$ establishes a physical-layer cryptographic gate, where high-fidelity retrieval is exclusively enabled by matching the dual basis; any parameter mismatch triggers deterministic inter-modal crosstalk that effectively neutralizes unauthorized access. Notably, this architecture intrinsically supports direct, iteration-free image retrieval across a wide range of sampling ratios, significantly reducing the computational overhead compared to conventional iterative reconstruction. We validate this framework on a single-pixel imaging platform, demonstrating a fundamental paradigm shift: by embedding security directly into the measurement physics, we transform image retrieval from a software-dependent task into a parameter-sensitive physical decryption process that ensures architecture-intrinsic confidentiality.
\end{abstract}


\section{Introduction}

Modern optical imaging and information theory are anchored in Hermitian physics, where observable operators govern physical states under strict metric orthogonality \cite{Bender1998PRL,Bender2002PRL,Mostafazadeh2002JMP,Brody2014JPA}. Within this paradigm, computational imaging has witnessed unprecedented growth, with representative methods including single-pixel imaging (SPI)~\cite{Edgar2019,Duarte2008}, ghost imaging~\cite{Shapiro2008,Gatti2004}, coded-aperture spectral imaging~\cite{Wagadarikar2008}, compressive imaging~\cite{Donoho2006,Candes2006}, various frequency-domain synthetic aperture modalities such as ptychography and Fourier ptychographic microscopy~\cite{Miao1999,Rodenburg2004,Zheng2013}, as well as, wavefront shaping through complex media~\cite{Popoff2010,Bertolotti2012,Rotter2017}, snapshot compressive imaging (SCI)~\cite{Llull2013OptExpress,Yuan2021SPM}, and non-line-of-sight imaging~\cite{Velten2012,OToole2018,Faccio2020}. Representative extensions of this general computational paradigm also include three-dimensional single-pixel imaging~\cite{Sun2013Science}, compressive spectral imaging~\cite{Gehm2007OptExpress}, compressive holography~\cite{Clemente2013OL}, deterministic phase retrieval and quantitative phase imaging~\cite{Teague1983JOSA,Park2018QPI}, and single-pixel ptychography~\cite{Li2021OL}. Despite their diverse implementations and specific hardware configurations, most of these schemes share a common, fundamental mechanism: physical scene information is encoded into multiplexed measurements and computationally retrieved via model-based inversion. Crucially, in these architectures, the physical encoding patterns and their computational decoding counterparts symmetrically share the same Hermitian Hilbert space, strictly adhering to the standard orthogonal metric $\langle\psi_m^{(\mathrm{H})}|\psi_n^{(\mathrm{H})}\rangle = \delta_{mn}$. Consequently, image reconstruction is bound to the Hermitian adjoint or pseudoinverse of the forward operator \cite{Candes2006,Donoho2006,CandesWakin2008IEEESPM,Duarte2008}, ensuring fidelity but simultaneously restricting the physical-layer degrees of freedom for decoding-selective information processing \cite{Refregier1995OptLett,Situ2004OptLett,Matoba2009ProcIEEE,Frauel2007OptExpress}. Because encoding and decoding share the same inner-product structure, any unauthorized reader possessing the physical measurement matrix can, in principle, invert the linear system to expose the underlying information \cite{Popoff2010,Popoff2010NatCommunImage,Yu2024NatCommun}. SPI has also been extended to optical information-security tasks, including single-pixel visual cryptography and coded information retrieval~\cite{Jiao2020OptExpress}. Existing physical-layer security schemes predominantly rely on structural randomness to expand the key space \cite{Refregier1995OptLett,Situ2004OptLett,Matoba2009ProcIEEE,Frauel2007OptExpress,Yu2024NatCommun}, yet these strategies merely obscure coefficients within the same Hermitian metric without altering the decoding rule; once the operator is leaked, the metric symmetry grants eavesdroppers identical decoding privileges as the authorized recipient\cite{Peng2006OptLett,Frauel2007OptExpress}. Achieving deterministic physical-layer decoding selectivity therefore requires a radical departure from metric symmetry, necessitating an asymmetric information space where accurate computational decoding terminates at a physical metric non-equivalent to the ordinary Hermitian adjoint of the encoding basis \cite{Mostafazadeh2002JMP,Brody2014JPA,Guo2009PRL,Rotter2009JPhysA,Lin2011PRL}.

Non-Hermitian wave physics provides a mathematically rigorous route to shatter this metric symmetry, reconfiguring the boundaries of information retrieval \cite{Bender1998PRL,Ruter2010NatPhys,Heiss2012JPA,ElGanainy2018NatPhys,Feng2017NatPhot,Miri2019Science,Ozdemir2019NatMater}. For a generic non-Hermitian operator $H_{\gamma}$ incorporating gain, loss, or complex potentials, spatial evolution and spectral projection are dictated by distinct right ($|\psi_n\rangle$) and left ($\langle\phi_n|$) eigenstates, defined by $H_{\gamma}|\psi_n\rangle = \lambda_n|\psi_n\rangle$ and $\langle\phi_n|H_{\gamma} = \lambda_n\langle\phi_n|$ \cite{Garrison1988PLA,Brody2014JPA}. Crucially, while these sets are asymmetric under ordinary Hermitian conjugation ($\langle\phi_n| \neq |\psi_n\rangle^\dagger$), they are uniquely bound through the biorthogonal completeness relation, $\langle\phi_m|\psi_n\rangle = \delta_{mn}$ \cite{Mostafazadeh2002JMP,Brody2014JPA}. This intrinsic asymmetry enables the design of a decoupled information architecture where the encoding basis is fundamentally detached from the Hermitian-adjoint decoding space \cite{Bender2002PRL,Rotter2009JPhysA}. Unlike unitary evolution, the non-Hermitian manifold allows for the construction of parameter-dependent operators where information is intrinsically protected by the topology of the biorthogonal space. This decoupling transcends the fundamental limits of conventional computational imaging: in standard Hermitian systems, the observer and the information are tethered by a shared adjoint metric, rendering the measurement matrix transparent to any party capable of matrix inversion \cite{Refregier1995OptLett,Frauel2007OptExpress,Yu2024NatCommun}. In contrast, the non-Hermitian framework permits the establishment of a decoding-selective channel where data retrieval is restricted to a specific biorthogonal metric, inherently routing unauthorized access attempts into the non-orthogonal leakage channels of the broader Hilbert space. By manipulating the biorthogonal coupling coefficients, one can effectively control the information flow, ensuring that only the authorized dual-basis observer can resolve the signal from the incoherent projection crosstalk 
\cite{Yu2024NatCommun}.

Building upon this non-Hermitian biorthogonal sensing architecture, we propose a generalized secure computational imaging framework. By transitioning the sensing process from a conventional Hermitian space into a dual-basis biorthogonal operator space, we map spatial information onto a non-orthogonal manifold governed by distinct left-basis $\langle\phi_{m}|$ and right-basis $|\psi_{n}\rangle$ modes. Within this regime, the system operates as a deterministic physical-layer cryptographic gate; unlike conventional methods that rely on software-side inversion, our approach enforces a strict biorthogonality condition ($\langle\phi_{m}|\psi_{n}\rangle = \delta_{mn}$) that acts as a native physical filter. Authorized retrieval is thus restricted exclusively to the dual-basis metric, while any unauthorized attempt at matrix inversion—blind to the underlying biorthogonal structure—inevitably triggers incoherent inter-modal crosstalk that manifests as deterministic noise. This framework signals a fundamental shift in imaging science: the reconstruction process is no longer merely a post-capture computational task, but a native, hardware-level biorthogonal projection. By replacing software-dependent randomization with inherent operator-space asymmetry, we encapsulate functional intelligence directly within the physics of measurement. Beyond the SPI platform demonstrated herein, this methodology functions as a universal protocol for high-throughput information acquisition, offering a transformative path toward next-generation sensing systems where security is an emergent property of the imaging physics itself.

\section{Principle}
\begin{figure}[htbp]
\centering
\includegraphics[width=\textwidth]{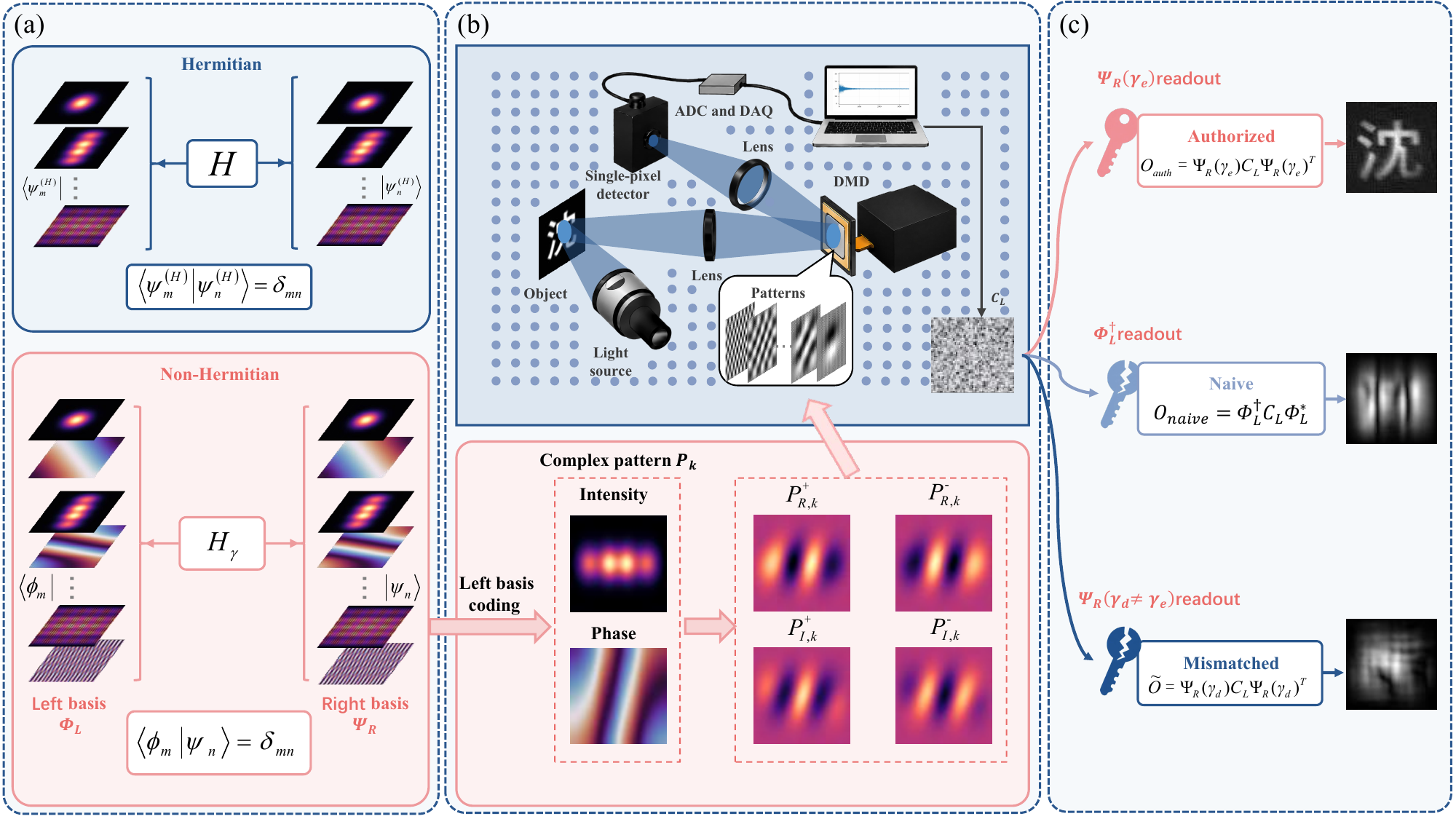}
\caption{Conceptual principle and passive SPI implementation: A non-Hermitian biorthogonal protocol. (a) Transition from a standard Hermitian orthogonal basis to a paired non-Hermitian left-right biorthogonal basis, governed by the completeness relation $\langle\phi_{m}|\psi_{n}\rangle=\delta_{mn}$.(b) Experimental setup for passive SPI. The scene's spatial information is directed onto a DMD, where the biorthogonal left-basis vectors $\langle\phi_{m}|$ are spatially imprinted as complex-valued modulation patterns $P_k$. These patterns, synthesized via a differential measurement protocol using non-negative intensity masks, function as structured mode-sampling templates that map the target scene into the operator space of the finite non-Hermitian operator $H_N(\gamma)$. (c) Biorthogonal decoding process. The measured coefficient matrix $C_L$ is decoded by projecting the signal stream onto the authorized dual-basis $|\psi_{n}\rangle$, which acts as a digital filter to extract structural information. This process exhibits inherent decoding selectivity: authorized recovery ($\Psi_R$) restores high-fidelity target topology, while naive Hermitian back-projection ($\Phi_L^\dagger$) or parameter-mismatched decoding ($\Psi_R(\gamma_d \neq \gamma_e)$) results in signal leakage into inter-modal crosstalk. 
    }
\label{fig:platform}
\end{figure}

As sketched in Fig.~\ref{fig:platform}, we implement this non-Hermitian biorthogonal computational framework by integrating a biorthogonal left-right basis with a passive SPI configuration. The optical field carrying the scene's spatial information is imaged onto a programmable spatial light modulator via a lens system. Notably, this architecture shares an identical hardware configuration with conventional Hermitian-based SPI systems, eliminating the need for specialized asymmetric optical components. Instead of actively modifying the illumination, the modulator sequentially imposes a set of left-basis patterns
$\langle\phi_m\vert{}$, physically mapping the scene into the non-Hermitian operator space. The total throughput is captured by a single-pixel detector to yield a sequence of scalar coefficients. Critically, this framework can function as a physical-layer cryptographic gate when the non-Hermitian control parameter $\gamma$ is accurately matched between encoding and decoding (see Supporting Information, Sec.~I): the original scene is inherently protected from unauthorized access due to inter-modal crosstalk, and image retrieval is performed through the matched right-basis decoding
$\vert{}\psi_n\rangle$, where the biorthogonality condition $\langle\phi_m\vert{}\psi_n\rangle = \delta_{mn}$ serves as the key for secure reconstruction. 

\begin{figure}[htbp]
\centering
\includegraphics[width=\textwidth]{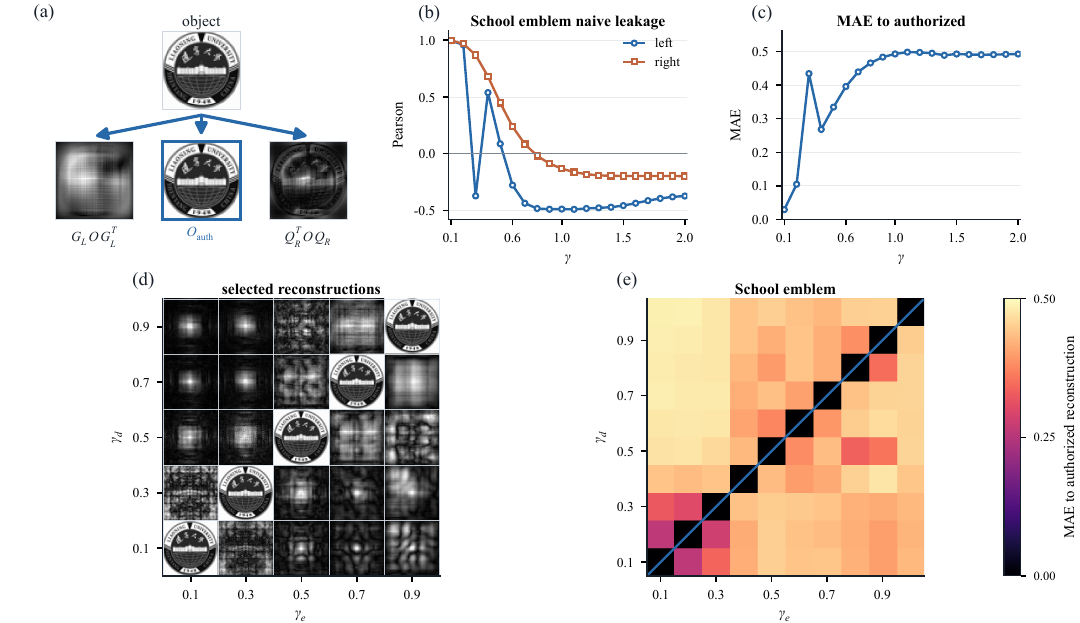}
\caption{Numerical demonstration of biorthogonal decoding selectivity.
(a) A $64 \times 64$ school-emblem object is encoded into the finite non-Hermitian operator space and reconstructed through different decoding metrics. The authorized dual-basis decoding $\Psi_R C_L \Psi_R^{T}$ restores the object through the identity channel, whereas ordinary Hermitian back-projection of the left or right basis redirects the corresponding encoded data into deterministic Gram-channel leakage, represented by $G_L O G_L^{T}$ and $Q_R^{T} O Q_R$.
(b) Pearson correlation between the naive and authorized reconstructions as the control parameter $\gamma$ is varied, revealing the tunable divergence between the correct biorthogonal metric and the wrong Hermitian metric.
(c) Mean absolute error (MAE) between the left-naive and authorized branches.
(d) Cross-parameter decoding results obtained by encoding at $\gamma_e$ and decoding with $\gamma_d$. Faithful recovery is concentrated on the matched condition $\gamma_d=\gamma_e$.
(e) MAE map of the parameter-mismatch channel, confirming that the identity-channel reconstruction is confined to the matched diagonal, while off-diagonal decoding is routed into inter-modal crosstalk.}
\label{fig:matched_naive}
\end{figure}

In the main protocol, we leverage the biorthogonal properties inherent to non-Hermitian systems. As illustrated in Fig. 1(a), the Hermitian protocol (top panel) relies on a single eigenbasis, whereas the non-Hermitian framework (bottom panel) requires two distinct eigenbases—the left $\Phi_L$ and right $\Psi_R$—that satisfy the biorthogonality condition $\langle \phi_m \vert{} \psi_n \rangle = \delta_{mn}$. Consequently, the modulation patterns are constructed from $\Phi_L$, while the corresponding computational decoding must be performed using $\Psi_R$. This intrinsic left-right pairing necessitates the use of two distinct operators: the physical encoding operator $A_\gamma$ and the authorized decoding operator $D_\gamma$. The vectorized sensing model adheres to the linear formalism common to conventional protocols and is expressed as

\begin{equation}
\mathbf{y} = A_\gamma \mathbf{x} + \mathbf{n},
\label{eq:spi_model}
\end{equation}
where $\mathbf{x} = \operatorname{vec}(O) \in \mathbb{R}^{N^2}$ represents the vectorized form of the target scene image $O \in \mathbb{R}^{N \times N}$, and $\mathbf{y} \in \mathbb{C}^{M}$ denotes the effective single-pixel measurement vector associated with the $M$ modulation modes. Here, $\mathbf{n} \in \mathbb{C}^{M}$ denotes the measurement noise, and $A_\gamma \in \mathbb{C}^{M\times N^2}$ denotes the physical encoding operator constructed from the finite-dimensional non-Hermitian operator for a given control parameter $\gamma$. Its specific realization also depends on the finite coordinate-window size, discretization size, mode ordering, and biorthogonal normalization convention.

The decoding process differs fundamentally from that of conventional Hermitian protocols. Unlike the latter, where reconstruction typically is achieved via a simple conjugate transpose rule ($\hat{\mathbf{x}} = A_\gamma^\dagger \mathbf{y}$), a distinct biorthogonal decoding rule is necessitated in our non-Hermitian framework. Accordingly, the authorized decoder is designated as $D_\gamma \in \mathbb{C}^{N^2 \times M}$, and the authorized reconstruction is defined as
\begin{equation}
\hat{\mathbf{x}}_{\rm auth} = D_\gamma \mathbf{y}.
\label{eq:auth_inverse}
\end{equation}
In the noiseless full-basis limit ($M = N^2$), biorthogonal completeness ensures $D_\gamma A_\gamma = I_{N^2}$. Notably, the authorized reconstruction is not defined by standard Hermitian back-projection but by the dual operator $D_\gamma$ matched to the non-Hermitian encoding basis. The central design principle is thus to enforce a strict mathematical inequivalence between the legitimate inverse $D_\gamma$ and the ordinary adjoint $A_\gamma^\dagger$.

In the present implementation, a finite-dimensional non-Hermitian oscillator is used as a generator for paired modulation and decoding patterns. The underlying complex-shifted oscillator and its finite-dimensional discretization follow established non-Hermitian and finite-difference formulations~\cite{Ahmed2001PLA,Fornberg1988MathComp}. The continuous Hamiltonian, finite-coordinate-window discretization, mode ordering, and biorthogonal normalization used to generate the paired bases in Fig. 1(a) are detailed in Supporting Information, Sec. I. This choice provides a compact and tunable realization of a non-Hermitian operator space with controllable left-right asymmetry. After discretization, the generator yields right and left mode matrices
$\Psi_R(\gamma)$ and $\Phi_L(\gamma)$ satisfying the finite-dimensional
biorthogonal completeness relation
$\Phi_L(\gamma)\Psi_R(\gamma)=I_N$, where $I_N$ denotes the
$N\times N$ identity matrix, whereas same-side overlaps remain
nonorthogonal under the ordinary Hermitian inner product.

To realize the physical encoding stage shown in Fig.~\ref{fig:platform}(b), the one-dimensional left modes contained in $\Phi_L(\gamma)$ are combined through separable outer products to generate two-dimensional complex left-basis modulation patterns. These patterns form the experimental DMD library and are sequentially displayed on the modulator to project the target scene onto the left basis. The resulting single-pixel measurements are assembled into the left-basis coefficient matrix $C_L=\Phi_L O\Phi_L^T$. Equivalently, using the standard vectorization--Kronecker identity~\cite{VanLoan2000JCAM}, the full separable encoding can be written as $\operatorname{vec}(C_L)=(\Phi_L\otimes\Phi_L)\operatorname{vec}(O)$, which connects the matrix representation directly to the vectorized sensing model in Eq.~\eqref{eq:spi_model}. The coefficient matrix $C_L$ is subsequently decoded using different decoding metrics and parameter settings, giving rise to the three distinct physical-layer reconstruction channels illustrated in Fig.~\ref{fig:platform}(c): the authorized, naive, and parameter-mismatched channels.

Importantly, the aforementioned reconstruction channels are highly sensitive to the parameter $\gamma$; a detailed analysis of this sensitivity is provided in Supporting Information, Secs.~II and III. Achieving robust security for the reconstruction relies on an appropriate selection of $\gamma$. To quantitatively characterize these channels,we formulate the reconstruction process as a mapping from the measurement matrix $C_L$ to the target $O$. Note that for full-sampling, $\Psi_R$ acts as an inverse operator, whereas for under-sampling, it represents a generalized reconstruction operator (e.g., the pseudo-inverse, an iterative regularized solver, or the truncated biorthogonal operator adopted here) designed to mitigate the ill-posed nature of the underdetermined system. (i) The authorized channel governs the authorized decoding, where the reconstruction matrix collapses to the target: $O_{\rm auth} = \Psi_R C_L \Psi_R^T = O$. (ii) The Gram channel, which intercepts a naive adversary, produces a distorted profile $O_{\rm naive}=\Phi_L^\dagger C_L\Phi_L^* =G_L O G_L^T$, where $G_L=\Phi_L^\dagger\Phi_L=U\Sigma^{-2}U^\dagger$ is the left-basis Gram operator obtained from the singular-value decomposition (SVD) $\Psi_R=U\Sigma W^\dagger$~\cite{TrefethenBau2022}. The complete singular-value derivation and the reciprocal weighting of the left- and right-basis naive channels are provided in Supporting Information, Sec.~I.4. (iii) The mismatch channel dictates the cross-parameter coordinate alignment, resulting in the misaligned estimate $\widetilde{O} = M_L(\gamma_d, \gamma_e) O M_L^T(\gamma_d, \gamma_e)$ with the parameter-sensitive operator $M_L(\gamma_d, \gamma_e) = \Psi_R(\gamma_d)\Phi_L(\gamma_e)$. Physically, while the authorized channel (i) yields a perfect deterministic reconstruction via biorthogonal cancellation, the naive Hermitian back-projection in channel (ii) forces the state into a wrong-metric projection that severely misaligns the mode spectrum. Concurrently, since the biorthogonal metric is acutely sensitive to the operator coordinates, even an infinitesimal parameter deviation ($\gamma_d \neq \gamma_e$) in channel (iii) induces a sharp collapse of spatial correlations.

It is important to note that, in the present implementation, the physical modes are ordered according to their eigenvalue-based modal energies, with lower-order modes prioritized in the projection. This ordering follows transform-domain and significance-ordered SPI, where low-frequency or structurally informative coefficients are preferentially acquired under partial sampling \cite{Zhang2015NatCommun,Sun2017SciRep,Yu2020SciRep}. Consequently, the reconstruction based on $\Psi_R$ forms a truncated biorthogonal projection, allowing structural consistency to be preserved by prioritizing the sampled lower-order basis components. This eigenvalue-dependent mode ordering acts as an implicit modal filter, allowing the system to preserve the principal object structure without the need for exhaustive iterative optimization .

Although the principles described above are demonstrated within a passive SPI configuration, this computational paradigm is universal and applicable to any computational imaging modality governed by the general linear inverse problem expressed in ~\eqref{eq:spi_model}. Moreover, the proposed paradigm offers an inherent layer of physical-layer security. Specifically, the authorized reconstruction is exclusively governed by the dual basis $D_{\gamma}$, which is essentially concealed from any unauthorized observer lacking precise knowledge of the non-Hermitian parameter $\gamma$ and the specific mode ordering. This security is further amplified under under-sampling conditions, where the reduction in measurement data exacerbates the ill-posedness for unauthorized observers, making the retrieval of target scene information significantly more difficult. Without these critical parameters, any attempts at reconstruction from the measurement vector $\mathbf{y}$ inevitably lead to significant inter-modal crosstalk and artifacts, effectively hindering unauthorized access. While a formal proof of unconditional security is beyond the scope of this work, this biorthogonal asymmetry provides a robust physical-layer mechanism that complicates unauthorized extraction.

\section{Numerical demonstration}

To validate the proposed non-Hermitian biorthogonal encoding paradigm and explicitly connect the numerical analysis with the reconstruction channels established above, we perform numerical simulations in a finite $N\times N=64\times64$ operator space. For each encoding parameter $\gamma_e$, the discretized non-Hermitian operator $H_N(\gamma_e)$ yields the paired left- and right-basis matrices, $\Phi_L(\gamma_e)$ and $\Psi_R(\gamma_e)$, satisfying the biorthogonal completeness relation $\Phi_L(\gamma_e)\Psi_R(\gamma_e)=I_N$. The target scene $O$ is chosen as a grayscale university-emblem object with the same $64\times64$ spatial dimension. Following the separable representation of the encoding process, the effective single-pixel coefficient matrix is evaluated directly as $C_L=\Phi_L(\gamma_e)O\Phi_L^{T}(\gamma_e)$, where each matrix element represents the modal coefficient associated with the corresponding separable two-dimensional left-basis projection. This direct matrix mapping constitutes the numerical counterpart of the physical left-basis encoding process illustrated in Fig.~\ref{fig:platform}(b) and is consistent with the sensing model in ~\eqref{eq:spi_model}. The same encoded coefficient matrix $C_L$ is subsequently decoded through the authorized, naive, and parameter-mismatched reconstruction channels defined in the Principle section. By holding the encoded coefficients fixed among the three branches, the comparison isolates the effects of the decoding metric and parameter alignment. This common-data protocol provides the basis for the full-sampling comparisons presented below. The complete basis construction, numerical workflow, metric definitions, additional-object tests, and extended mismatch scans are provided in Supporting Information, Sec.~III.

Figure~\ref{fig:matched_naive}(a) compares the authorized reconstruction with two naive Hermitian reconstructions obtained from the same encoded coefficient matrix. The authorized channel preserves the target morphology because the reconstruction operator is the paired dual of the physical encoding basis. By contrast, enforcing an ordinary Hermitian back-projection on either the left or right basis breaks the biorthogonal cancellation condition and replaces the identity channel by a Gram-channel transformation. The resulting images are not random noise, but deterministic leakage morphologies governed by the wrong metric. This behavior is important for the physical-layer interpretation of the protocol: unauthorized decoding does not simply produce a lower-quality version of the correct image, but routes the encoded information into a different operator channel.

The $\gamma$-dependent behavior in Figs.~\ref{fig:matched_naive}(b) and \ref{fig:matched_naive}(c) demonstrates that the authorized/naive separation is a tunable security boundary governed by the non-Hermitian control parameter. As $\gamma$ increases, the bifurcation of left and right eigenvectors provides a physical mechanism for information masking. The two naive branches exhibit distinct leakage morphologies: the left-naive channel generates structured artifacts by enhancing near-singular modal components, while the right-naive channel suppresses fine details. These artifacts are not merely incidental noise but act as physical noise signatures that scramble unauthorized extraction. This confirms that the observed leakage is intrinsically tied to the left-right metric asymmetry, transforming $\gamma$ into an intrinsic hardware-level security token that ensures information is cryptographically inaccessible without the correct biorthogonal key.

We further investigate $\gamma$ as a physical decoding key. Figures~\ref{fig:matched_naive}(d) and \ref{fig:matched_naive}(e) illustrate the encoding/decoding parameter space. A high-fidelity reconstruction is achieved only when $\gamma_d = \gamma_e$ (the identity channel), establishing a parameter-sensitive authentication protocol. Once the decoding coordinate is detuned ($\gamma_d \neq \gamma_e$), the system forces the signal into a cross-parameter mismatch channel $M_L(\gamma_d, \gamma_e) O M_L^{T}(\gamma_d, \gamma_e)$. This leads to a deterministic misdirection: the system does not simply degrade image quality but maps the encoded data into a physically distinct manifold, rendering the raw information cryptographically useless to any observer lacking the specific physical key $\gamma$.

\begin{figure}[!t]
\centering
\includegraphics[width=1\textwidth]{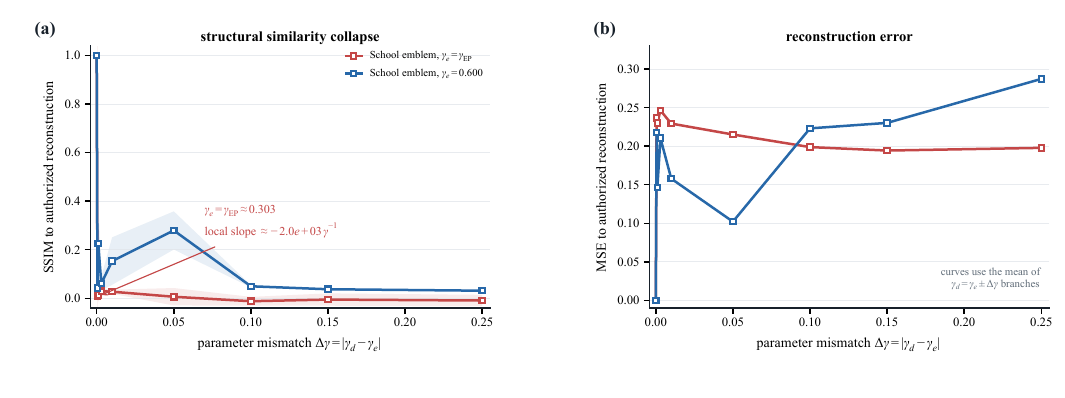}
\caption{Parameter-mismatch sensitivity of the biorthogonal decoding gate.
(a) Structural similarity index (SSIM) between the parameter-mismatched reconstruction and the authorized reconstruction as a function of $\Delta\gamma=|\gamma_d-\gamma_e|$. Near the finite-matrix exceptional point, $\gamma_e=\gamma_{\rm EP}\simeq0.303$, the SSIM collapses sharply under a small parameter deviation, revealing enhanced parameter sensitivity of the non-Hermitian decoding coordinate.
(b) Mean-square error (MSE) as a function of $\Delta\gamma$. The curves are obtained by averaging over the two mismatch branches $\gamma_d=\gamma_e\pm\Delta\gamma$. These results show that parameter-mismatched decoding is not merely a gradually degraded reconstruction channel, but is routed into an ill-conditioned inverse channel whose sensitivity is amplified near the exceptional point.}
\label{fig:param_mismatch}
\end{figure}

To quantify this decoding-coordinate sensitivity, Fig.~\ref{fig:param_mismatch} evaluates the local response of the mismatch channel using the structural similarity index (SSIM)~\cite{Wang2004TIP} and reconstruction error as the decoding coordinate deviates from the encoding coordinate. The exactly matched case reaches unit structural similarity and zero reconstruction error, whereas a small parameter detuning rapidly drives the reconstruction away from the authorized solution. The collapse is especially sharp near $\gamma_e=\gamma_{\rm EP}$, where the finite non-Hermitian response matrix becomes strongly ill-conditioned and the local slope of the structural-similarity curve reaches approximately $-2.0\times10^{3}\gamma^{-1}$. Such enhanced parametric response near an exceptional point is consistent with previous studies of EP-assisted sensitivity~\cite{Wiersig2014PRL,Chen2017Nature}; importantly, the associated nonorthogonality can also amplify noise and impose a practical sensitivity limit~\cite{Wang2020NatCommun}. These results show that the biorthogonal metric provides the physical-layer decoding gate, while the exceptional point defines a high-sensitivity region in which parameter mismatch and noise response are amplified in the finite operator space.

Taken together, these results complete the security evidence chain of the proposed paradigm. By forcing data into an incorrect manifold through parameter detuning, we ensure that the measurement matrix $C_L$ serves as a cryptographically bound cipher-text rather than a transparent signal. Unauthorized decoding performs a deterministic misdirection of information, effectively neutralizing unauthorized access at the physics layer. This concludes that the non-Hermitian biorthogonal encoding paradigm provides an inherent, hardware-level security defense, where the parameter $\gamma$ functions as an intrinsic token to protect against unauthorized data reconstruction.

Following the demonstration of the secure evidence chain under full-sampling, we now evaluate whether the security robustness of our paradigm persists in the partial-sampling regime. This assessment is critical for performance validation in bandwidth-constrained or noisy environments, where an attacker might only gain access to a truncated modal subset. As shown in Supporting Information, Sec.~V and Figs.~S12--S17, the authorized channel functions as a selective decryption filter, successfully recovering the dominant object structure from ordered modes. In contrast, both the naive and parameter-mismatched channels remain trapped in a state of high-fidelity distortion—an inherent consequence of wrong-metric crosstalk—even when provided with the identical modal subset. This transitions the system from an exact identity mapping to a truncated biorthogonal projection, confirming that decoding-basis consistency is the primary bottleneck for unauthorized extraction. The resulting enhanced sensitivity creates an information-theoretic gap, ensuring that the 'modal key' is rendered useless without underlying coordinate synchronization. Thus, our paradigm provides an inherent, hardware-layer defense, effectively neutralizing partial-data reconstruction attacks at the physical level.

\section{Experimental demonstration}

To verify the physical feasibility of the proposed decoding-selective protocol, we implemented the non-Hermitian biorthogonal encoding scheme on a passive SPI experimental platform. As shown in Fig.~\ref{fig:platform}(b), the hardware employs a standard intensity-only SPI architecture: light reflected from the target is relayed onto a digital micromirror device (DMD), spatially modulated by our pre-designed biorthogonal encoding masks, collected by a single-pixel detector (Thorlabs DET36A2), and digitized by a data-acquisition card (USB DAQ-610). Notably, this experimental setup does not require gain/loss media or specialized non-Hermitian materials; instead, the non-Hermitian operator space is synthesized through a library of programmable complex-valued masks, while physical-layer selectivity is implemented through the paired mechanism of left-basis encoding and right-basis computational decoding.

To implement this secure mapping within a standard SPI architecture, we employ a multi-frame differential measurement strategy, following the general principles of differential bucket detection and intensity-only implementation of complex-valued single-pixel sensing coefficients~\cite{Ferri2010PRL,Pastuszczak2016AO}. The DMD operates on a $1024\times768$ canvas, where each $64\times64$ computational mode is expanded into a $768\times768$ active region, corresponding to $12\times12$ micromirrors per image pixel. For a complete separable $64\times64$ basis, the protocol involves 4,096 two-dimensional modes. Given the DMD's limitation to binary amplitude modulation, each pre-normalization complex mode $\widetilde{P}_k^L$ is normalized as $P_k^L=\widetilde{P}_k^L/\alpha_k$, where $\alpha_k=\max_{x,y}|\widetilde{P}_k^L(x,y)|$. The normalized mode is written as $P_k^L=P_{R,k}+iP_{I,k}$ and decomposed into positive and negative components. Specifically, the real and imaginary parts are expressed as $P_{R,k}=P_{R,k}^{+}-P_{R,k}^{-}$ and $P_{I,k}=P_{I,k}^{+}-P_{I,k}^{-}$, respectively. These four non-negative components are sequentially displayed on the DMD. By collecting the corresponding four bucket signals ($B_{R,k}^{+}$, $B_{R,k}^{-}$, $B_{I,k}^{+}$, and $B_{I,k}^{-}$), the calibrated complex coefficient $c_{L,k}$ is obtained from the differential measurements [see the bottom of Fig.~\ref{fig:platform}(b)] as
\begin{equation}
c_{L,k}
=
\alpha_k
\left[
(B_{R,k}^{+}-B_{R,k}^{-})
+i(B_{I,k}^{+}-B_{I,k}^{-})
\right].
\end{equation}
Here, multiplication by $\alpha_k$ restores the mode-dependent scale removed during display normalization. The complete set of calibrated coefficients is then assembled into the measured coefficient matrix $C_L$ according to the fixed mode ordering. This differential strategy suppresses common-mode noise and circumvents the hardware constraint of binary-only modulation. The DMD update rate is set to $f_{\mathrm{DMD}}=20~\mathrm{Hz}$, and the detector output is digitized at $f_s=1000~\mathrm{samples/s}$. After excluding the transient response of the DMD, each bucket value is obtained by averaging the signal over a steady-state temporal window. Further details on mask-library construction, pattern normalization, four-mask differential acquisition, and the experimental reconstruction pipeline are provided in Sec.~IV of the Supporting Information.

\begin{figure}[!t]
\centering
\includegraphics[width=0.90\textwidth]{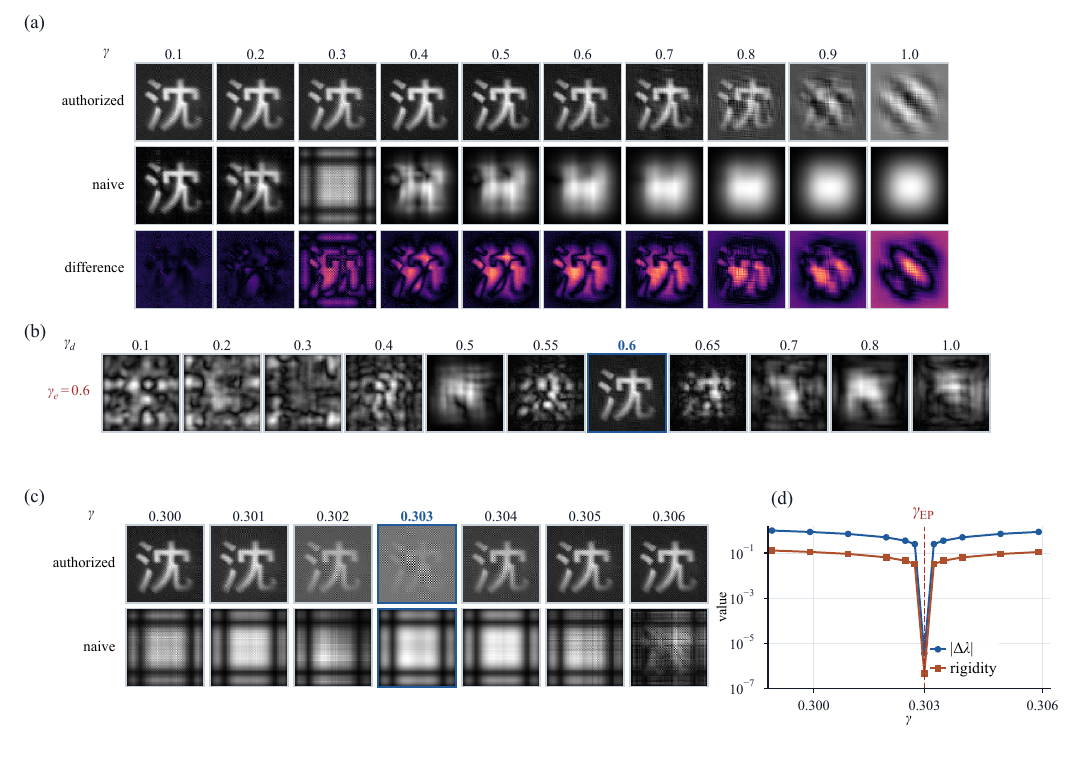}
\caption{Experimental validation of biorthogonal decoding selectivity and finite-matrix exceptional-point response.
(a) Passive SPI reconstructions of a reflective Chinese-character target under left-basis encoding for $\gamma=0.1$--$1.0$. Authorized dual-basis decoding preserves the target topology across the tested $\gamma$ values, whereas naive Hermitian back-projection produces $\gamma$-dependent wrong-metric leakage; the difference maps visualize the spatially structured divergence between the two channels.
(b) Decoding-coordinate scan at a fixed encoding parameter $\gamma_e=0.6$. The target is recovered only when $\gamma_d=\gamma_e$, demonstrating parameter-selective access to the matched non-Hermitian coordinate system.
(c) Fine $\gamma$ scan near the finite-matrix exceptional point. Around $\gamma_{\mathrm{EP}}\simeq0.303$, the authorized channel becomes more susceptible to perturbation-induced artifacts, while the naive channel is dominated by near-singular Gram-channel crosstalk.
(d) Eigenvalue splitting $|\Delta\lambda|$ and phase rigidity extracted from the finite $N=64$ operator, identifying $\gamma_{\mathrm{EP}}$ as a sensitivity boundary that amplifies both parameter selectivity and noise response.}
\label{fig:exp_ep}
\end{figure}

Figure~\ref{fig:exp_ep}(a) presents a comparative analysis between the authorized decoding channel and the naive (Hermitian) back-projection channel across a range of non-Hermitian operator parameters $\gamma$ (0.1 to 1.0). In the authorized channel, the consistent high-fidelity reconstruction of the Chinese-character target—maintained throughout the parameter range—validates the efficacy of the biorthogonal basis in capturing projected spatial information. Specifically, by aligning the decoding basis precisely with the encoding parameter $\gamma$, both the global topology and delicate stroke features are recovered. This alignment is critical: the parameter $\gamma$ defines the geometric structure of the non-Hermitian Hilbert space, acting as a physical-layer decryption key. Any deviation from the correct $\gamma$ value results in an orthogonal mismatch, effectively scrambling the projected information. In stark contrast, the naive channel, utilizing standard Hermitian back-projection, fails to properly resolve the target. As $\gamma$ increases, this channel exhibits severe signal degradation, manifesting as blurred, $\gamma$-dependent leakage patterns characterized by contrast loss, stroke distortion, and significant grid-like crosstalk. These artifacts confirm that without the precise knowledge of the encoding parameter $\gamma$, the measurement data cannot be coherently mapped back to the object space. Furthermore, the "difference" maps (bottom row) quantify this divergence, revealing that the discrepancy is not merely stochastic noise but a spatially structured corruption. This structured variance underscores how projecting onto an incorrect Hilbert space effectively "scrambles" the target information, providing visual confirmation of the protocol's physical-layer security. These experimental observations are consistent with the authorized--naive channel separation found in the numerical simulations.

Figure~\ref{fig:exp_ep}(b) illustrates the experimental performance of the physical decryption gate by scanning the decoding coordinate $\gamma_d$ against a fixed encoding coordinate $\gamma_e = 0.6$. The reconstruction results demonstrate an extreme sensitivity to the parameter match: a faithful and clear representation of the target is achieved exclusively when $\gamma_d = \gamma_e = 0.6$ (highlighted in blue). As $\gamma_d$ departs from the encoded value, the reconstructed image rapidly degrades into unintelligible artifacts and spatial noise. Even subtle deviations (e.g., $\gamma_d = 0.55$ or $0.65$) lead to a catastrophic collapse of the structural information, confirming that the information is "locked" within the specific Hilbert space defined by $\gamma_e$. This sharp transition behavior serves as a visual validation of the system's role as a physical-layer authentication mechanism, where the precise alignment of the dual-basis parameters acts as the essential cryptographic key for successful information retrieval.

Figures~\ref{fig:exp_ep}(c) and \ref{fig:exp_ep}(d) investigate the system response in the vicinity of the finite-matrix EP. For the discretized $N=64$ operator space, the tracked eigenvalue pair approaches an EP at $\gamma_{\mathrm{EP}} \simeq 0.303$, where the eigenvalue splitting $\vert{}\Delta\lambda\vert{}$ and phase rigidity simultaneously collapse, as shown in Fig. 4(d). In this high-sensitivity regime, the biorthogonal basis becomes severely ill-conditioned. Consequently, as illustrated in Fig.~\ref{fig:exp_ep}(c), the authorized reconstruction exhibits an increased susceptibility to experimental noise, while the naive branch displays heightened sensitivity to metric-mismatch artifacts. The experimental results reveal a "singular" degradation—rather than a gradual one—occurring precisely at the EP, where the structural fidelity of the authorized channel diminishes, and the naive channel’s crosstalk becomes significantly enhanced. These observations demonstrate that the observed image degradation is not merely a technical limitation, but a fundamental consequence of the conditioning of the finite non-Hermitian state space near the EP.

\begin{figure}[htbp]
\centering
\includegraphics[width=\textwidth]{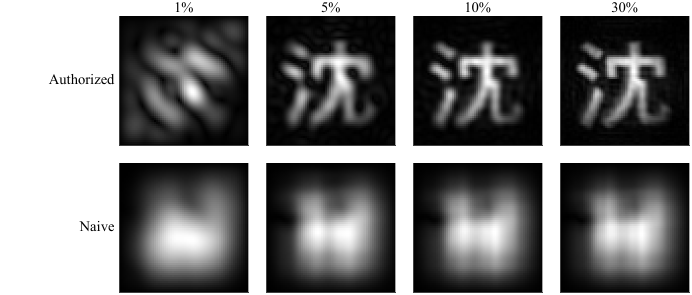}
\caption{Low-sampling experimental comparison between authorized and naive decoding. The upper and lower rows show reconstructions obtained using the authorized biorthogonal decoder and the naive adjoint decoder, respectively, at sampling ratios of $1\%$, $5\%$, $10\%$, and $30\%$.}
\label{fig:subsampling}
\end{figure}

Figure~\ref{fig:subsampling} demonstrates the robustness of our biorthogonal protocol under conditions of partial data acquisition, conducted at a fixed non-Hermitian control parameter of $\gamma=0.6$. The partial measurement datasets are generated by retaining specified fractions (1\%, 5\%, 10\%, and 30\%) of the complete, mode-ordered measurement sequence obtained from the full-basis acquisition. As illustrated in the top row of Fig. \ref{fig:subsampling}, even with severe sub-sampling, the authorized channel successfully reconstructs the target’s morphology. Mathematically, this resilience stems from the fact that the biorthogonal basis $\{\Phi_L(\gamma),\Psi_R(\gamma)\}$ satisfies the biorthogonality condition $\Phi_L(\gamma)\Psi_R(\gamma)=I_N$. Although sub-sampling renders the inverse problem formally underdetermined, the reconstruction $O_{\mathrm{auth}}^{(K)}
=
\Psi_R(\gamma)C_L^{(K)}\Psi_R^T(\gamma)$ effectively projects the sparse measurement coefficients $C_L^{(K)}$ onto the target-specific non-Hermitian manifold defined by the "basis-key" $\gamma$. In this framework, the biorthogonal projection acts as a physical regularization operator: it constrains the solution space to the manifold dictated by $\gamma$, allowing the system to recover the target image by leveraging the inherent structural correlation of the basis rather than requiring full-rank data acquisition. Crucially, this inherent underdetermined nature serves as an active layer of defense. For an unauthorized user, the absence of the correct $\gamma$ transforms the reconstruction into an ill-posed inverse problem with an infinite search space. Without the physical constraint of the non-Hermitian manifold, any attempt to resolve the target from partial data results in a degenerate solution dominated by structured noise. In stark contrast, the naive channel (bottom row) fails across all sampling ratios. Without the correct basis $\Psi_R(\gamma)$, the naive adjoint decoder performs a projection onto an incorrect Hilbert space, i.e., $O_{\mathrm{naive}}^{(K)}
=
\Phi_L^\dagger(\gamma)C_L^{(K)}\Phi_L^*(\gamma)$. Because $\Phi_L^\dagger(\gamma)\neq\Psi_R(\gamma)$ in our non-Hermitian system, the mismatch between the measurement and the decoding space results in featureless, $\gamma$-dependent leakage patterns rather than the target image. This confirms that the target information is not merely obscured but is intrinsically locked within the biorthogonal geometry. Consequently, the "basis-key" $\gamma$ is an indispensable requirement for decryption; the underdetermined nature of the acquisition process ensures that partial data provides no information advantage to unauthorized users, as they lack the physical "template" required to collapse the underdetermined solution space into the correct target morphology. Further details and experimental results on low-sampling biorthogonal projection are provided in the Supporting Information, Sec.~V.

\section{Discussion}
The non-Hermitian biorthogonal sensing architecture presented in this work fundamentally redefines the relationship between physical encoding and computational decoding. Unlike conventional Hermitian paradigms, where the observer and information are tethered by a shared adjoint metric—rendering the system transparent to any party capable of matrix inversion—our framework enforces a strict metric inequivalence. By mapping spatial information into a biorthogonal operator space, we transform the imaging process from a passive software-inversion task into an active, hardware-level projection governed by distinct left and right eigenbases. 
At the core of this paradigm is the "basis-key" $\gamma$, which functions as an intrinsic physical token. Our analysis confirms that the security of this protocol is an emergent property of the system's topological geometry. When the decoding basis matches the encoding basis ($\gamma_d = \gamma_e$), the biorthogonal completeness relation ensures high-fidelity identity reconstruction. Conversely, any parameter detuning forces the information into a mismatched manifold, resulting in deterministic, structured crosstalk that effectively masks the target. Notably, the underdetermined nature of the measurement process under partial data acquisition—e.g., retaining 1\%, 5\%, 10\%, or 30\% of the data—serves as an active defense. The absence of the correct basis $\Psi_R(\gamma)$ transforms the inverse problem into an ill-posed search within an infinite Hilbert space. Without the physical "template" provided by $\gamma$, unauthorized users cannot collapse the solution space into the correct target morphology, rendering partial data acquisition completely uninformative.

Beyond its immediate security applications, the topological nature of our non-Hermitian biorthogonal framework offers a transformative approach to imaging through scattering media, where wavefront control is conventionally used to compensate for complex multiple scattering~\cite{Vellekoop2007OptLett}. In traditional systems, scattering events typically decohere the wavefront, leading to a catastrophic loss of information. However, the biorthogonal manifold—characterized by its unique non-Hermitian spectral properties—exhibits robustness against certain classes of spatial perturbations. We anticipate that by embedding the "biorthogonal footprint" into the scattering field, it may be possible to establish a "topological transmission channel." In this scenario, the biorthogonal eigenmodes could potentially act as robust carriers, where the spatial information is protected by the geometric phase of the non-Hermitian system, partially mitigating the scrambling effects of random scattering media.

Furthermore, this approach necessitates future experimental verification across diverse computational imaging architectures. Key challenges include: (1) Paradigm Generalization: Evaluating the versatility of this biorthogonal framework across varying modalities, such as high-throughput holographic, hyperspectral, or dynamic compressive sensing systems, to determine the universality and scalability of the non-Hermitian projection; (2) Information Retrieval Limits: Characterizing the fundamental trade-off between the security-induced "biorthogonal footprint" and the reconstruction fidelity, specifically regarding how signal-to-noise ratios degrade when operating near the information-theoretic limit of the biorthogonal manifold; and (3) Adversarial Resilience: Quantifying the system’s boundary conditions against advanced computational attacks, such as iterative blind deconvolution or machine-learning-based optimization, that attempt to reconstruct target information from deterministic leakage patterns. 
These investigations will define the operational limits of our biorthogonal architecture and validate its robustness for deployment in non-cooperative, complex physical environments. Ultimately, by transitioning from static reconstruction to geometry-aware computational imaging, this paradigm establishes a foundational shift in sensing technology. We envision a future where confidentiality and signal robustness are no longer extrinsic computational overheads, but are intrinsic, inseparable properties of the imaging architecture itself. By embedding these functionalities directly into the system's biorthogonal design, we move toward a new generation of sensors that achieve high-fidelity reconstruction even in the most challenging and non-cooperative physical conditions.

\section{Conclusion}
In conclusion, we have proposed and experimentally demonstrated a generalized secure computational imaging framework based on non-Hermitian biorthogonal symmetry breaking. Taking a passive SPI platform as a practical exemplar, we have shown that spatial information can be mapped into a biorthogonal operator space through left-basis encoding and retrieved solely via matched right-basis dual decoding. Within this architecture, the non-Hermitian control parameter $\gamma$ acts as an intrinsic hardware-level security key that defines the measurement manifold, establishing a deterministic physical-layer cryptographic gate. The authorized channel ensures an identity reconstruction pathway, whereas any parameter-mismatched decoding forces the data into deterministic cross-channel crosstalk, effectively rendering the information cryptographically useless and inherently suppressing unauthorized access. Furthermore, this architecture serves as a universal protocol that intrinsically supports direct, iteration-free image retrieval across a wide range of sampling ratios, enabling high-throughput imaging with significantly reduced computational overhead compared to conventional iterative methods. Beyond the specific SPI implementation, these results establish non-Hermitian left-right asymmetry as a programmable physical resource. This framework facilitates a fundamental paradigm shift, transforming computational imaging from a software-dependent post-processing task into a parameter-sensitive physical decryption process, thereby offering a transformative path toward next-generation sensing systems that encapsulate functional intelligence directly within the physics of measurement.

\section*{Conflicts of Interest}
The authors declare no conflicts of interest.

\section*{Data Availability Statement}
The data that support the findings of this study are available from the corresponding author upon reasonable request.

\renewcommand{\refname}{References}

\clearpage
\phantomsection
\begin{center}
{\LARGE\bfseries Supporting Information\par}
\vspace{0.75em}
{\Large A Non-Hermitian Biorthogonal Encoding Paradigm for Physical-Layer Secure Computational Imaging\par}
\vspace{0.75em}
{\normalsize Xi-Hao Chen, Kan-Xu Jia, En-Rui Zhang, Yi-Zhu Zhang, Xin-Peng Wei, Bu-Ran Yu, Qian-Qian Bao, and Shao-Ying Meng\par}
\vspace{0.4em}
{\small Key Laboratory of Optoelectronic Devices and Detection Technology, School of Physics, Liaoning University, Shenyang 110036, China\par}
{\small Corresponding author: Shao-Ying Meng (\href{mailto:mengshaoying@163.com}{mengshaoying@163.com})\par}
\end{center}
\vspace{1em}

\setcounter{section}{0}
\setcounter{subsection}{0}
\setcounter{figure}{0}
\setcounter{table}{0}
\setcounter{equation}{0}
\renewcommand{\thesection}{\Roman{section}}
\renewcommand{\thesubsection}{\thesection.\arabic{subsection}}
\renewcommand{\thefigure}{S\arabic{figure}}
\renewcommand{\thetable}{S\arabic{table}}
\renewcommand{\theequation}{S\arabic{equation}}
\renewcommand{\theHsection}{SI.\arabic{section}}
\renewcommand{\theHsubsection}{SI.\arabic{section}.\arabic{subsection}}
\renewcommand{\theHfigure}{SI.\arabic{figure}}
\renewcommand{\theHtable}{SI.\arabic{table}}
\renewcommand{\theHequation}{SI.\arabic{equation}}

\section{Theoretical Framework and Biorthogonal Imaging Channels}
\label{sec:supp_nh_basis_channels}

This section establishes the operator framework underlying the encoding and decoding channels used in the main text. We first derive the continuous left and right eigenmodes, then construct their finite-dimensional representation and fixed modal ordering, and finally use the singular-value spectrum to explain the distinct leakage morphologies produced by naive left- and right-basis decoding.
\subsection{Theoretical derivation}

We first formulate the imaging basis in the continuous spatial Hilbert space $\mathcal L^2(\mathbb R)$, the space of square-integrable functions on the real line. In dimensionless units, the one-dimensional operator is
\begin{equation}
\hat H_{\gamma}=-\frac{1}{2}\frac{d^2}{dx^2}+\frac{x^2}{2}+i\gamma x ,
\end{equation}
where $x\in\mathbb R$ is the spatial coordinate, $d^2/dx^2$ is the second-derivative operator, $i^2=-1$, and $\gamma\in\mathbb R$ controls the strength of the imaginary linear potential. The first two terms are the kinetic and harmonic-confinement contributions, respectively, whereas $i\gamma x$ introduces the left-right asymmetry used by the imaging protocol. For $\gamma\neq0$, $\hat H_{\gamma}\neq \hat H_{\gamma}^{\dagger}$, so its right and left eigenstates must be defined separately~\cite{SI-Garrison1988PLA,SI-Mostafazadeh2002JMP,SI-Brody2014JPA}:
\begin{equation}
\hat H_{\gamma}|\psi_n\rangle=\lambda_n|\psi_n\rangle,\qquad
\langle\phi_n|\hat H_{\gamma}=\lambda_n\langle\phi_n|.
\end{equation}
Here $n=0,1,2,\ldots$ is the modal index, $\lambda_n$ is the corresponding eigenvalue, $|\psi_n\rangle$ is a right eigenstate, and $\langle\phi_n|$ is its paired left eigenstate. The dagger denotes the Hermitian adjoint. Although a general non-Hermitian operator may have complex eigenvalues, the continuous shifted oscillator considered here has the real spectrum derived below.

Introducing the complex coordinate $z=x+i\gamma$ converts the potential into a shifted harmonic oscillator,
\begin{equation}
\frac{x^2}{2}+i\gamma x=\frac{(x+i\gamma)^2}{2}+\frac{\gamma^2}{2}.
\end{equation}
In the coordinate representation, the right eigenstate takes the analytically continued Hermite--Gaussian form
\begin{equation}
\psi_n(x;\gamma)=\mathcal N_n^{\psi} H_n(x+i\gamma)
\exp\left[-\frac{(x+i\gamma)^2}{2}\right],
\end{equation}
with eigenvalues
\begin{equation}
\lambda_n=n+\frac{1}{2}+\frac{\gamma^2}{2}.
\label{eq:supp_continuous_spectrum}
\end{equation}
Here $H_n$ is the physicists' Hermite polynomial of order $n$, and $\mathcal N_n^{\psi}$ is a normalization constant. The complex shift does not destroy square integrability: for real $\gamma$, the Gaussian magnitude is proportional to $\exp(-x^2/2)$, while $H_n(x+i\gamma)$ grows only polynomially. Hence $\psi_n\in\mathcal L^2(\mathbb R)$ despite its complex argument~\cite{SI-Ahmed2001PLA}.

The relation between the two mode families is fixed by parity rather than by an informal ``counterpart'' identification. Let $\hat P$ be the parity operator, $(\hat P f)(x)=f(-x)$. For real $\gamma$,
\begin{equation}
\hat H_{\gamma}^{\dagger}
=
\hat H_{-\gamma}
=
\hat P\hat H_{\gamma}\hat P .
\label{eq:supp_continuous_parity_relation}
\end{equation}
Consequently, the adjoint eigenket associated with $\langle\phi_n|$ can be chosen as $|\phi_n(\gamma)\rangle\propto\hat P|\psi_n(\gamma)\rangle$, or equivalently as the $\gamma\rightarrow-\gamma$ continuation of the right eigenmode. Its coordinate wavefunction $\phi_n(x;\gamma)=\langle x|\phi_n\rangle$ is
\begin{equation}
\phi_n(x;\gamma)=\mathcal N_n^{\phi} H_n(x-i\gamma)
\exp\left[-\frac{(x-i\gamma)^2}{2}\right].
\end{equation}
The parity factor $(-1)^n$ has been absorbed into the normalization constant $\mathcal N_n^{\phi}$. The same Gaussian argument shows that $\phi_n\in\mathcal L^2(\mathbb R)$.

The constants $\mathcal N_n^{\psi}$ and $\mathcal N_n^{\phi}$ are chosen jointly, rather than as two independent Euclidean normalizations, so that the left and right families satisfy
\begin{equation}
\langle\phi_m|\psi_n\rangle=\delta_{mn}.
\label{eq:supp_continuous_biorthogonality}
\end{equation}
Here $\delta_{mn}$ is the Kronecker delta. In coordinate form, the overlap is $\int_{-\infty}^{\infty}\phi_m^*(x;\gamma)\psi_n(x;\gamma)\,dx$, and the normalization condition can be written as
\begin{equation}
(\mathcal N_m^{\phi})^*\mathcal N_n^{\psi}
\int_{-\infty}^{\infty}
H_m(x+i\gamma)H_n(x+i\gamma)
e^{-(x+i\gamma)^2}\,dx
=
\delta_{mn}.
\label{eq:supp_continuous_overlap_integral}
\end{equation}
The integrand is entire, and its Gaussian decay allows the contour to be shifted from $\mathbb R+i\gamma$ back to the real axis. The usual Hermite orthogonality therefore remains valid under this analytic continuation; for example, one may choose $\mathcal N_n^{\phi}=\mathcal N_n^{\psi}=(2^n n!\sqrt{\pi})^{-1/2}$ with a consistent phase convention~\cite{SI-Ahmed2001PLA}.

These two mode families define different but mutually dual encoding and decoding spaces. Physical encoding by one family is inverted by projection onto its paired family through the biorthogonal metric above. An ordinary same-side Hermitian back-projection does not implement this inverse. This mismatch between the physical encoding basis and a Hermitian decoding basis is precisely what necessitates the biorthogonal metric encoding framework developed in the main text~\cite{SI-Brody2014JPA}.

\subsection{Finite matrix representation}

The continuous solution establishes the left-right pairing, but the simulations and displayed modulation patterns require a finite pixel basis. We therefore sample the dimensionless coordinate on a window of width $\ell$ using $N$ grid points,
\begin{equation}
x_j=-\frac{\ell}{2}+(j-1)\Delta x,\qquad
\Delta x=\frac{\ell}{N-1},\qquad j=1,\ldots,N .
\end{equation}
Here $x_j$ is the coordinate of sample $j$, $\Delta x$ is the grid spacing, and $N$ is both the one-dimensional basis size and the matrix dimension. For a sampled field vector $\bm u=(u_1,\ldots,u_N)^T$, where $T$ denotes ordinary transpose, each $u_j$ represents the complex amplitude sampled at $x_j$, effectively decoupling the binary amplitude modulation components. The second derivative is approximated by the three-point, second-order finite-difference matrix~\cite{SI-Fornberg1988MathComp}. As this is a finite-matrix approximation, the resulting operator constitutes a discrete non-Hermitian system
\begin{equation}
D_2
=
\frac{1}{\Delta x^2}
\operatorname{tridiag}(1,-2,1),
\qquad
(D_2\bm u)_j
=
\frac{u_{j+1}-2u_j+u_{j-1}}{\Delta x^2}.
\label{eq:supp_second_difference}
\end{equation}
Here $\text{tridiag}(1, -2, 1)$ denotes the $N \times N$ matrix with $-2$ on the main diagonal and $1$ on the adjacent diagonals. The first and last stencil rows use a homogeneous Dirichlet closure: the unavailable samples immediately outside the retained grid are set to zero. Stating this closure is essential because the boundary treatment is part of the finite operator and affects its high-order spectrum. Consequently, this specific boundary-informed construction is integral to the non-Hermitian nature of the discrete operator within our finite-matrix framework.

The resulting $N\times N$ matrix representation of the continuous operator is

\begin{equation}
\label{eq:finite_HN}
H_N(\gamma)=-\frac{1}{2}D_2+
\operatorname{diag}\left(\frac{x_j^2}{2}+i\gamma x_j\right).
\end{equation}
Here $H_N(\gamma)\in\mathbb C^{N\times N}$ denotes a matrix, rather than the infinite-dimensional operator $\hat H_{\gamma}$; $\operatorname{diag}(\cdot)$ places the $N$ sampled potential values on the main diagonal. We retain the symbol without a hat throughout the finite-dimensional analysis to make this distinction explicit.

Unless otherwise stated, all reported calculations use $N=64$ and $\ell=6$. These are protocol parameters, not neutral numerical settings: the finite window, grid spacing, and boundary closure modify the eigenvalues, eigenvectors, Gram matrices, and conditioning. In particular, the continuous shifted oscillator has the equally spaced spectrum in ~\eqref{eq:supp_continuous_spectrum} and no eigenvalue coalescence for real $\gamma$~\cite{SI-Ahmed2001PLA}, whereas its finite matrix representation can exhibit pair-dependent exceptional points generated by truncation and boundary effects. The reported exceptional-point location is strictly a property of the specified $N = 64, \ell = 6$ discrete operator, arising from truncation and boundary effects, and should not be conflated with any intrinsic feature of the infinite-domain analytic model.

\begin{figure}[htbp]
\centering
\includegraphics[width=0.92\textwidth]{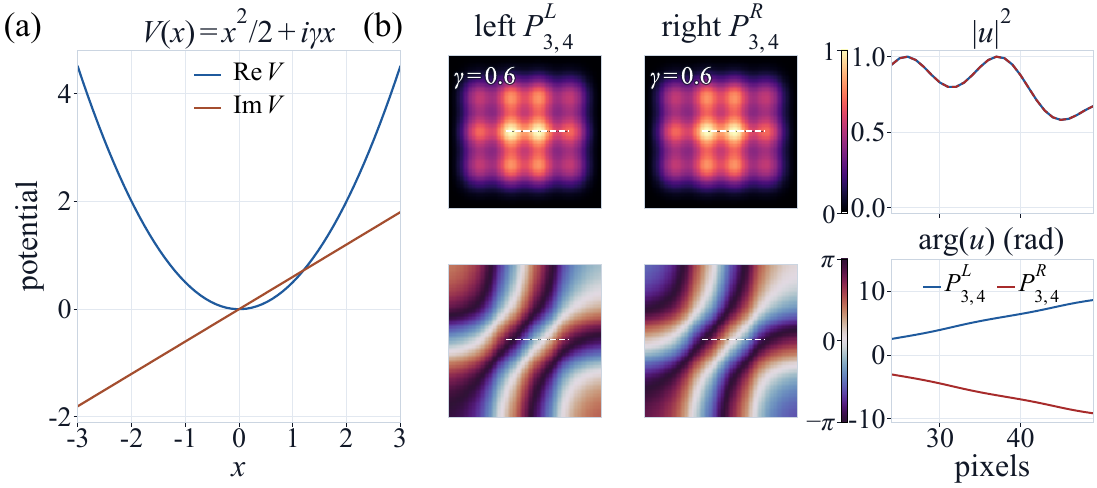}
\caption{Finite non-Hermitian potential and paired complex basis patterns.
(a) Real and imaginary parts of the finite non-Hermitian potential $V(x)=x^2/2+i\gamma x$.
(b) Representative left and right complex basis patterns for $N=64$ and $\gamma=0.6$, shown through intensity and phase maps, together with line cuts of $|u|^2$ and $\arg(u)$, where $u$ denotes the complex modal amplitude.}
\label{fig:s1_modes_dmd}
\end{figure}

The implementation of the biorthogonal imaging protocol relies on a discrete non-Hermitian operator, whose complex potential landscape and resulting modal structure are visualized in Fig.~\ref{fig:s1_modes_dmd}. Panel (a) plots the real and imaginary parts of the sampled complex potential $V(x)=x^2/2+i\gamma x$. Panel (b) shows the intensity and phase maps of a representative paired left and right mode for $N=64$ and $\gamma=0.6$, together with one-dimensional line cuts. The real quadratic potential provides spatial confinement, whereas the imaginary linear term imposes the $\gamma$-dependent left-right asymmetry. The paired patterns consequently retain similar intensity envelopes but acquire different phase structures. This phase-induced asymmetry must be compensated by the biorthogonal metric to ensure accurate signal decoding. Thus, the figure illustrates the mapping between the theoretical non-Hermitian operator and the spatially discretized complex basis patterns implemented in the experimental imaging protocol.
For the finite matrix, we relabel the $N$ retained modes by $n=1,\ldots,N$. Let the right eigenvectors be arranged as columns of $\Psi_R$ and the left eigenvectors as rows of $\Phi_L$:
\begin{equation}
\Psi_R=(|\psi_1\rangle,\ldots,|\psi_N\rangle),\qquad
\Phi_L=
\begin{bmatrix}
\langle\phi_1|\\
\vdots\\
\langle\phi_N|
\end{bmatrix}.
\end{equation}
Here $\Psi_R,\Phi_L\in\mathbb C^{N\times N}$ contain the complete ordered one-dimensional bases. After each left-right eigenvector pair has been scaled according to ~\eqref{eq:supp_continuous_biorthogonality}, the matrices satisfy
\begin{equation}
\Phi_L\Psi_R=I_N,\qquad \Psi_R\Phi_L=I_N .
\end{equation}
The symbol $I_N$ denotes the $N\times N$ identity matrix. These equalities express completeness and biorthogonal inversion for the full square basis: $\Phi_L=\Psi_R^{-1}$. They do not imply ordinary Hermitian orthogonality within either mode family~\cite{SI-Brody2014JPA}. To expose that distinction, we define the same-side Hermitian Gram matrices
\begin{equation}
S_L
\equiv
\Phi_L\Phi_L^\dagger
\neq I_N,
\qquad
S_R
\equiv
\Psi_R^\dagger\Psi_R
\neq I_N .
\label{eq:supp_same_side_gram}
\end{equation}
Here $\dagger$ denotes conjugate transpose, while $S_L$ and $S_R$ quantify the ordinary within-family overlaps of the left and right bases, respectively. Equivalently, for modal indices $i,j=1,\ldots,N$,
\begin{equation}
(\Phi_L\Phi_L^\dagger)_{ij}
\not\equiv
\delta_{ij},
\qquad
(\Psi_R^\dagger\Psi_R)_{ij}
\not\equiv
\delta_{ij},
\qquad
i,j=1,\ldots,N .
\label{eq:supp_same_side_gram_components}
\end{equation}
Only in the Hermitian limit, where the paired left basis becomes the Hermitian adjoint of the right basis, do both same-side matrices reduce to the identity. The finite basis used here is therefore complete and biorthogonal, but neither family is generally orthonormal under the ordinary Hermitian inner product. The off-diagonal entries of $S_L$ and $S_R$ quantify the deterministic within-family cross-mode couplings associated with naive same-side decoding.

To separate the reconstruction effects of same-side nonorthogonality from numerical inaccuracies in the paired inverse, we evaluate the biorthogonality error as
\begin{equation}
\varepsilon_{\rm bio}=\|\Phi_L\Psi_R-I_N\|_F .
\end{equation}
Here, $\|\cdot\|_F$ denotes the Frobenius matrix norm, and $\varepsilon_{\rm bio}$ measures the accumulated departure from exact biorthogonality. For $N=64$, $\varepsilon_{\rm bio}$ remains near numerical precision over the $\gamma$ range used in the main text. The authorized inverse is therefore numerically well defined, whereas the artifacts in naive decoding arise from applying a same-side Hermitian metric rather than from a failure of the paired inverse.

\begin{figure}[htbp]
\centering
\includegraphics[width=1\textwidth]{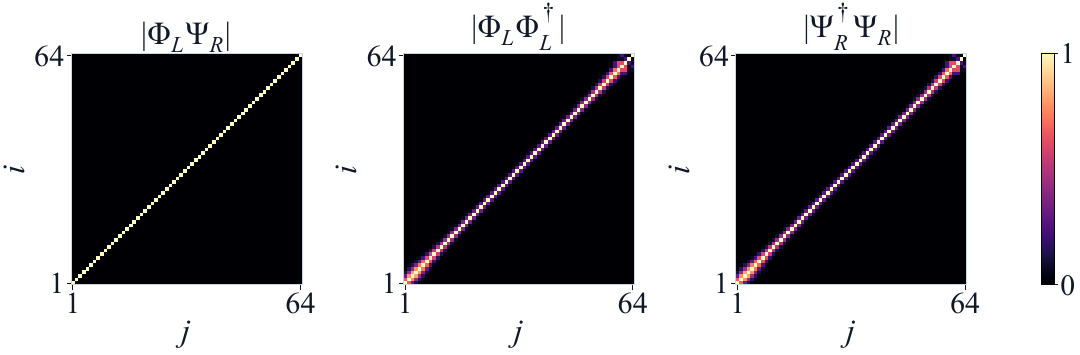}
\caption{Metric structure of the finite non-Hermitian basis for $N=64$. The left panel shows the matched left--right product $|\Phi_L\Psi_R|$, whereas the middle and right panels show the same-side Gram matrices $|\Phi_L\Phi_L^\dagger|$ and $|\Psi_R^\dagger\Psi_R|$, respectively. The color scale spans matrix magnitudes from 0 to 1.}
\label{fig:s2_metric}
\end{figure}

Fig~\ref{fig:s2_metric} provides a direct matrix-level comparison of paired left--right biorthogonality and same-side modal overlaps in the finite basis. Consistent with the small $\varepsilon_{\rm bio}$ reported above, the identity-like diagonal structure of $|\Phi_L\Psi_R|$ confirms the paired biorthogonal relation between $\Phi_L$ and $\Psi_R$. In contrast, the prominent off-diagonal elements in the same-side Gram matrices reveal finite intermodal overlaps within each family. Consequently, neither family alone forms an orthonormal basis under the standard Hermitian inner product. Together, however, they form a complete biorthogonal pair through their mutual duality.

This distinction between paired and same-side metrics directly determines the subsequent reconstruction behavior. Authorized left--right decoding follows the near-identity channel and preserves the correspondence between modal coefficients and their paired spatial modes. Naive same-side back-projection instead routes the coefficients through non-identity Gram channels, whose off-diagonal elements introduce deterministic cross-mode coupling. This metric mismatch explains the later differences between authorized and naive reconstructions.

\subsection{Mode ordering}

The finite matrix supplies a complete modal set, but its numerical eigensolver does not assign a physically meaningful order to that set. A fixed order is nevertheless required to generate reproducible pattern libraries, preserve left--right pairing, and define which modes are retained under partial acquisition. Let $\Psi_R^{(0)}\in\mathbb C^{N\times N}$ and $\Phi_L^{(0)}\in\mathbb C^{N\times N}$ denote the paired right- and left-basis matrices before ordering, with $\Phi_L^{(0)}\Psi_R^{(0)}=I_N$. If $F_{\mathrm{ord}}\in\mathbb R^{N\times N}$ is the permutation matrix that implements the selected one-dimensional order, the ordered matrices are
\begin{equation}
\Psi_R=\Psi_R^{(0)}F_{\mathrm{ord}},\qquad
\Phi_L=F_{\mathrm{ord}}^T\Phi_L^{(0)}  .
\end{equation}
Here, each column of $F_{\mathrm{ord}}$ contains one unit entry,
$F_{\mathrm{ord}}^T F_{\mathrm{ord}}=I_N$, and the superscript $T$ denotes the ordinary transpose. Applying the inverse permutation to the rows of the left basis preserves the original eigenvalue matching and hence the biorthogonal relation,
\begin{equation}
\Phi_L\Psi_R
=F_{\mathrm{ord}}^T
\Phi_L^{(0)}\Psi_R^{(0)}
F_{\mathrm{ord}}
=I_N .
\end{equation}
The authorized, naive, and $\gamma$-mismatched reconstructions all use the same paired order. For full sampling, a simultaneous permutation of the encoding and decoding libraries leaves the reconstructed image unchanged. For partial sampling, however, the order determines which coefficients are acquired. We therefore rank each separable two-dimensional mode by the real energy coordinate $\eta_{i_y i_x}
=
\operatorname{Re}\lambda_{i_y}
+
\operatorname{Re}\lambda_{i_x}$ and retain modes in ascending $\eta_{i_y i_x}$, with stable index ordering used to break exact ties. This two-dimensional acquisition rule is defined explicitly in Sec.~\ref{sec:partial_sampling}. Thus, an alternative order would define a different partial-acquisition protocol rather than a harmless post-processing convention.

\subsection{Singular-value origin of naive-decoding distortions}\label{sec:supp_singular_value}

To explain the distinct leakage morphologies generated by naive
decoding, we analyze the singular-value spectrum of the ordered and
biorthogonally normalized finite basis. The same biorthogonal pair
supports two reciprocal imaging configurations. In the left-encoding
configuration, the object is encoded with the left basis and decoded
with the paired right basis. In the reciprocal configuration, the
object is encoded with the right basis and decoded with the paired left
basis. We denote the corresponding reconstruction channels by the
superscripts $(L)$ and $(R)$ for left- and right-basis encoding,
respectively.

The ordered matrices $\Phi_L$ and $\Psi_R$ collect the left eigenbras
as rows and their paired right eigenkets as columns, respectively. The
left-encoding branch coincides with the protocol adopted in the main
text, so
$O_{\rm auth}^{(L)}\equiv O_{\rm auth}$ and
$O_{\rm naive}^{(L)}\equiv O_{\rm naive}$.
For the complete finite basis, biorthogonality and completeness give
\begin{equation}
\Phi_L\Psi_R=\Psi_R\Phi_L=I_N,
\qquad
\Phi_L=\Psi_R^{-1},
\end{equation}
where $I_N$ denotes the $N\times N$ identity matrix.

This inverse relation allows the two reciprocal channels to be
compared through a single singular-value decomposition. The
singular-value decomposition of $\Psi_R$~\cite{SI-TrefethenBau2022} is
\begin{equation}
\Psi_R=U\Sigma W^\dagger,
\qquad
\Sigma=
\operatorname{diag}
(\sigma_1,\sigma_2,\ldots,\sigma_N).
\end{equation}
The matrices $U,W\in\mathbb C^{N\times N}$ contain the left and right
singular vectors of $\Psi_R$, respectively. The matrix
$\Sigma\in\mathbb R^{N\times N}$ is diagonal, and its entries
$\sigma_i>0$ are the singular values. The dagger denotes the Hermitian
adjoint. The paired left-basis matrix is therefore
\begin{equation}
\Phi_L
=
\Psi_R^{-1}
=
W\Sigma^{-1}U^\dagger.
\end{equation}

In the Hermitian limit, the basis can be chosen unitary, all singular
values satisfy $\sigma_i=1$, and the Hermitian adjoint coincides with
the biorthogonal dual. In the non-Hermitian case, the singular-value
spectrum is generally nonuniform. Replacing the paired dual basis with
a same-side Hermitian back-projection therefore introduces
deterministic modal reweighting.

Let $O\in\mathbb R^{N\times N}$ denote the sampled object. The
coefficient matrices of the two reciprocal encoding branches are
defined as
\begin{equation}
C_L
=
\Phi_L O\Phi_L^T,
\qquad
C_R
=
\Psi_R^T O\Psi_R.
\end{equation}
Here, $C_L,C_R\in\mathbb C^{N\times N}$, and the superscript $T$
denotes transpose without complex conjugation. The matrix $C_L$
corresponds to the left-basis encoding used in the main text, whereas
$C_R$ corresponds to reciprocal right-basis encoding.

Applying the paired decoding basis in each branch gives
\begin{equation}
\begin{aligned}
O_{\rm auth}^{(L)}
&=
\Psi_R C_L\Psi_R^T
=
O,\\
O_{\rm auth}^{(R)}
&=
\Phi_L^T C_R\Phi_L
=
O.
\end{aligned}
\end{equation}
Both equalities follow from $\Psi_R\Phi_L=I_N$ and its transpose.
Thus, the two reciprocal configurations produce the same identity
channel under paired decoding.

For the left-encoding branch, the same-side Hermitian back-projection
uses $\Phi_L^\dagger$ along the first spatial axis and $\Phi_L^*$ along
the second. Here, $*$ denotes elementwise complex conjugation. The
resulting naive reconstruction is
\begin{equation}
O_{\rm naive}^{(L)}
=
\Phi_L^\dagger C_L\Phi_L^*
=
G_L O G_L^T,
\end{equation}
where
\begin{equation}
G_L
=
\Phi_L^\dagger\Phi_L
=
U\Sigma^{-2}U^\dagger.
\end{equation}
The matrix $G_L\in\mathbb C^{N\times N}$ is the left same-side Gram
operator. It weights the $i$th singular direction by $\sigma_i^{-2}$.
Directions associated with small singular values are therefore
amplified. Their overlap with the object spectrum produces mode mixing
and localized leakage, often at high spatial frequencies.
\begin{figure}[htbp]
\centering
\includegraphics[width=1\textwidth]{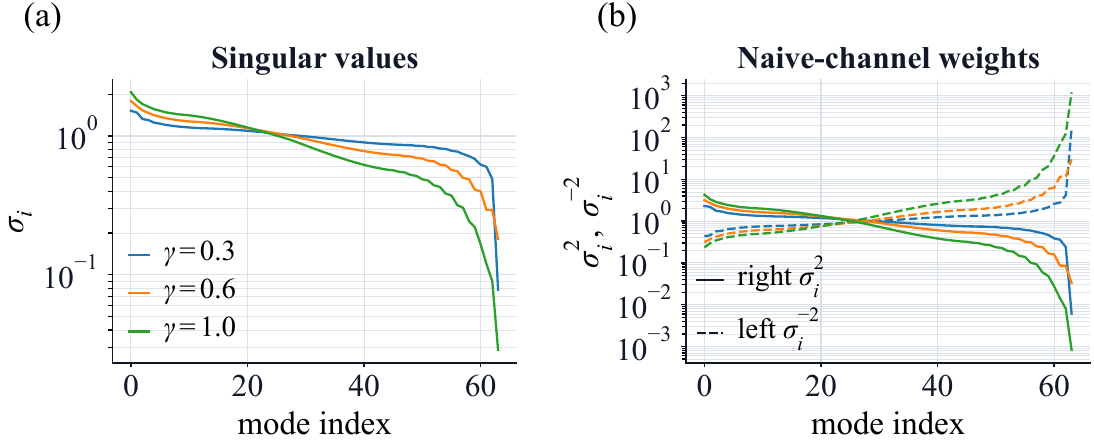}
\caption{Singular-value origin of the naive-decoding distortions. (a) Ordered singular-value spectra $\sigma_i$ of the $N=64$, $\ell=6$ finite right basis for representative values of $\gamma$, highlighting the pivot behavior across mode indices. (b) Corresponding single-axis weights of the two same-side Gram channels, where the left-naive channel carries $\sigma_i^{-2}$ with exponential amplification at high modes, and the right-naive channel carries $\sigma_i^{2}$ with progressive suppression; the separable two-dimensional weights are products of these one-axis factors.}
\label{fig:s3_singular_weights}
\end{figure}
For the reciprocal right-encoding branch, the same-side Hermitian
back-projection gives
\begin{equation}
O_{\rm naive}^{(R)}
=
\Psi_R^* C_R\Psi_R^\dagger
=
Q_R^T O Q_R,
\end{equation}
where
\begin{equation}
Q_R
=
\Psi_R\Psi_R^\dagger
=
U\Sigma^{2}U^\dagger.
\end{equation}
The matrix $Q_R\in\mathbb C^{N\times N}$ is the right same-side Gram
operator. It weights the same singular directions by $\sigma_i^{2}$.
Small-singular-value components are therefore suppressed rather than
amplified. For the tested targets, this suppression removes fine
spatial details and leaves predominantly low-order, contour-like
remnants. Note that the earlier matrices $S_L=\Phi_L\Phi_L^\dagger$ and $S_R=\Psi_R^\dagger\Psi_R$ quantify same-family modal overlaps, whereas $G_L=\Phi_L^\dagger\Phi_L$ and $Q_R=\Psi_R\Psi_R^\dagger$ are the Gram operators that act directly on the object in the corresponding naive reconstruction channels.
The two same-side Gram operators are exact inverses. From their
singular-value representations,
\begin{equation}
G_LQ_R=Q_RG_L=I_N,
\qquad
G_L=Q_R^{-1}.
\end{equation}
The left- and right-naive branches therefore do not represent two
unrelated reconstruction errors. Instead, they apply reciprocal metric
filters along the same singular directions: a direction weighted by
$\sigma_i^{-2}$ in the left-naive channel is weighted by $\sigma_i^{2}$
in the right-naive channel. The distinct leakage morphologies thus
originate from two inverse wrong-metric transformations of the same
biorthogonal basis.

Because the two-dimensional image basis is separable, the one-dimensional modal weights act independently along the row and column axes. Each two-dimensional mode is a tensor product of the $i$th one-dimensional singular mode along the rows and the $j$th one-dimensional singular mode along the columns, and is therefore indexed by the pair $(i,j)$. The resulting modal weights for the two naive reconstruction channels are
\begin{equation}
\mathcal{W}_{ij}^{(L)}
=
\sigma_i^{-2}\sigma_j^{-2},
\qquad
\mathcal{W}_{ij}^{(R)}
=
\sigma_i^{2}\sigma_j^{2}.
\end{equation}
Here, $\mathcal{W}_{ij}^{(L)}$ and $\mathcal{W}_{ij}^{(R)}$ denote
the modal weights of the left-naive and right-naive Hermitian
back-projection channels, respectively.
Figure~\ref{fig:s3_singular_weights} illustrates the quantitative behavior of the singular-value and weight spectra: panel~(a) shows the ordered singular-value spectra $\{\sigma_i\}_{i=1}^{N}$ for representative values of $\gamma$, revealing a distinct crossover pivot where increasing $\gamma$ slightly enhances low-order components while causing a sharp, accelerated drop-off at high mode indices. Panel~(b) compares the corresponding single-axis weights, demonstrating that the left-naive channel ($\sigma_i^{-2}$, dashed curves) suffers from several-orders-of-magnitude runaway amplification as the mode index approaches $N=64$ (surging past $10^3$). This severe high-index over-weighting directly accounts for the localized high-frequency leakage observed in left-naive reconstructions. Conversely, the right-naive channel ($\sigma_i^2$, solid curves) aggressively quenches these high-order components down toward zero, leaving predominantly low-order structural remnants.

The reciprocal trends in panel (b) show that the two naive channels do not generate unstructured numerical error. They represent two deterministic but different wrong-metric transformations of the same object: the left-naive channel experiences runaway amplification near singular directions, whereas the right-naive channel aggressively suppresses them. This singular-value picture explains the distinct leakage morphologies analyzed numerically in Sec. ~\ref{sec:supp_simulation} and also anticipates the enhanced sensitivity of the left-naive branch near an exceptional point.

\section{Exceptional-Point Diagnostics and Reconstruction Conditioning}
This section identifies the pairwise exceptional point (EP) governing the reported finite-basis response and explains why this spectral transition matters for image reconstruction. The analysis proceeds in three steps. We first derive the discrete spectral symmetry and locate the tracked EP. We then use the left--right overlap and phase rigidity to quantify the loss of eigenvector independence. Finally, we separate two consequences for the imaging protocol: perturbation amplification in the authorized channel and deterministic inverse-overlap amplification in the  left-naive Gram channel. This distinction is essential because both reconstructed images can change near the same spectral coordinate, although the two changes originate from different mechanisms.

\subsection{Finite-matrix spectral symmetry}

We first establish the symmetry that constrains the finite-matrix spectrum and makes the pairwise transition identifiable. Let $\Pi\in\mathbb R^{N\times N}$ be the reversal matrix with elements
\begin{equation}
\Pi_{jk}=\delta_{j,N+1-k}.
\end{equation}
Here, $j,k\in\{1,\ldots,N\}$ are the row and column indices, and $\delta_{j,N+1-k}$ is the Kronecker delta. Thus, $\Pi_{jk}=1$ when $k=N+1-j$ and $\Pi_{jk}=0$ otherwise. The nonzero entries lie on the anti-diagonal. Multiplication by $\Pi$ reverses the order of a sampled vector, so $\Pi$ implements spatial parity on the discrete grid. It is real, symmetric, and involutory,
\begin{equation}
\Pi^T=\Pi,
\qquad
\Pi^2=I_N .
\label{eq:supp_reversal_properties}
\end{equation}
where the superscript $T$ denotes matrix transpose and $I_N$ is the $N\times N$ identity matrix. Applying parity twice therefore returns the original sampled vector.

The discrete position operator is $X=\operatorname{diag}(x_1,\ldots,x_N)$, where the symmetric coordinate grid satisfies $x_{N+1-j}=-x_j$. The centered second-difference matrix $D_2$ represents the discrete second derivative introduced in Sec.~I.2. Under reversal of the grid,
\begin{equation}
\Pi X\Pi=-X,
\qquad
\Pi D_2\Pi=D_2.
\end{equation}
The first identity states that the position operator is odd under parity. The second states that the discrete kinetic operator is even. Consequently, parity reverses the sign of the imaginary linear potential $i\gamma X$ but leaves both the kinetic term and the real quadratic potential unchanged.

For real $\gamma$, complex conjugation changes $i\gamma X$ into $-i\gamma X$. Combining complex conjugation with the parity operation therefore restores the original finite operator,
\begin{equation}
\Pi H_N^*(\gamma)\Pi=H_N(\gamma),
\end{equation}
where the asterisk denotes elementwise complex conjugation. This equation is the discrete parity--time ($\mathcal{PT}$) symmetry of $H_N(\gamma)$. Its spectral consequence follows from
\begin{equation}
H_N(\gamma)|\psi\rangle
=
\lambda|\psi\rangle
\quad\Longrightarrow\quad
H_N(\gamma)\Pi|\psi^*\rangle
=
\lambda^*\Pi|\psi^*\rangle .
\label{eq:supp_pt_spectral_pairing}
\end{equation}
Therefore, whenever $\lambda$ is an eigenvalue, $\lambda^*$ is also an eigenvalue. An isolated branch may remain real in a locally unbroken region. After symmetry breaking, the corresponding eigenvalues must occur as a complex-conjugate pair. Tracking the change between these two configurations provides the spectral criterion used below to identify the EP of a selected pair~\cite{SI-Bender1998PRL,SI-ElGanainy2018NatPhys,SI-Feng2017NatPhot}.

\subsection{Identification of the tracked pairwise EP}

The finite matrix $H_N(\gamma)$ can contain several EPs as $\gamma$ varies. Each EP is associated with a particular pair of eigenvalues and eigenvectors, so a quoted value of $\gamma_{\mathrm{EP}}$ must be tied to a specified pair and discretization. To state the algebraic condition, we define the characteristic polynomial
\begin{equation}
p(\lambda,\gamma)=\det\!\left[\lambda I_N-H_N(\gamma)\right].
\end{equation}
Here, $\lambda\in\mathbb C$ is the spectral variable, $\gamma$ is the non-Hermitian control parameter, and $\det(\cdot)$ denotes the matrix determinant. The roots of $p(\lambda,\gamma)$ at fixed $\gamma$ are the eigenvalues of $H_N(\gamma)$. A second-order EP at $(\lambda_{\mathrm{EP}},\gamma_{\mathrm{EP}})$ satisfies~\cite{SI-Heiss2012JPA}
\begin{equation}
p(\lambda_{\mathrm{EP}},\gamma_{\mathrm{EP}})=0,
\qquad
\partial_\lambda p(\lambda_{\mathrm{EP}},\gamma_{\mathrm{EP}})=0,
\end{equation}
where $\partial_\lambda$ denotes differentiation with respect to the spectral variable $\lambda$. The first equality places $\lambda_{\mathrm{EP}}$ on the spectrum. The second makes it a repeated root of the characteristic polynomial and therefore gives algebraic multiplicity two. An EP also requires the rank condition
\begin{equation}
\operatorname{rank}\!\left[H_N(\gamma_{\mathrm{EP}})-\lambda_{\mathrm{EP}}I_N\right]=N-1,
\end{equation}
where $\operatorname{rank}(\cdot)$ is the number of linearly independent rows or columns. This condition leaves a one-dimensional null space and therefore only one independent eigenvector for the repeated eigenvalue. The resulting geometric multiplicity is one. This eigenvector defect distinguishes an EP from an ordinary Hermitian degeneracy, where two independent eigenvectors can remain available at the same eigenvalue.

Although these equations define the EP exactly, direct expansion of a high-order determinant is not the most stable numerical procedure. We instead track the two eigenvalue branches $\lambda_1(\gamma)$ and $\lambda_2(\gamma)$ associated with the reported near-EP imaging response. Their complex separation and a real-valued branch diagnostic are defined as
\begin{equation}
\Delta \lambda(\gamma)=\lambda_1(\gamma)-\lambda_2(\gamma),
\qquad
F(\gamma)=\operatorname{Re}\!\left[(\Delta \lambda(\gamma))^2\right].
\end{equation}
Here, $\Delta \lambda$ measures the instantaneous separation of the tracked eigenvalues, and $\operatorname{Re}(\cdot)$ extracts the real part. On the locally unbroken branch, $\lambda_1$ and $\lambda_2$ are both real. Their difference is real, so $(\Delta \lambda)^2>0$ and $F(\gamma)>0$. After the transition, the two eigenvalues form a complex-conjugate pair. Their difference is then purely imaginary, so $(\Delta \lambda)^2<0$ and $F(\gamma)<0$. At their coalescence, $\Delta \lambda=0$, and the EP of this tracked pair is identified by
\begin{equation}
F(\gamma_{\mathrm{EP}})=0.
\end{equation}
The sign change of $F$ therefore distinguishes the real and complex-conjugate branches without requiring a direct determinant expansion.

For the $N=64$, $\ell=6$ finite operator used in the simulations and basis-pattern construction, interpolation across the sign change gives
\begin{equation}
\gamma_{\mathrm{EP}}\approx0.303,
\qquad
\lambda_{\mathrm{EP}}\approx220.074.
\end{equation}
The first value is the control coordinate at which the tracked pair coalesces, and the second is the corresponding common eigenvalue. Both values belong to the specified finite matrix. In particular, $\gamma_{\mathrm{EP}}\approx0.303$ is not the unique EP of the complete finite spectrum and is not a universal constant of the continuous oscillator. Changing the matrix size $N$, coordinate-window size $\ell$, grid spacing, or boundary treatment changes the finite spectrum and can shift this pairwise transition.

\begin{figure}[htbp]
\centering
\includegraphics[width=1\linewidth]{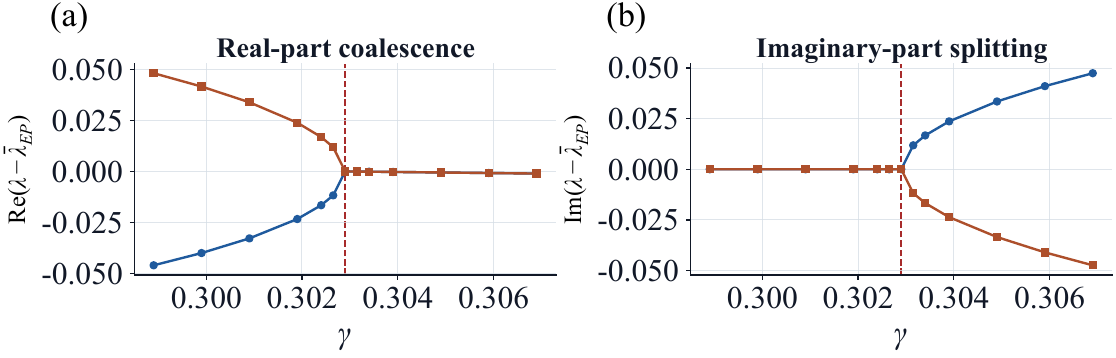}
\caption{Spectral identification of the tracked pairwise exceptional point in the finite non-Hermitian operator. The eigenvalues are centered by subtracting the reference coalescence value $\bar{\lambda}_{\mathrm{EP}}$, so the vertical axes display deviations from the EP eigenvalue rather than its absolute value. (a) Real parts of the two tracked branches. They approach from opposite sides and coalesce at the vertical dashed line. (b) Imaginary parts of the same branches. They remain zero before the transition and split into equal-magnitude, opposite-sign branches after it. Circle and square markers distinguish the two tracked branches. The dashed line marks $\gamma_{\mathrm{EP}}\approx0.303$ for the $N=64$, $\ell=6$ finite operator used in the simulations and basis-pattern construction.}
\label{fig:supp_ep_spectrum}
\end{figure}

Figure~\ref{fig:supp_ep_spectrum} provides the direct spectral evidence for the tracked transition over the narrow interval surrounding $\gamma\approx0.303$. The plotted ordinates are centered by subtracting the reference coalescence value $\bar{\lambda}_{\mathrm{EP}}$. This shift places the transition near zero and improves visibility, but it does not change the eigenvalue separation or the inferred EP coordinate. The circle- and square-marked curves represent the two tracked eigenvalue branches.

In panel (a), the vertical axis gives $\operatorname{Re}(\lambda-\bar{\lambda}_{\mathrm{EP}})$ and the horizontal axis gives $\gamma$. Below the dashed line, the two real parts approach zero from opposite sides as $\gamma$ increases. At the marked coordinate, they merge. Above it, the real parts remain nearly equal, which is the expected behavior after the pair has entered the complex-conjugate branch. Panel (b) plots $\operatorname{Im}(\lambda-\bar{\lambda}_{\mathrm{EP}})$ on the same $\gamma$ interval. Both imaginary parts remain zero below the transition. They then split into positive and negative branches of approximately equal magnitude above the dashed line. The dashed line in both panels marks $\gamma_{\mathrm{EP}}\approx0.303$ for the specified $N=64$, $\ell=6$ operator.

The simultaneous real-part coalescence and opposite-sign imaginary-part splitting are mutually consistent with the discrete $\mathcal{PT}$ symmetry derived in Sec.~II.1. Together, the two panels distinguish the locally unbroken and broken branches of the tracked pair. They also establish the spectral reference coordinate used in the phase-rigidity and near-EP reconstruction analyses. Figure~\ref{fig:supp_ep_spectrum} does not itself measure reconstruction quality. Instead, it identifies the finite-model coordinate against which the image-domain anomalies and perturbation sensitivity are compared later. The interpretation must therefore remain local to the selected pair and the specified discretization.

\subsection{Biorthogonal normalization and phase rigidity}

The eigenvalue coalescence is accompanied by a loss of linear independence between the corresponding left and right eigenvectors. Let the $n$th right and unnormalized left eigenstates satisfy
\begin{equation}
H_N|\psi_n\rangle
=
\lambda_n|\psi_n\rangle,
\qquad
\langle\widetilde\phi_n|H_N
=
\lambda_n\langle\widetilde\phi_n| .
\label{eq:supp_raw_eigenproblems}
\end{equation}

Here, $|\psi_n\rangle$ is a column right eigenvector, and
$\langle\widetilde\phi_n|$ is a row left eigenvector before
biorthogonal normalization. The tilde distinguishes this unnormalized
left state from the normalized state introduced below. To fix the
otherwise arbitrary eigenvector gauge in the finite model, each raw
right eigenvector and each raw adjoint eigenvector is first scaled to
unit Euclidean norm,
$\|\psi_n\|_2=\|\widetilde{\phi}_n\|_2=1$,
before the left row is rescaled by $1/s_n$. All singular values,
spectral norms, and condition numbers reported below refer to this
fixed normalization convention. The pairwise overlap of the raw states
is
\begin{equation}
s_n=\langle\widetilde\phi_n|\psi_n\rangle.
\end{equation}

The complex scalar $s_n$ measures the mutual overlap of the paired left and right states under the biorthogonal pairing. Near an EP, the coalescing states become self-orthogonal in this pairing, so the magnitude $|s_n|$ decreases. Away from the exact singular point, where $s_n\neq0$, the normalized left state is defined by
\begin{equation}
\langle\phi_n|=
\frac{\langle\widetilde\phi_n|}{s_n}.
\end{equation}
This normalization is chosen so that $\langle\phi_n|\psi_n\rangle=1$. For distinct paired modes, the corresponding cross-overlaps vanish, yielding $\langle\phi_m|\psi_n\rangle=\delta_{mn}$. At the exact EP, $s_n=0$ and this normalization becomes singular, which reflects the loss of a complete eigenvector basis.

The same normalization can be written for the complete finite basis. Collecting the unnormalized left eigenvectors as the rows of $\widetilde\Phi_L\in\mathbb C^{N\times N}$ and the right eigenvectors as the columns of $\Psi_R\in\mathbb C^{N\times N}$ gives
\begin{equation}
\widetilde\Phi_L\Psi_R=S,
\qquad
S=\operatorname{diag}(s_1,\ldots,s_N).
\end{equation}
The diagonal matrix $S$ contains the pairwise overlaps. Its $n$th diagonal entry is $s_n$, while the off-diagonal entries vanish under the adopted modal pairing. The normalized left-basis matrix is therefore
\begin{equation}
\Phi_L=S^{-1}\widetilde\Phi_L.
\end{equation}
Left multiplication by $S^{-1}$ rescales each row of $\widetilde\Phi_L$ by the inverse overlap $1/s_n$. Substitution into the preceding equation gives $\Phi_L\Psi_R=I_N$, which is the finite-dimensional biorthogonal identity used by the authorized decoder. This construction is valid only while $S$ remains invertible.

Because the numerical value of $s_n$ depends on the arbitrary scale assigned to the two unnormalized eigenvectors, we also use the phase rigidity~\cite{SI-Heiss2012JPA,SI-Miri2019Science,SI-Ozdemir2019NatMater},
\begin{equation}
r_n=
\frac{|\langle\widetilde\phi_n|\psi_n\rangle|}
{\sqrt{\langle\psi_n|\psi_n\rangle
\langle\widetilde\phi_n|\widetilde\phi_n\rangle}}
\,.
\end{equation}
The numerator is the magnitude of the biorthogonal overlap. The two factors in the denominator are the ordinary Hermitian squared norms of the right and unnormalized left states. Their product removes the arbitrary eigenvector scales, so $r_n$ is dimensionless and satisfies $0\le r_n\le1$. A value near unity indicates well-separated paired states under the adopted normalization, whereas a value approaching zero signals eigenvector coalescence and self-orthogonality. For the tracked pair, $r_n$ decreases near $\gamma_{\mathrm{EP}}$ because the numerator $|s_n|$ becomes small relative to the ordinary state norms.

The relation $\Phi_L\Psi_R=I_N$ can remain algebraically valid near, but not at, the singular point. However, enforcing this identity requires the factors $1/s_n$ in $S^{-1}$ to grow as the overlap decreases. The transform is therefore exact in ideal arithmetic but increasingly ill-conditioned. This distinction is central to interpreting the reconstructed images. In the authorized channel, the large dual factors cancel for exact coefficients but can amplify finite measurement perturbations. In the  left-naive channel, the paired cancellation is absent and the inverse-overlap factors remain in the reconstruction operator. The next two subsections derive these mechanisms separately.

\subsection{Perturbation amplification in the authorized channel}

In the noiseless full-basis limit, the authorized dual-basis channel
forms an exact identity mapping. In the left-basis protocol used in the
main text, the object $O\in\mathbb R^{N\times N}$ is mapped to
$C_L=\Phi_L O\Phi_L^T$, where $C_L\in\mathbb C^{N\times N}$ and the
superscript $T$ denotes transpose without complex conjugation. Applying
the paired right basis along both spatial coordinates gives
$O_{\rm auth}^{(L)}=\Psi_R C_L\Psi_R^T=O$. Because the complete square
bases are mutual inverses away from the exact singular point,
$\Psi_R\Phi_L=I_N$, the authorized identity channel remains exact in
ideal arithmetic at every nonsingular value of $\gamma$. Proximity to
the finite-matrix exceptional point therefore does not by itself break
the biorthogonal cancellation condition.

Experimentally recovered coefficients are not exact. Let
$C_L^{\rm meas}=C_L+\delta C_L$, where $\delta C_L$ collects detector
noise, finite DMD modulation accuracy, illumination and timing
fluctuations, calibration errors, and numerical roundoff. The
corresponding authorized reconstruction error is
\begin{equation}
\delta O_{\rm auth}^{(L)}
=
O_{\rm auth,noisy}^{(L)}-O
=
\Psi_R\delta C_L\Psi_R^T.
\end{equation}
Using the Frobenius norm $\|\cdot\|_F$, the spectral norm
$\|\cdot\|_2$, the identity
$\|\Psi_R^T\|_2=\|\Psi_R\|_2$, and the submultiplicative inequality for
matrix products gives
\begin{equation}
\|\delta O_{\rm auth}^{(L)}\|_F
\leq
\|\Psi_R\|_2^2\|\delta C_L\|_F.
\end{equation}
The direct coefficient-to-image perturbation gain is therefore bounded
by $\sigma_{\max}^2(\Psi_R)$ under the fixed normalization convention.

The relative end-to-end sensitivity is characterized by the spectral
condition number
$\kappa_2(\Psi_R)=\sigma_{\max}(\Psi_R)/
\sigma_{\min}(\Psi_R)$. Since $\Psi_R^{-1}=\Phi_L$, it can equivalently
be written as
$\kappa_2(\Psi_R)=\|\Psi_R\|_2\|\Phi_L\|_2$. The encoding relation
$C_L=\Phi_L O\Phi_L^T$ satisfies
$\|C_L\|_F\leq\|\Phi_L\|_2^2\|O\|_F$. Combining this encoding bound with
the preceding decoder bound gives, for a nonzero object,
\begin{equation}
\frac{\|\delta O_{\rm auth}^{(L)}\|_F}{\|O\|_F}
\leq
\left[\kappa_2(\Psi_R)\right]^2
\frac{\|\delta C_L\|_F}{\|C_L\|_F}.
\end{equation}
The squared condition number arises because the basis acts independently
along the two spatial coordinates.

The absolute and relative bounds describe different levels of
sensitivity and should not be conflated. Under the fixed unit-norm
convention, coalescence of the tracked right eigenvectors drives
$\sigma_{\min}(\Psi_R)$ toward zero, while
$\sigma_{\max}(\Psi_R)$ remains bounded. Consequently, the direct
coefficient-to-image gain $\sigma_{\max}^2(\Psi_R)$ does not diverge as
the tracked EP is approached. By contrast, the relative end-to-end
sensitivity contains
$\kappa_2^2(\Psi_R)\propto\sigma_{\min}^{-2}(\Psi_R)$ and therefore
increases sharply as the basis becomes nearly linearly dependent.

The coefficient-domain perturbation $\delta C_L$ is also affected by
the physical realization of the left-basis patterns. In the four-mask
protocol described in Sec.~IV, each recovered coefficient has the form
$c_{L,k}=\alpha_k\Delta B_k$, where $\Delta B_k$ denotes the complex
differential bucket combination and $\alpha_k$ restores the
mode-dependent scale removed before DMD display. A bucket-domain
perturbation $\delta(\Delta B_k)$ therefore contributes
$\alpha_k\delta(\Delta B_k)$ to the recovered coefficient error.
Because the biorthogonally normalized left modes contain inverse-overlap
factors, the restoration factor $\alpha_k$ can increase near the
tracked EP, converting a fixed bucket fluctuation into a larger entry of
$\delta C_L$.

The near-EP degradation of the authorized reconstruction should
therefore be attributed to the combined effect of basis-dependent
coefficient formation and the increasing relative condition number of
the complete encoding--decoding map. It is not caused by a breakdown of
the matched biorthogonal identity channel, nor does it imply a divergent
absolute gain of the right-basis decoder alone. Instead, a finite
measurement perturbation is first mapped into the coefficient domain
through the normalized left-basis acquisition and is then propagated
through a nearly singular biorthogonal coordinate system
\cite{SI-TrefethenBau2022}.

\subsection{Deterministic amplification in the left-naive channel}

The left-naive reconstruction uses the same left-basis coefficient
matrix $C_L=\Phi_L O\Phi_L^T$ as the authorized channel, but replaces
the paired right-basis decoder with a same-side Hermitian
back-projection. In the noiseless full-basis limit, this reconstruction
can be written as
\begin{equation}
\begin{aligned}
O_{\rm naive}^{(L)}
&=
\Phi_L^\dagger C_L\Phi_L^*
=
\left(\Phi_L^\dagger\Phi_L\right)
O
\left(\Phi_L^\dagger\Phi_L\right)^T \\
&=
G_L O G_L^T,
\qquad
G_L=\Phi_L^\dagger\Phi_L .
\end{aligned}
\label{eq:supp_left_naive_gram}
\end{equation}
Here, the superscripts $\dagger$ and $*$ denote the Hermitian adjoint
and elementwise complex conjugation, respectively, and
$G_L\in\mathbb C^{N\times N}$ is the left same-side Gram operator.
Unlike the paired dual-basis identity channel, $G_L$ is generally
nonidentity because the normalized left modes are not orthonormal under
the ordinary Hermitian inner product. The left-naive reconstruction is
therefore governed by a wrong-metric Gram-channel transformation acting
independently along both image coordinates.

Substituting the biorthogonal normalization
$\Phi_L=S^{-1}\widetilde\Phi_L$ derived in the preceding subsection into
$G_L=\Phi_L^\dagger\Phi_L$ gives
\begin{equation}
\begin{aligned}
G_L
&=
\widetilde\Phi_L^\dagger
(S^{-1})^\dagger S^{-1}
\widetilde\Phi_L,\\
(S^{-1})^\dagger S^{-1}
&=
\operatorname{diag}
\left(
|s_1|^{-2},\ldots,|s_N|^{-2}
\right).
\end{aligned}
\label{eq:supp_left_inverse_overlap}
\end{equation}
Thus, the rank-one contribution associated with the $n$th
unnormalized left mode enters $G_L$ with the inverse-overlap weight
$|s_n|^{-2}$. Because $G_L$ acts on both sides of the object in
~\eqref{eq:supp_left_naive_gram}, a separable two-dimensional
contribution indexed by modes $(i,j)$ carries the product weight
$|s_i|^{-2}|s_j|^{-2}$. If the same near-coalescent mode contributes
along both coordinates, the corresponding prefactor can scale as
$|s_n|^{-4}$. This scaling describes the weight of the associated
modal contribution; the final image also depends on the target
coefficients and on interference among the nonorthogonal modal terms.

The inverse-overlap factors in
~\eqref{eq:supp_left_inverse_overlap} cancel in the authorized
dual-basis channel but remain explicitly in the left-naive Gram
channel. As the tracked eigenvector pair approaches the finite-matrix
EP, the corresponding $|s_n|$ decreases and the affected modal
contributions are increasingly weighted. Consequently, the
left-naive reconstruction can develop a pronounced response even when
the coefficient matrix is exact, namely when $\delta C_L=0$. This
response is deterministic: it arises from the same-side metric
mismatch rather than from additive measurement noise or numerical
coefficient perturbations.

\begin{figure}[htbp]
\centering
\includegraphics[width=1\linewidth]{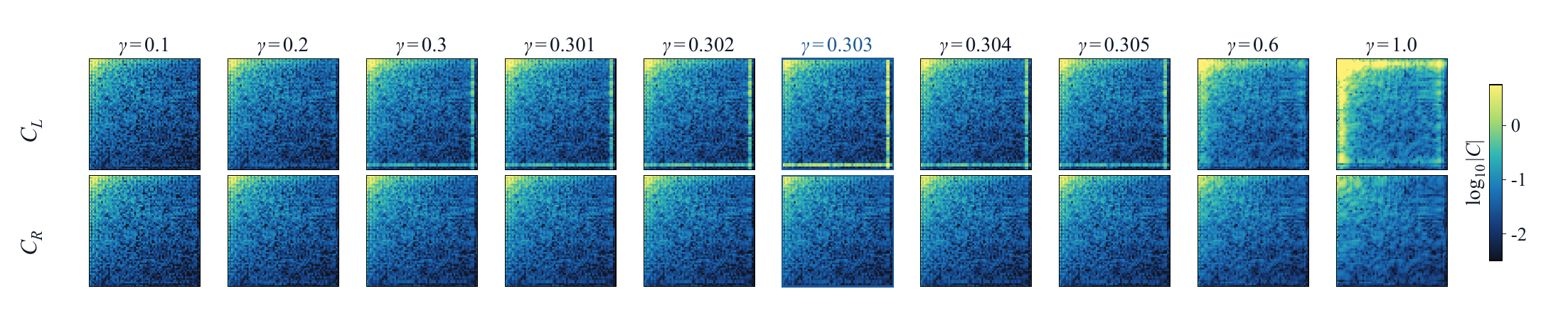}
\caption{Eigenvalue-ordered modal coefficient spectra for left- and
right-basis encoding of the same $N=64$ Cameraman target. The upper and
lower rows show $\log_{10}(|C_L|+10^{-12})$ for
$C_L=\Phi_L O\Phi_L^T$ and $\log_{10}(|C_R|+10^{-12})$ for
$C_R=\Psi_R^T O\Psi_R$, respectively, with the modes ordered along both
axes according to the real parts of their corresponding eigenvalues.
All maps share the same logarithmic color scale. At $\gamma=0.303$,
marked by the blue title and frame and representing the closest sampled
coordinate to $\gamma_{\rm EP}\approx0.302902$, $C_L$ develops
pronounced bright row-and-column bands, whereas $C_R$ remains
comparatively diffuse. This spectral contrast is the target-dependent
coefficient-space manifestation of inverse-overlap weighting near the
EP.}
\label{fig:supp_energy_sorted_coefficients}
\end{figure}

The coefficient spectra in
Fig.~\ref{fig:supp_energy_sorted_coefficients} provide a
coefficient-space representation of this inverse-overlap
redistribution. Using the normalization established above, the
left-basis coefficient matrix can be written directly as
$C_L=S^{-1}\widetilde\Phi_L O\widetilde\Phi_L^T(S^{-1})^T$, whereas the
right-basis coefficient matrix is $C_R=\Psi_R^T O\Psi_R$. Because $S$
is diagonal, the individual left-basis coefficients satisfy
$(C_L)_{ij}=s_i^{-1}s_j^{-1}
(\widetilde\Phi_L O\widetilde\Phi_L^T)_{ij}$, where $i$ and $j$ denote
the modal indices along the row and column coordinates, respectively.
A small $|s_i|$ therefore enhances the corresponding row of $C_L$,
while a small $|s_j|$ enhances the corresponding column. When both
indices belong to the tracked near-coalescent pair, their intersection
receives the combined inverse-overlap weighting. By contrast, $C_R$
contains no explicit $S^{-1}$ prefactors and provides a right-basis
comparison in coefficient space.

For the same $N=64$ Cameraman target, the modal indices in both rows of
Fig.~\ref{fig:supp_energy_sorted_coefficients} are ordered according to
the real parts of their corresponding eigenvalues, and all coefficient
maps are displayed using the same logarithmic color scale. The sampled
sequence combines the global reference coordinates
$\gamma=0.1$, $0.2$, $0.3$, $0.6$, and $1.0$ with refined sampling from
$\gamma=0.301$ to $0.305$ in increments of $0.001$. The coordinate
$\gamma=0.303$ is the sampled value closest to the tracked
finite-matrix EP at $\gamma_{\rm EP}\approx0.302902$. The exact EP is
excluded because $S$ becomes singular there and a complete
biorthogonally normalized eigenvector basis is no longer available.

Near $\gamma=0.303$, the left-basis coefficient matrix $C_L$ develops
pronounced bright row-and-column bands, whereas $C_R$ remains
comparatively diffuse. The operator relation in
~\eqref{eq:supp_left_inverse_overlap} establishes the inverse-overlap
mechanism, while the spectral maps show how this mechanism is populated
by the specific target. The observed contrast is therefore the
target-dependent coefficient-space manifestation of the enhanced
weighting of the near-coalescent left-basis directions.

The distinction from the authorized channel is essential. In the
authorized branch, paired right-basis decoding cancels the
inverse-overlap factors exactly when $\delta C_L=0$, whereas finite
relative coefficient perturbations remain subject to the squared basis
condition number derived in the preceding subsection. In the
left-naive branch, the same-side Gram operator retains and compounds
the inverse-overlap weights even for exact coefficients. The two
channels can consequently exhibit their strongest responses near the
same spectral coordinate, but the underlying mechanisms are different:
perturbation sensitivity in the authorized channel and deterministic
wrong-metric amplification in the left-naive channel.

\FloatBarrier

\section{Numerical Validation and Extended Decoding-Channel Analysis}
\label{sec:supp_simulation}
This section translates the operator relations established above into controlled tests of image reconstruction. After defining the common workflow and metrics, we use two targets and both encoding families to distinguish ordinary wrong-metric decoding, cross-parameter $\gamma$ mismatch, and the additional conditioning response near the finite-matrix EP. The left-basis and right-basis simulations use the same $N=64$, $\ell=6$ operator, modal order, target preprocessing, and display metrics. Within every comparison, the encoded coefficient matrix is held fixed so that differences in the reconstructed images arise solely from the decoding rule.
\subsection{Numerical workflow}

The numerical demonstrations use the finite operator space, modal ordering, and normalization convention defined in Sec.~I. Unless otherwise stated, each target is resized to $64\times64$ pixels, the coordinate window has total width $\ell=6$, and the one-dimensional matrix is constructed from ~\eqref{eq:finite_HN}. No learned prior, regularizer, or iterative inverse solver is introduced. The forward coefficient matrix is evaluated once, after which only the decoding metric or decoding coordinate is changed; the resulting branches are therefore direct tests of the corresponding linear operators.

The full simulation procedure is as follows.
\begin{enumerate}
\item The grayscale object is represented by $O\in\mathbb R^{N\times N}$ on an $N\times N$ grid with $N=64$ and is normalized to the interval $[0,1]$ for display-domain comparison.

\item For each selected real coordinate $\gamma$, the finite matrix $H_N(\gamma)$ is diagonalized. Its right eigenvectors are stored as the columns of $\Psi_R(\gamma)\in\mathbb C^{N\times N}$, and the matched left eigenvectors are stored as the rows of $\Phi_L(\gamma)\in\mathbb C^{N\times N}$.

\item The left and right bases are paired by eigenvalue matching and then biorthogonally normalized so that
\begin{equation}
\Phi_L(\gamma)\Psi_R(\gamma)=I_N,
\end{equation}
within numerical precision.

\item Each separable two-dimensional mode is indexed by $(i_y,i_x)$ and ordered by ascending $\eta_{i_y i_x}=\operatorname{Re}\lambda_{i_y}+\operatorname{Re}\lambda_{i_x}$. The same stable ordering is used for all authorized, naive, and wrong-$\gamma$ calculations.

\item For left-basis encoding, the coefficient matrix is calculated as
\begin{equation}
C_L(\gamma_{\mathrm e})
=
\Phi_L(\gamma_{\mathrm e})O\Phi_L^T(\gamma_{\mathrm e}),
\label{eq:supp_left_coeff}
\end{equation}
where $C_L\in\mathbb C^{N\times N}$ and $\gamma_{\mathrm e}$ is the encoding coordinate. The superscript $T$ is the ordinary transpose, as required by the separable bilinear projection along the two spatial axes.

\item Holding $C_L(\gamma_{\mathrm e})$ fixed, the authorized image, strict same-side Hermitian image, and wrong-$\gamma$ image are evaluated as
\begin{equation}
O_{\rm auth}^{(L)}
=
\Psi_R(\gamma_{\mathrm e})C_L(\gamma_{\mathrm e})\Psi_R^T(\gamma_{\mathrm e}),
\label{eq:supp_left_auth}
\end{equation}
\begin{equation}
O_{\rm naive}^{(L)}
=
\Phi_L^\dagger(\gamma_{\mathrm e})C_L(\gamma_{\mathrm e})\Phi_L^*(\gamma_{\mathrm e}),
\label{eq:supp_left_naive}
\end{equation}
and
\begin{equation}
\widetilde O^{(L)}(\gamma_{\mathrm d},\gamma_{\mathrm e})
=
\Psi_R(\gamma_{\mathrm d})C_L(\gamma_{\mathrm e})\Psi_R^T(\gamma_{\mathrm d}).
\label{eq:supp_wrong_gamma_direct}
\end{equation}
Here, $\gamma_{\mathrm d}$ is the decoding coordinate, the dagger denotes the Hermitian adjoint, and the asterisk denotes elementwise complex conjugation.
\end{enumerate}

The wrong-$\gamma$ image can also be written as a cross-parameter mismatch channel,
\begin{equation}
\widetilde O^{(L)}(\gamma_{\mathrm d},\gamma_{\mathrm e})
=
M_L(\gamma_{\mathrm d},\gamma_{\mathrm e})O
M_L^T(\gamma_{\mathrm d},\gamma_{\mathrm e}),
\qquad
M_L(\gamma_{\mathrm d},\gamma_{\mathrm e})
=
\Psi_R(\gamma_{\mathrm d})\Phi_L(\gamma_{\mathrm e}).
\label{eq:supp_mismatch_operator}
\end{equation}
The matrix $M_L\in\mathbb C^{N\times N}$ maps coefficient coordinates defined at $\gamma_{\mathrm e}$ into object coordinates decoded at $\gamma_{\mathrm d}$. In the matched case, $M_L(\gamma_{\mathrm e},\gamma_{\mathrm e})=I_N$; otherwise it is generally nonidentity.

For right-basis encoding, the coefficient matrix is
\begin{equation}
C_R(\gamma_{\mathrm e})
=
\Psi_R^T(\gamma_{\mathrm e})O\Psi_R(\gamma_{\mathrm e}),
\label{eq:supp_right_coeff}
\end{equation}
where $C_R\in\mathbb C^{N\times N}$. Its authorized and naive reconstructions are, respectively,
\begin{equation}
O_{\rm auth}^{(R)}=\Phi_L^T C_R\Phi_L,
\qquad
O_{\rm naive}^{(R)}=\Psi_R^* C_R\Psi_R^\dagger,
\end{equation}
with every basis evaluated at $\gamma_{\mathrm e}$. For parameter-mismatched right-basis decoding, the same $C_R(\gamma_{\mathrm e})$ is held fixed while the paired left basis is evaluated at $\gamma_{\mathrm d}$:
\begin{equation}
\widetilde O^{(R)}(\gamma_{\mathrm d},\gamma_{\mathrm e})
=
\Phi_L^T(\gamma_{\mathrm d})C_R(\gamma_{\mathrm e})\Phi_L(\gamma_{\mathrm d})
=
M_R^T(\gamma_{\mathrm d},\gamma_{\mathrm e})O
M_R(\gamma_{\mathrm d},\gamma_{\mathrm e}),
\label{eq:supp_right_wrong_gamma_direct}
\end{equation}
where
\begin{equation}
M_R(\gamma_{\mathrm d},\gamma_{\mathrm e})
=
\Psi_R(\gamma_{\mathrm e})\Phi_L(\gamma_{\mathrm d}).
\label{eq:supp_right_mismatch_operator}
\end{equation}
Here $M_R\in\mathbb C^{N\times N}$ is the right-encoding cross-parameter mismatch operator. Its first argument denotes the decoding coordinate, whereas its second argument denotes the encoding coordinate. In the matched case, $M_R(\gamma_{\mathrm e},\gamma_{\mathrm e}) =\Psi_R(\gamma_{\mathrm e})\Phi_L(\gamma_{\mathrm e})=I_N$, and the paired decoder recovers the object in the noiseless full-basis limit. Away from the matched diagonal, $M_R$ is generally nonidentity. These definitions verify that both left- and right-basis protocols possess an exact paired identity channel, while their same-side branches are governed by the reciprocal Gram operators derived in Sec.~\ref{sec:supp_singular_value}.

In summary, the authorized branch tests the biorthogonal identity channel, the naive branch tests a same-side Hermitian Gram channel, and the wrong-$\gamma$ branch tests a cross-parameter mismatch channel. For left encoding, the  naive channel is weighted by $\sigma_i^{-2}$ along each one-dimensional singular direction; for right encoding, the corresponding weight is $\sigma_i^{2}$. Holding the encoded data fixed isolates these operator actions from changes in either the target or the reconstruction procedure.

\subsection{Metrics and display convention}

All quantitative comparisons are performed on real display arrays normalized to $[0,1]$. For authorized images we normalize the real part, whereas for complex leakage images we normalize the modulus, following the display operation used to generate the figures. Let $A,R\in[0,1]^{N\times N}$ denote a displayed test image and its displayed reference, respectively. The reference is the full-sampling authorized result in the decoding-channel scans and the normalized target in the partial-sampling scans. The mean absolute error (MAE) is
\begin{equation}
{\rm MAE}
=
\frac{1}{N^2}
\sum_{i=1}^{N}\sum_{j=1}^{N}
\left|A_{ij}-R_{ij}\right|.
\label{eq:supp_mae}
\end{equation}
The normalized mean-squared error (NMSE), used in the partial-sampling source data, is
\begin{equation}
{\rm NMSE}
=\frac{\sum_{i,j}(A_{ij}-R_{ij})^2}
{\max\!\left[\sum_{i,j}R_{ij}^2,\epsilon_{\rm num}\right]},
\end{equation}
where $\epsilon_{\rm num}$ is machine precision and prevents division by zero. The Pearson correlation coefficient is
\begin{equation}
r
=
\frac{
\sum_{i,j}
\left(A_{ij}-\overline A\right)
\left(R_{ij}-\overline R\right)
}{
\sqrt{
\sum_{i,j}
\left(A_{ij}-\overline A\right)^2
}
\sqrt{
\sum_{i,j}
\left(R_{ij}-\overline R\right)^2
}
}.
\label{eq:supp_pearson}
\end{equation}
Here, $i$ and $j$ index pixels, while $\overline A=N^{-2}\sum_{i,j}A_{ij}$ and $\overline R=N^{-2}\sum_{i,j}R_{ij}$ are spatial means. If either centered image has zero norm, the numerical implementation reports $r=0$.

For the low-sampling curves, the mean-squared error and peak
signal-to-noise ratio are defined as
\begin{equation}
{\rm MSE}
=
\frac{1}{N^2}
\sum_{i,j}(A_{ij}-R_{ij})^2,
\qquad
{\rm PSNR}
=
10\log_{10}\!\left(\frac{I_{\max}^2}{{\rm MSE}}\right)
=
-10\log_{10}({\rm MSE})
\quad {\rm dB},
\end{equation}
where $I_{\max}=1$ because the displayed images are normalized to
the range $[0,1]$. Thus, the second equality is the standard PSNR
definition specialized to the normalized display range. Exact equality
gives ${\rm PSNR}=+\infty$; for visualization only, values above
$60$ dB are displayed at the $60$ dB ceiling.
We also calculate the structural similarity index~\cite{SI-Wang2004TIP}
\begin{equation}
{\rm SSIM}
=\frac{1}{N^2}\sum_{p}
\frac{(2\mu_A(p)\mu_R(p)+C_1)(2\sigma_{AR}(p)+C_2)}
{(\mu_A^2(p)+\mu_R^2(p)+C_1)(\sigma_A^2(p)+\sigma_R^2(p)+C_2)},
\end{equation}
where $p$ indexes image pixels; $\mu_A(p)$ and $\mu_R(p)$ are local means; $\sigma_A^2(p)$ and $\sigma_R^2(p)$ are local variances; and $\sigma_{AR}(p)$ is the local covariance. These statistics are computed with a $7\times7$ uniform window and reflected image boundaries, using $C_1=0.01^2$ and $C_2=0.03^2$. All images within a panel group use the same orientation and the stated display convention. Difference maps show $|A-R|$ and are used to localize deviations; they do not introduce an additional reconstruction operation.

\subsection{Cross-object validation of the left-basis Gram channel}

\begin{figure}[htbp]
\centering
\includegraphics[width=1\linewidth]{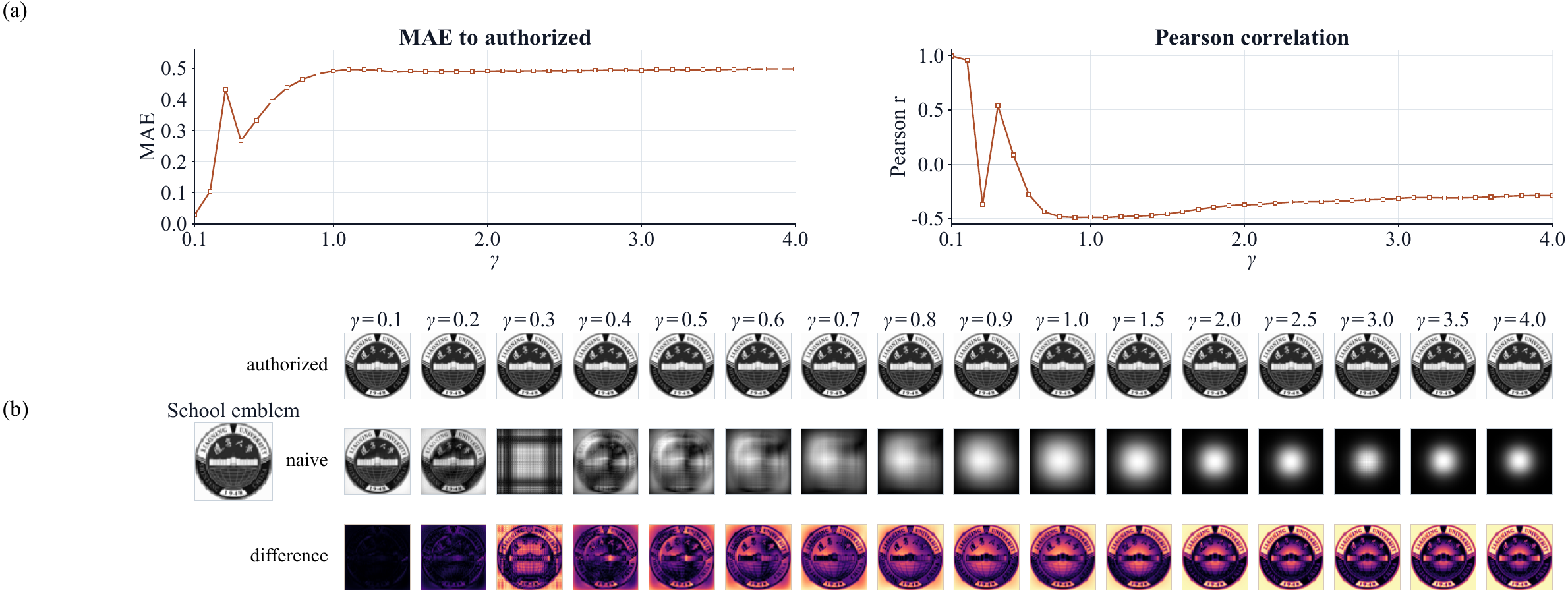}
\caption{Left-basis identity- and Gram-channel comparison for the school-emblem target. (a) Display-domain MAE and Pearson correlation of the left-naive reconstruction relative to the authorized reconstruction over $0.1\le\gamma\le4.0$. (b) Normalized target followed by authorized reconstructions $\Psi_RC_L\Psi_R^T$, left-naive reconstructions $\Phi_L^\dagger C_L\Phi_L^*$, and absolute display-difference maps. The displayed values span $0.1\leq\gamma\leq4.0$, sampled in increments of 0.1 up to $\gamma=1.0$ and 0.5 thereafter. Each column uses one common coefficient matrix $C_L=\Phi_LO\Phi_L^T$. The noiseless calculation uses the full $N=64$, $\ell=6$ basis.}
\label{fig:supp_school_readout}
\end{figure}

\begin{figure}[htbp]
\centering
\includegraphics[width=1\linewidth]{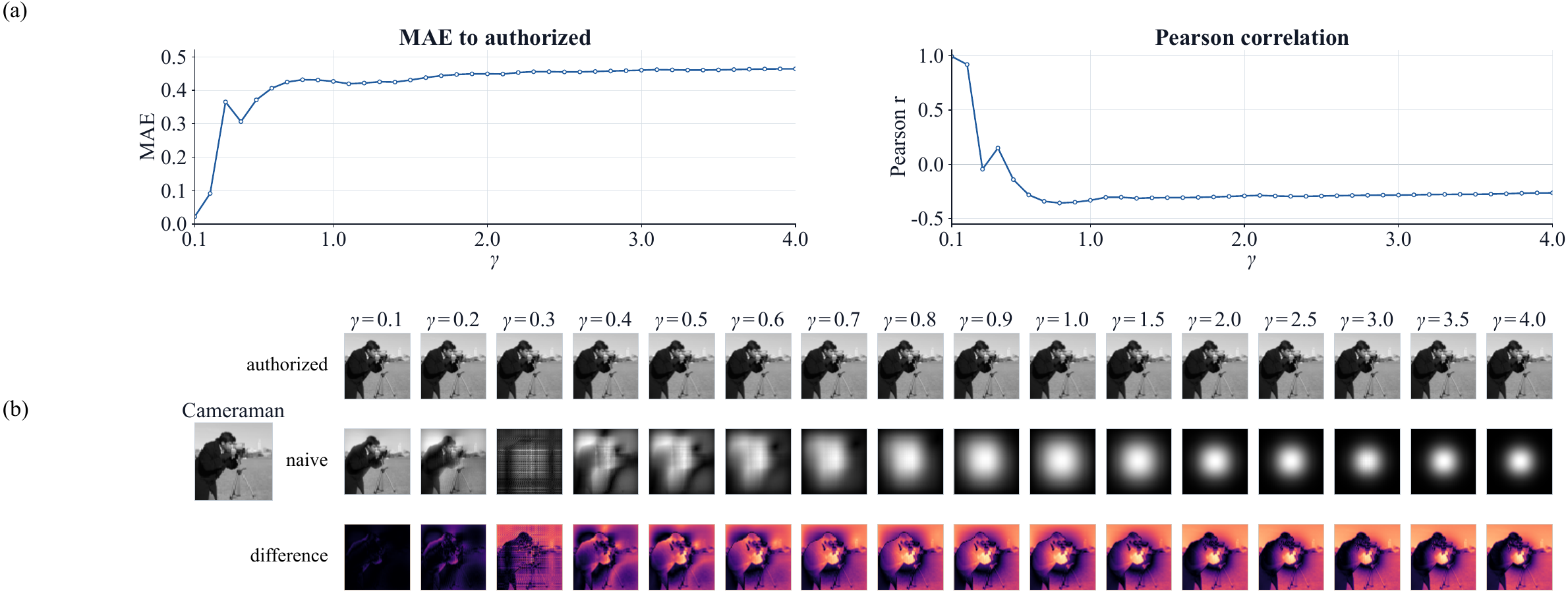}
\caption{Left-basis identity- and Gram-channel comparison for the Cameraman target. (a) Display-domain MAE and Pearson correlation of the left-naive reconstruction relative to the authorized reconstruction over $0.1\le\gamma\le4.0$. (b) Normalized target followed by authorized reconstructions $\Psi_RC_L\Psi_R^T$, left-naive reconstructions $\Phi_L^\dagger C_L\Phi_L^*$, and absolute display-difference maps. The displayed columns and numerical conditions are identical to Fig.~\ref{fig:supp_school_readout}. Within each column, both rows use the same $C_L=\Phi_LO\Phi_L^T$.}
\label{fig:supp_cameraman_readout}
\end{figure}

We first test whether the separation between paired and same-side decoding persists for targets with different spatial statistics. Figures~\ref{fig:supp_school_readout} and \ref{fig:supp_cameraman_readout} use the same left-basis forward model, $C_L=\Phi_LO\Phi_L^T$, for two $64\times64$ targets. The school emblem has sparse high-contrast boundaries and fine text, whereas the Cameraman image contains continuous tones, extended regions, and texture. The comparison therefore probes whether the channel-level conclusion depends on a single object spectrum.

In each figure, panel (a) contains two quantitative scans evaluated on a uniform grid with $\Delta\gamma=0.1$ over $0.1\le\gamma\le4.0$. The left plot gives the display-domain MAE of the left-naive image relative to the authorized image, and the right plot gives their Pearson correlation. Panel (b) places the normalized target at the left and then shows authorized, left-naive, and absolute-difference rows. To retain the rapid low-$\gamma$ evolution, including the EP-associated excursion sampled near $\gamma\simeq0.3$ and the subsequent crossover, the displayed image columns use $\Delta\gamma=0.1$ over $0.1\le\gamma\le1.0$. Once the response enters the broad large-$\gamma$ plateau, the morphology changes only gradually, so a coarser interval of $\Delta\gamma=0.5$ is used for $1.0<\gamma\le4.0$ to avoid redundant, nearly unchanged images. This nonuniform selection applies only to the representative image columns in panel (b); no points are omitted from the quantitative scans in panel (a). Within each column, both decoders receive exactly the same full coefficient matrix $C_L(\gamma)$.

At $\gamma=0.1$, the left-naive image remains close to the school emblem, and panel (a) correspondingly shows a small MAE and a Pearson coefficient near unity. The response is not monotonic in $\gamma$. Around $\gamma=0.3$, close to the tracked pairwise EP, the naive image changes abruptly into a gridlike square pattern. The MAE rises sharply and the Pearson coefficient changes sign. At the neighboring sampled coordinate $\gamma=0.4$, part of the emblem reappears and the metric separation temporarily decreases. This local excursion is consistent with the narrow near-EP response resolved later in Fig.~\ref{fig:supp_near_ep_recon_school_emblem}.

Beyond the local transition, the naive branch progressively loses the emblem boundaries and text. At larger $\gamma$, its independently normalized display is dominated by a compact central lobe, whereas the authorized row continues to reproduce the target. The broad MAE plateau and negative Pearson values show that this lobe is not a low-contrast copy of the emblem. It is the display-normalized morphology selected by the inverse-weighted Gram operator. The corresponding difference maps retain the missing circular boundary, lettering, and internal structures, thereby locating the object information rejected or remixed by the same-side decoder.

The Cameraman result separates the same two regimes. The left-naive image initially retains the main silhouette, but it develops a pronounced gridlike anomaly near $\gamma=0.3$ and loses continuous-tone detail as $\gamma$ increases. At larger coordinates, the naive display again approaches a localized central lobe. Its MAE remains high and its Pearson coefficient remains negative, showing that the same-side output is spatially dissimilar to the authorized image even after independent intensity normalization. The difference maps identify errors across both the extended background and the high-frequency camera and facial features.

The two targets give the same operator-level conclusion but not identical leakage images. Exact recovery in every authorized column follows from $\Psi_R\Phi_L=I_N$ and is independent of the target spectrum. By contrast, the naive result is $G_LOG_L^T$, so the common Gram operator acts on different modal content in the two targets. Object-dependent structure is therefore expected at small and intermediate $\gamma$. The increasingly similar central-lobe morphology at large $\gamma$ indicates that a small set of strongly weighted Gram directions dominates both targets after display normalization. These broad scans also show that the near-EP excursion is a local feature superimposed on a wider wrong-metric response; it is not the sole cause of authorized--naive separation.

\subsection{Gamma-mismatch simulation for left-basis encoding}

\begin{figure}[htbp]
\centering
\includegraphics[width=1\linewidth]{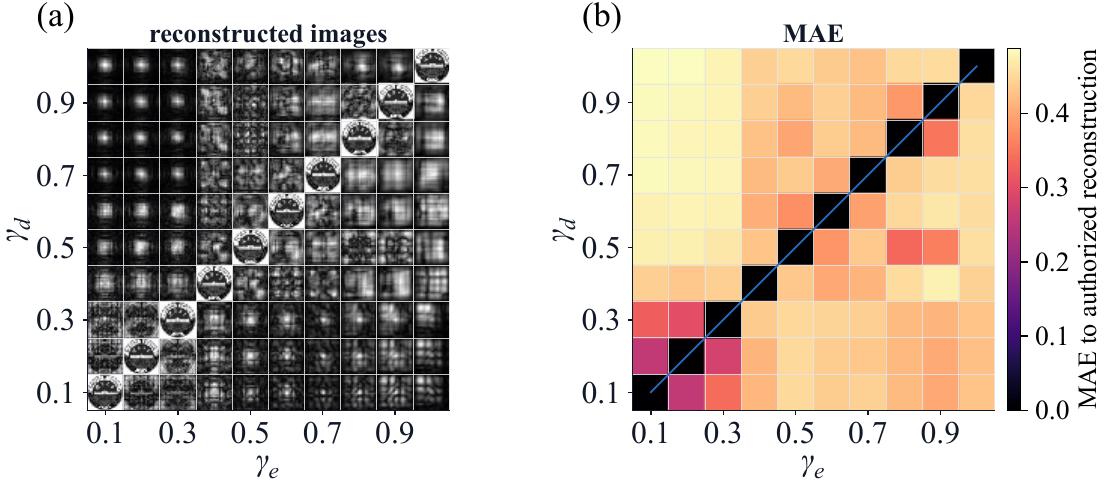}
\caption{Full cross-parameter response of left-basis encoding for the school-emblem target. (a) Reconstruction matrix obtained by fixing $C_L(\gamma_{\mathrm e})=\Phi_L(\gamma_{\mathrm e})O\Phi_L^T(\gamma_{\mathrm e})$ within each column and applying $\Psi_R(\gamma_{\mathrm d})C_L(\gamma_{\mathrm e})\Psi_R^T(\gamma_{\mathrm d})$ along the rows. (b) Display-domain MAE relative to the authorized image in the same $\gamma_{\mathrm e}$ column. The blue diagonal and black cells mark $\gamma_{\mathrm d}=\gamma_{\mathrm e}$, where $M_L(\gamma_{\mathrm e},\gamma_{\mathrm e})=I_N$. The calculation uses $N=64$, $\ell=6$, and full sampling, with $\gamma_{\mathrm e}$ and $\gamma_{\mathrm d}$ independently varied from 0.1 to 1.0 in increments of 0.1.}
\label{fig:gamma_mismatch_full_school_emblem}
\end{figure}

\begin{figure}[htbp]
\centering
\includegraphics[width=1\linewidth]{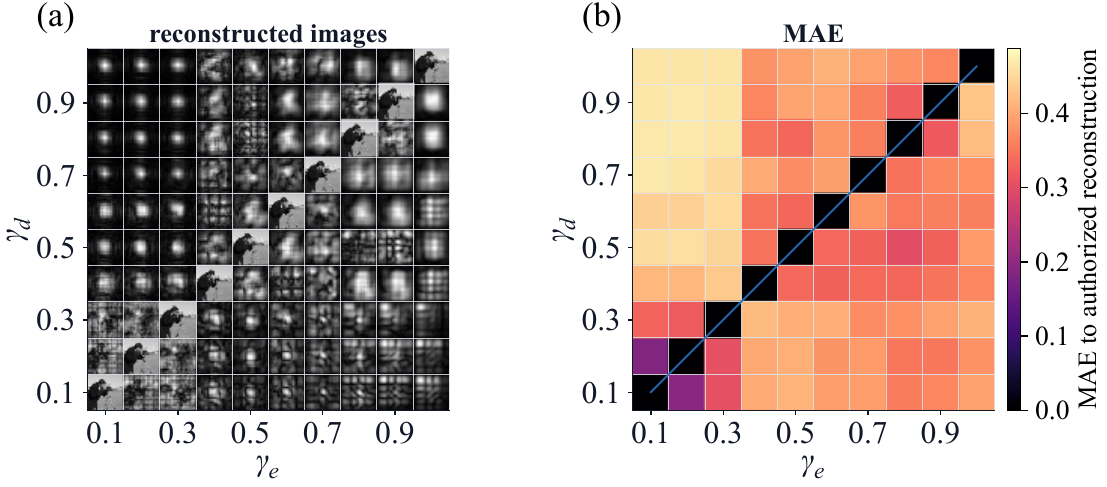}
\caption{Full cross-parameter response of left-basis encoding for the Cameraman target. (a) Reconstruction matrix obtained by fixing $C_L(\gamma_{\mathrm e})=\Phi_L(\gamma_{\mathrm e})O\Phi_L^T(\gamma_{\mathrm e})$ within each column and applying $\Psi_R(\gamma_{\mathrm d})C_L(\gamma_{\mathrm e})\Psi_R^T(\gamma_{\mathrm d})$ along the rows. (b) Display-domain MAE relative to the authorized image in the same $\gamma_{\mathrm e}$ column. The blue diagonal and black cells mark $\gamma_{\mathrm d}=\gamma_{\mathrm e}$, where $M_L(\gamma_{\mathrm e},\gamma_{\mathrm e})=I_N$. The calculation uses $N=64$, $\ell=6$, and full sampling, with $\gamma_{\mathrm e}$ and $\gamma_{\mathrm d}$ independently varied from 0.1 to 1.0 in increments of 0.1.}
\label{fig:gamma_mismatch_full_cameraman}
\end{figure}

Having established the same-coordinate Gram-channel response, we next examine a distinct decoding mechanism: whether $\gamma$ defines a coordinate-specific decoding condition rather than merely modifying the spatial structure of the basis patterns. The encoder remains the left basis evaluated at $\gamma_{\mathrm e}$, whereas decoding uses the paired right-basis family evaluated independently at $\gamma_{\mathrm d}$; this comparison therefore changes the operator coordinate rather than the decoding metric family. In Figs.~\ref{fig:gamma_mismatch_full_school_emblem} and \ref{fig:gamma_mismatch_full_cameraman}, the encoding and decoding coordinates, $\gamma_{\mathrm e}$ and $\gamma_{\mathrm d}$, are varied independently from 0.1 to 1.0 in increments of 0.1. In panel (a), columns correspond to the encoding coordinate $\gamma_{\mathrm e}$ and rows correspond to the decoding coordinate $\gamma_{\mathrm d}$. For a given column, $C_L(\gamma_{\mathrm e})$ is calculated once. Moving vertically through that column changes only $\Psi_R(\gamma_{\mathrm d})$ in \eqref{eq:supp_wrong_gamma_direct}.

Panel (b) uses the same row--column convention and reports the display-domain MAE relative to the authorized image in each $\gamma_{\mathrm e}$ column. The blue diagonal marks $\gamma_{\mathrm d}=\gamma_{\mathrm e}$. Its black cells correspond to zero MAE because $M_L(\gamma_{\mathrm e},\gamma_{\mathrm e})=I_N$ there. The off-diagonal colors measure deterministic coordinate mismatch under full sampling and noiseless arithmetic; they do not contain measurement noise or coefficient truncation.

The school emblem is recovered along the full aligned diagonal of Fig.~\ref{fig:gamma_mismatch_full_school_emblem}(a), rather than at one preferred value of $\gamma$. The diagonal should therefore not be interpreted as a resonance centered at a particular operating coordinate. Instead, it is the locus along which the encoding basis and the paired decoding basis are mutual inverses, $M_L(\gamma_{\mathrm e},\gamma_{\mathrm e})=I_N$. Parameter matching, rather than the absolute value of $\gamma$, establishes the exact identity channel. The zero-MAE diagonal in panel (b) is the quantitative image-domain expression of this operator identity.

Away from the diagonal, the reconstructions do not fade uniformly. Instead, they evolve through asymmetric grids, localized lobes, and mixed remnants of the emblem. These are structured outputs of $M_L(\gamma_{\mathrm d},\gamma_{\mathrm e})O M_L^T(\gamma_{\mathrm d},\gamma_{\mathrm e})$, because each nonidentity mismatch operator acts along coordinate-dependent directions. Consequently, cells with the same $|\gamma_{\mathrm d}-\gamma_{\mathrm e}|$ need not have the same morphology or MAE: $M_L$ depends on both absolute coordinates and the direction of the detuning, so the image structures can differ above and below the diagonal. This asymmetry makes the matrix grid a direct image-domain signature of coordinate mismatch. It also distinguishes mismatch from naive decoding. The mismatch branch uses the correct right-basis family at the wrong coordinate, whereas the naive branch uses the left basis itself under the Hermitian metric at the encoding coordinate.

The Cameraman scan has the same exact diagonal but different off-diagonal content. On the diagonal, the silhouette, background, and grayscale structure are recovered because the basis product is the identity. Away from it, the mismatch operator redistributes continuous-tone and high-frequency components into blurred lobes and gridlike mixtures. The associated MAE remains nonzero throughout the off-diagonal cells. The common diagonal for two dissimilar targets follows from the operator identity, whereas the distinct off-diagonal morphologies follow from applying the nonidentity $M_L$ to different object spectra.

Together, Figs.~\ref{fig:gamma_mismatch_full_school_emblem} and
\ref{fig:gamma_mismatch_full_cameraman} isolate the
parameter-alignment requirement at both the operator and image levels.
The common zero-error diagonal across two structurally dissimilar
targets shows that exact recovery is imposed by the basis identity,
whereas the distinct off-diagonal morphologies remain dependent on the
object spectrum. Because every cell already uses the complete $N^2$
coefficient set, the off-diagonal errors cannot be attributed to
coefficient omission, measurement noise, or insufficient sampling.
Exact recovery requires the decoding coordinate to reproduce the
inverse of the basis used during encoding.

Within this noiseless full-sampling model, $\gamma$ can therefore serve
as a coordinate-specific physical-layer decoding parameter with
key-like selectivity: parameter alignment opens the exact identity
channel, whereas detuning routes the same complete coefficient matrix
through a nonidentity cross-parameter transformation. Additional
coefficients cannot repair this transformation because coefficient
availability is already complete. This conclusion concerns
deterministic channel isolation; the tolerance of the matched channel
to finite parameter errors and experimental perturbations is governed
by the conditioning of the basis pair.

The present scan uses a parameter spacing of $\Delta\gamma=0.1$ and
therefore establishes the global alignment structure rather than the
local response near the tracked finite-matrix EP. The next subsection
resolves that narrow interval using a finer parameter grid.

\FloatBarrier

\subsection{Near-EP numerical check}

\begin{figure}[htbp]
\centering
\includegraphics[width=0.8\linewidth]{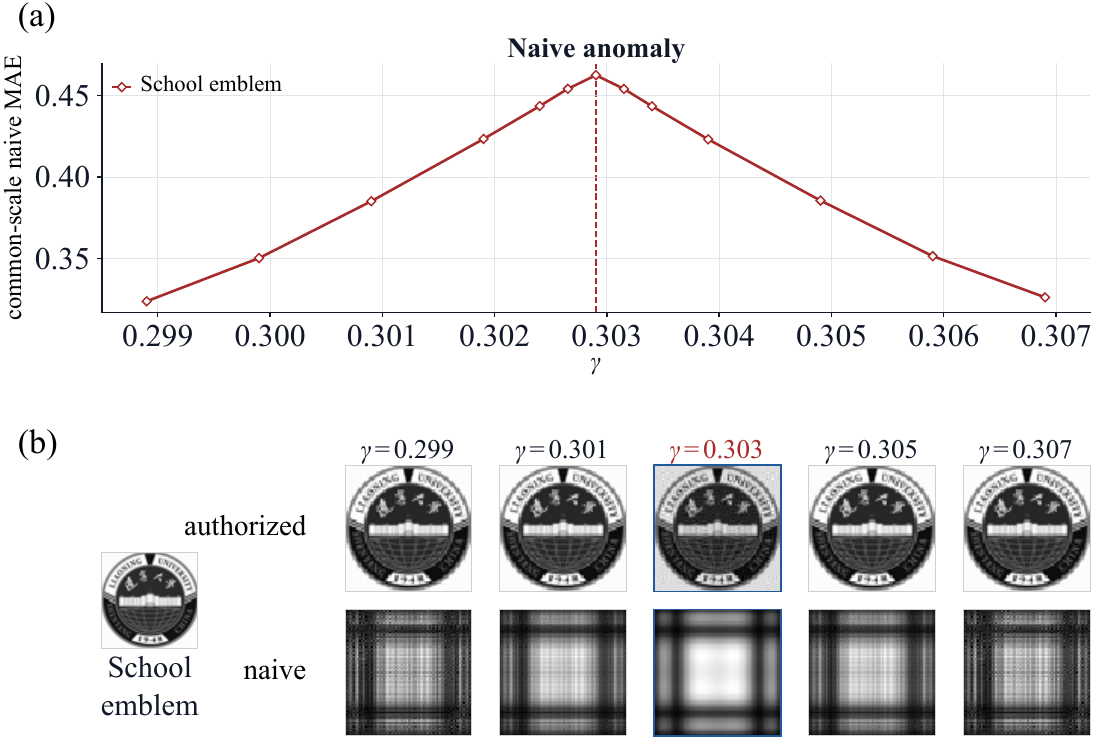}
\caption{Near-EP response of the left-basis protocol for the school-emblem target. (a) Display-domain MAE of the left-naive reconstruction relative to the authorized result over $0.299\le\gamma\le0.307$. The dashed line marks the rounded coordinate $\gamma=0.303$ near the tracked finite-matrix EP. (b) Normalized target and authorized and left-naive reconstructions at $\gamma=0.299$, $0.301$, $0.303$, $0.305$, and $0.307$. The highlighted $\gamma=0.303$ column is the closest displayed coordinate to the inferred transition. The noiseless full-sampling calculation uses $N=64$ and $\ell=6$.}
\label{fig:supp_near_ep_recon_school_emblem}
\end{figure}

\begin{figure}[htbp]
\centering
\includegraphics[width=0.8\linewidth]{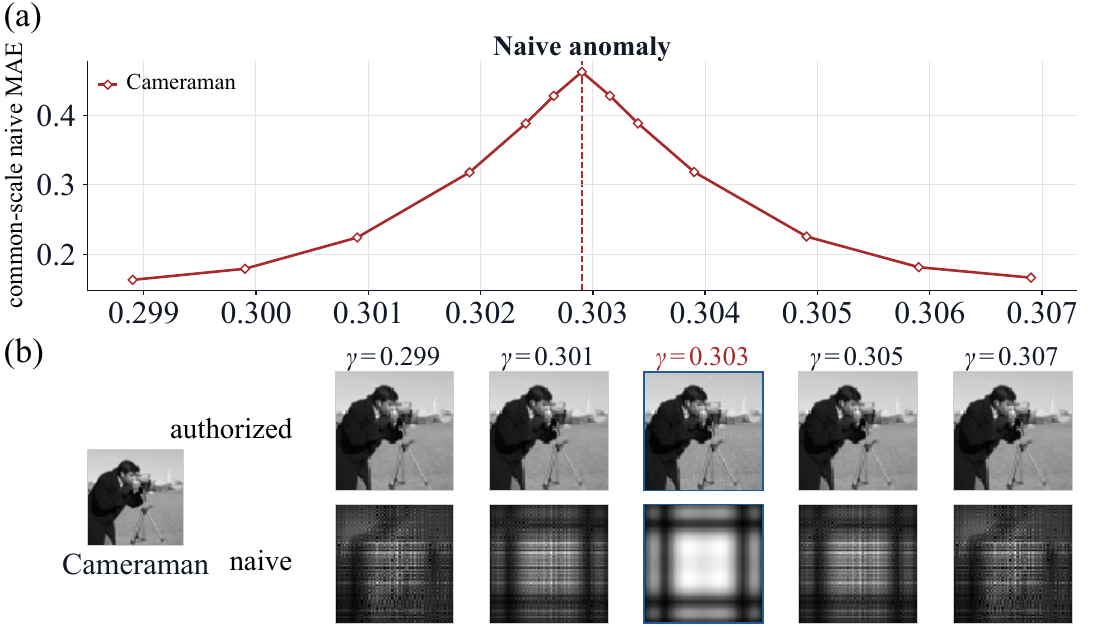}
\caption{Near-EP response of the left-basis protocol for the Cameraman target. (a) Display-domain MAE of the left-naive reconstruction relative to the authorized result over $0.299\le\gamma\le0.307$. (b) Normalized target and authorized and left-naive reconstructions at the same five coordinates used in Fig.~\ref{fig:supp_near_ep_recon_school_emblem}. The dashed line and highlighted column identify the rounded coordinate $\gamma=0.303$ near the tracked finite-matrix EP. The calculation is noiseless, fully sampled, and uses $N=64$, $\ell=6$.}
\label{fig:supp_near_ep_recon_cameraman}
\end{figure}

The broad same-coordinate Gram-channel scans in Figs.~\ref{fig:supp_school_readout} and \ref{fig:supp_cameraman_readout} reveal a nonmonotonic excursion near the coarsely sampled coordinate $\gamma=0.3$, but they do not resolve its local parameter dependence. We therefore examine the narrower interval $0.299\leq\gamma\leq0.307$ around the tracked pairwise EP of the $N=64$, $\ell=6$ finite matrix. Figures~\ref{fig:supp_near_ep_recon_school_emblem} and \ref{fig:supp_near_ep_recon_cameraman} use noiseless full sampling and compare the authorized and left-naive branches at the same coordinate. Panel (a) plots the display-domain naive MAE relative to the authorized image. Panel (b) shows representative reconstructions at $\gamma=0.299$, $0.301$, $0.303$, $0.305$, and $0.307$.

The tracked finite-matrix EP identified in Sec.~II.2 is located at $\gamma_{\mathrm{EP}}\approx0.302902$. The dashed line marks the displayed scan coordinate $\gamma=0.303$, which is the closest labeled coordinate to this value rather than the exactly defective point itself. At the exact EP, the eigenvector basis is incomplete and the biorthogonal normalization is singular. All displayed reconstructions are evaluated at nearby numerically invertible matrices, for which the paired basis remains well defined.

For the school emblem, the naive MAE rises toward a sharp local maximum near the dashed line and then decreases on the opposite side. The selected images reveal the corresponding morphological sequence. Gridlike structures at $\gamma=0.299$ and $0.301$ collapse into a smooth square-lobe pattern near $\gamma=0.303$, after which the finer grid structure reappears at $\gamma=0.305$ and $0.307$. The authorized row remains equal to the emblem throughout the interval. The peak is therefore not caused by a change in the encoded target or by a failure of the paired algebra in exact arithmetic.

The Cameraman target exhibits the same peak coordinate and the same transition toward a square-lobe naive image near $\gamma=0.303$. Away from the peak, its gridlike leakage contains target-dependent intensity and texture. Thus, the location of the anomaly is fixed by the common basis, while its metric amplitude and residual spatial content depend on the object. The similar square-lobe morphology at the center of both scans indicates that a strongly weighted Gram direction dominates the independently normalized display near the tracked transition.

This local anomaly is consistent with the spectral evidence in Fig.~\ref{fig:supp_ep_spectrum}. There, the tracked eigenvalues coalesce in real part and split in imaginary part near the same coordinate. Sec.~II.5 shows that the left-naive Gram operator retains inverse-overlap factors $|s_n|^{-2}$. As the tracked overlap becomes small, these factors enhance the associated modal contributions along both image coordinates. Because the same-side Gram operator acts independently on the two coordinates, a separable contribution involving the same near-coalescent mode along both axes can carry the compounded weight $|s_n|^{-4}$, as derived in Sec.~II.5. This two-axis weighting provides the operator-level origin of the sharply localized square-lobe response. The resulting peak is deterministic and remains present even when the coefficient perturbation $\delta C_L$ is zero.

The authorized and naive near-EP responses have different origins. In the present noiseless calculation, the authorized decoder remains the identity channel. Its paired factors cancel on the numerically invertible side of the transition. In an experiment, finite errors in $C_L$ can be amplified by the separate conditioning mechanism derived in Sec.~II.4. The authorized image may therefore also degrade near the same coordinate. Figures~\ref{fig:supp_near_ep_recon_school_emblem} and \ref{fig:supp_near_ep_recon_cameraman} isolate only the deterministic wrong-metric contribution.

Taken together, the numerical results establish three specific conclusions within the stated noiseless full-basis model. Matched left--right decoding recovers both targets through the full-basis identity channel. Same-side Hermitian decoding routes the same coefficients through a target-dependent Gram filter, with a distinct local anomaly near the tracked finite-matrix EP. Paired decoding at an incorrect $\gamma_{\mathrm d}$ instead produces a cross-parameter mismatch channel whose exact diagonal is fixed by $\gamma_{\mathrm d}=\gamma_{\mathrm e}$. The EP therefore acts as a local sensitivity boundary within a broader biorthogonal decoding landscape; it is not the sole origin of parameter selectivity or wrong-metric leakage. These noiseless numerical results therefore provide the reference
against which the measured authorized and naive responses are
interpreted in the following experimental section.

\FloatBarrier

\section{Experimental Implementation and Validation}
\label{sec:supp_experiment}

\subsection{Optical setup and acquisition}
This section connects the finite-operator model to its implementation in the passive single-pixel imaging (SPI) experiment. Main-text Fig.~1(b) summarizes the acquisition sequence: white-light illumination, reflection from the target, DMD modulation, single-pixel detection, and digitization. Its lower inset illustrates how a complex left-basis pattern is decomposed into four non-negative DMD frames. Main-text Fig.~1(c) shows the three decoding branches applied to the measured coefficient matrix: paired right-basis decoding, naive reconstruction, and parameter-mismatched decoding. We first describe the hardware and acquisition timing, and then define the DMD-mask construction, coefficient recovery, and reconstruction conventions used for main-text Fig.~4.

The hardware is a conventional passive, intensity-only SPI chain. A white LED illuminates the reflective Chinese-character target, and a lens system relays the light reflected or scattered by that target onto a programmable digital micromirror device (DMD). The DMD sequentially displays binary masks derived from the complex non-Hermitian left-basis modes. A Thorlabs DET36A2 single-pixel detector integrates the modulated intensity, and an AUMANYU USB DAQ-610 card digitizes the detector output. No non-Hermitian material is placed in the optical path. Instead, the physical encoding is determined by the programmable mask library, and the corresponding dual basis is applied computationally after acquisition~\cite{SI-Duarte2008,SI-Edgar2019}.

The DMD uses a $1024\times768$ canvas. Each $64\times64$ numerical pattern is expanded by a factor of 12 to form a $768\times768$ active region, so that one object pixel corresponds to a fixed $12\times12$ micromirror block. This region occupies rows 1--768 and columns 129--896. The complete separable basis contains $64^2=4096$ complex modes and, because each mode requires four binary intensity masks, $4096\times4=16384$ formal DMD frames.

The DMD update rate is $f_{\rm DMD}=20$ Hz, and the detector output is sampled at $f_s=1000$ samples/s. Thus, $N_s=f_s/f_{\rm DMD}=50$ digitized voltage samples are recorded during each displayed frame, where $f_{\rm DMD}$ is the pattern rate, $f_s$ is the digitizer sampling rate, and $N_s$ is the number of samples per frame. For every formal mask, the bucket value is the mean of a 10-sample window beginning 20 samples after the frame onset; the initial samples are thereby excluded from the average to reduce DMD switching transients. The same window is applied to every mask, mode, and value of $\gamma$, so the decoding branches receive the same temporally processed data.

\subsection{Construction of left-basis two-dimensional speckles}

The experimental mask library is constructed from the same finite matrix and modal order used in the numerical model. After $H_N(\gamma)$ is diagonalized and biorthogonally normalized, each formal mode index $k\in\{1,\ldots,N^2\}$ is mapped to a pair $(i_y,i_x)$ of one-dimensional left-mode indices. If $\phi_{i_y}(y;\gamma)$ and $\phi_{i_x}(x;\gamma)$ are the corresponding sampled left eigenfunctions along the row coordinate $y$ and column coordinate $x$, their separable outer product defines the ideal complex two-dimensional speckle
\begin{equation}
\widetilde{P}_k^L(y,x;\gamma)
=
\phi_{i_y}(y;\gamma)\phi_{i_x}(x;\gamma).
\end{equation}
Equivalently, if $\phi_{i_y}$ and $\phi_{i_x}$ denote the associated rows of $\Phi_L$, the sampled pattern matrix is
\begin{equation}
\widetilde{P}_k^L
=
\phi_{i_y}\left(\phi_{i_x}\right)^T .
\end{equation}
Here, $\widetilde{P}_k^L\in\mathbb C^{N\times N}$ is the ideal complex encoding speckle before display normalization, not a pattern reconstructed from four intensity measurements. Before DMD conversion, it is normalized by its maximum modulus,
\begin{equation}
\alpha_k
=
\max_{x,y}
\left|
\widetilde{P}_k^L(x,y)
\right|,
\qquad
P_k^L
=
\frac{\widetilde{P}_k^L}{\alpha_k},
\end{equation}
where $\alpha_k>0$ is the largest pixel modulus of mode $k$, and $P_k^L\in\mathbb C^{N\times N}$ therefore satisfies $|P_k^L(x,y)|\le1$. The mode-dependent factor $\alpha_k$ is stored and restored after differential bucket formation. This operation keeps every displayed component within the available modulation range without changing the coefficient scale used to assemble $C_L$.

\subsection{Four-mask DMD realization of complex speckles}

The lower part of main-text Fig.~1(b) first shows the quantities used in the complex-pattern conversion. The left side contains the modulus $|P_k^L|$ and phase $\arg(P_k^L)$ of a representative normalized mode; the right side contains the two complementary real-component masks and two complementary imaginary-component masks. This four-frame decomposition is required because the DMD modulates non-negative intensity rather than a complex field. We write
\begin{equation}
P_k^L=P_{R,k}+iP_{I,k},
\end{equation}
where $P_{R,k}=\operatorname{Re}P_k^L$ and $P_{I,k}=\operatorname{Im}P_k^L$ are real arrays with values in $[-1,1]$. They are represented by complementary non-negative pairs,
\begin{equation}
P_{R,k}^{\pm}=\frac{1\pm P_{R,k}}{2},
\qquad
P_{I,k}^{\pm}=\frac{1\pm P_{I,k}}{2}.
\end{equation}
Each array $P_{R,k}^{\pm}$ or $P_{I,k}^{\pm}$ lies in $[0,1]$, and the differences satisfy $P_{R,k}^{+}-P_{R,k}^{-}=P_{R,k}$ and $P_{I,k}^{+}-P_{I,k}^{-}=P_{I,k}$.
For mode $k$, the four frames are displayed in the order $P_{R,k}^+$, $P_{R,k}^-$, $P_{I,k}^+$, and $P_{I,k}^-$. Their temporally averaged bucket values are
\begin{equation}
B_{R,k}^+,\qquad B_{R,k}^-,
\qquad B_{I,k}^+,\qquad B_{I,k}^-,
\end{equation}
where, for example, $B_{R,k}^{+}$ is the detector average obtained while displaying $P_{R,k}^{+}$.

Following differential bucket measurements for common-mode and
background rejection~\cite{SI-Ferri2010PRL}, together with
intensity-only implementations of complex-valued single-pixel
coefficients using binary or phase-shifted patterns~\cite{SI-Pastuszczak2016AO,SI-Zhang2015NatCommun},
the four measurements give the calibrated complex coefficient
\begin{equation}
c_{L,k}
=
\alpha_k
\left[
(B_{R,k}^+-B_{R,k}^-)
+i(B_{I,k}^+-B_{I,k}^-)
\right].
\label{eq:four_mask}
\end{equation}
Here, $c_{L,k}\in\mathbb C$ is the measured coefficient associated with formal mode $k$. The two bucket differences cancel the common offset of each complementary pair, while multiplication by $\alpha_k$ restores the scale removed during display normalization. ~\eqref{eq:four_mask} therefore connects the pre-normalization speckle $\widetilde{P}_k^L$ to one complex measurement coefficient. Once all modes have been measured, the coefficients are returned to their $(i_y,i_x)$ positions to form $C_L\in\mathbb C^{N\times N}$.

Each continuous non-negative mask is converted into a one-bit amplitude pattern by deterministic pulse-density encoding within its $12\times12$ micromirror block. A fixed pseudorandom threshold tile is reused for every object pixel and mode. For a signed component value $p\in[-1,1]$, the two complementary masks have intended duty fractions $(p+1)/2$ and $(1-p)/2$; their block-averaged difference therefore approximates $p$, with finite quantization set by the 144-micromirror block. The real and imaginary components are processed independently. Figure~1(b) consequently illustrates a physical four-measurement realization of one complex projection, rather than direct modulation of optical phase.

\subsection{Experimental reconstruction pipeline and limitations}
After differential formation and normalization compensation, each
$c_{L,k}$ is returned to the modal position specified by the fixed
ordering map, producing the measured complex matrix
$C_L(\gamma_{\mathrm e})$. The three reconstruction branches are

\begin{equation}
\begin{aligned}
O_{\rm auth}^{\rm (exp)}
&=
\Psi_R(\gamma_{\mathrm e})
C_L(\gamma_{\mathrm e})
\Psi_R^T(\gamma_{\mathrm e}),
\\
O_{\rm naive}^{\rm (exp)}
&=
\Phi_L^\dagger(\gamma_{\mathrm e})
C_L(\gamma_{\mathrm e})
\Phi_L^*(\gamma_{\mathrm e}),
\\
\widetilde O^{\rm (exp)}
(\gamma_{\mathrm d},\gamma_{\mathrm e})
&=
\Psi_R(\gamma_{\mathrm d})
C_L(\gamma_{\mathrm e})
\Psi_R^T(\gamma_{\mathrm d}).
\end{aligned}
\label{eq:supp_exp_decoders}
\end{equation}

Here, the dagger denotes the Hermitian adjoint and the asterisk
denotes elementwise complex conjugation. The naive reconstruction
has the same definition in the numerical and experimental analyses.
For parameter-mismatched reconstruction, the measured
$C_L(\gamma_{\mathrm e})$ remains fixed and only the computational
right basis is evaluated at $\gamma_{\mathrm d}$.

Main-text Fig.~4 first presents the experimental image comparisons.
Panel (a) contains the authorized, naive, and absolute-difference
images over $0.1\leq\gamma\leq1.0$. Panel (b) fixes
$\gamma_{\mathrm e}=0.6$ and displays reconstructions obtained with
different decoding coordinates $\gamma_{\mathrm d}$. In panel (a),
the authorized reconstruction preserves the Chinese-character
strokes through the better-conditioned part of the scan, whereas
the naive reconstruction evolves from a partially recognizable
image into gridlike, stripe-like, and broad-lobe structures. At
larger $\gamma$, the authorized reconstruction also degrades because
finite coefficient errors are transformed by the increasingly
ill-conditioned basis pair. In panel (b), the matched column gives
the most faithful recovery, while the mismatched columns contain
structured outputs governed by the cross-parameter operator in the
last line of ~\eqref{eq:supp_exp_decoders}.

Panels (c) and (d) connect the image response to the finite-matrix
EP. Panel (c) contains the authorized and naive reconstructions over
$0.300\leq\gamma\leq0.306$, while panel (d) shows the tracked
eigenvalue gap and phase rigidity over the same interval. The
authorized reconstruction exhibits its strongest degradation near
$\gamma\simeq0.303$, consistent with measurement perturbations being
transformed by a poorly conditioned basis pair. The naive
reconstruction develops a localized square-like feature near the
same coordinate. Because the numerical and experimental naive
reconstructions use the same operator, panel (c) provides an
experimental counterpart to the near-EP Gram-channel response.
Differences in detailed morphology arise from measurement
perturbations and display normalization.

The dominant experimental perturbations are DMD block quantization, finite contrast, illumination drift, detector noise, timing jitter, and residual error within the selected bucket window. These perturbations enter through $C_L^{\rm meas}=C_L+\delta C_L$ and are subsequently transformed by the selected decoder. This separation motivates the partial-acquisition study below: reducing the number of retained modes changes the available coefficient subspace, whereas changing the decoder changes how the same retained coefficients are mapped to the image domain.

\section{Low-Sampling Biorthogonal Projection}
\label{sec:supp_low_sampling}

The preceding sections establish complete-basis encoding, its experimental realization, and the distinction among paired, naive, and parameter-mismatched decoding. We now retain only part of the ordered measurement sequence and examine the resulting truncated biorthogonal projection. We first define the selection operator and zero-filled coefficient matrix, then compare authorized and naive numerical reconstructions at regular and near-EP operating points, isolate parameter mismatch under fixed partial sampling, and finally test the same mechanisms with measured coefficients. This progression separates truncation error, decoding-metric error, and experimental perturbations while keeping the acquired data identical within each comparison.

\subsection{Ordered partial acquisition}
\label{sec:partial_sampling}

For an $N\times N$ object, the complete separable encoding basis contains $N^{2}$ complex modal patterns. We use column-major vectorization, for which $\operatorname{vec}(BXC^T)=(C\otimes B)\operatorname{vec}(X)$. The full coefficient-domain measurement model is
\begin{equation}
\mathbf{y}
=
A_{\gamma_{\mathrm{e}}}\mathbf{x}
+
\mathbf{n},
\label{eq:supp_ls_forward}
\end{equation}
where $\mathbf{x}=\operatorname{vec}(O)\in\mathbb R^{N^2}$ is the vectorized object; $\mathbf{y}=\operatorname{vec}(C_L)\in\mathbb C^{N^2}$ is the full complex coefficient vector; $A_{\gamma_{\mathrm e}}\in\mathbb C^{N^2\times N^2}$ is the complete left-basis encoding operator; and $\mathbf n\in\mathbb C^{N^2}$ represents coefficient-domain noise. The separable form of the operator is
\begin{equation}
A_{\gamma_{\mathrm{e}}}
=
\Phi_{L}(\gamma_{\mathrm{e}})
\otimes
\Phi_{L}(\gamma_{\mathrm{e}}).
\end{equation}
Here, $\otimes$ denotes the Kronecker product, and the identity follows directly from $C_L=\Phi_LO\Phi_L^T$.

Let $\Omega_K\subset\{1,\ldots,N^2\}$ be the set of $K$ sampled-mode indices, with $|\Omega_K|=K$, and let $S_{\Omega_K}\in\{0,1\}^{K\times N^2}$ be the corresponding row-selection matrix. The partially acquired vector is
\begin{equation}
\mathbf{y}_{\Omega_{K}}
=
S_{\Omega_{K}}
A_{\gamma_{\mathrm{e}}}
\mathbf{x}
+
\mathbf{n}_{\Omega_{K}}.
\end{equation}
Here, $\mathbf y_{\Omega_K}\in\mathbb C^K$ and $\mathbf n_{\Omega_K}=S_{\Omega_K}\mathbf n\in\mathbb C^K$. The transpose $S_{\Omega_K}^T$ reinserts these $K$ entries at their original positions in an otherwise zero $N^2$-component vector. The authorized reconstruction then applies the paired right-basis decoder,
\begin{equation}
\mathbf{x}_{\mathrm{auth}}^{(K)}
=
D_{\gamma_{\mathrm e}}
S_{\Omega_K}^{T}
\mathbf{y}_{\Omega_K},
\qquad
D_{\gamma_{\mathrm e}}
=
\Psi_R(\gamma_{\mathrm e})
\otimes
\Psi_R(\gamma_{\mathrm e}),
\label{eq:supp_ls_auth}
\end{equation}
where $\mathbf x_{\mathrm{auth}}^{(K)}\in\mathbb C^{N^2}$ is the vectorized reconstruction and $D_{\gamma_{\mathrm e}}\in\mathbb C^{N^2\times N^2}$ is the separable right-basis synthesis operator. Ignoring noise to expose the deterministic mapping, the effective object-to-reconstruction channel is
\begin{equation}
T_{\Omega_{K}}
=
D_{\gamma_{\mathrm{e}}}
S_{\Omega_{K}}^{T}
S_{\Omega_{K}}
A_{\gamma_{\mathrm{e}}}.
\label{eq:supp_ls_effective}
\end{equation}

At complete sampling, $K=N^{2}$ and $S_{\Omega_{K}}^{T}S_{\Omega_{K}}=I_{N^2}$. Because $\Phi_L\Psi_R=I_N$ and the matrices are square, $\Psi_R\Phi_L=I_N$ as well. The Kronecker-product relation then gives
\begin{equation}
T_{\Omega_{N^{2}}}
=
D_{\gamma_{\mathrm{e}}}
A_{\gamma_{\mathrm{e}}}
=
I_{N^2},
\end{equation}
so the paired right basis inverts the complete left-basis encoding operator. Under partial sampling, the diagonal mask $S_{\Omega_{K}}^{T}S_{\Omega_{K}}$ sets every unmeasured coefficient to zero, and $T_{\Omega_{K}}\in\mathbb C^{N^2\times N^2}$ becomes a truncated biorthogonal projection rather than the identity.

The sampled modes are selected according to the fixed eigenvalue-based ordering introduced in Sec.~I.3. For the mode indexed by $(i_y,i_x)$, the ordering quantity is
\begin{equation}
\eta_{i_y i_x}
=
\operatorname{Re}(\lambda_{i_y})
+
\operatorname{Re}(\lambda_{i_x}),
\label{eq:supp_ls_order}
\end{equation}
 where $i_y,i_x\in\{1,\ldots,N\}$, $\lambda_{i_y}$ and $\lambda_{i_x}$ are the corresponding one-dimensional eigenvalues, and $\operatorname{Re}$ denotes the real part. Under column-major vectorization, the corresponding flattened modal index is $q(i_y,i_x)=i_y+(i_x-1)N$. Let $\pi=(\pi_1,\ldots,\pi_{N^2})$ be the permutation of column-major flattened modal indices that sorts $\eta_{i_y i_x}$ in ascending order; stable original-index ordering breaks exact ties. Let $\rho\in(0,1]$ denote the retained fraction. The retained mode count and sampled index set are

\begin{equation}
K
=
\max\left\{
1,
\operatorname{round}\left(\rho N^{2}\right)
\right\},
\qquad
\Omega_{K}
=
\left\{
\pi_1,
\pi_2,
\ldots,
\pi_K
\right\}.
\end{equation}
Here, $\operatorname{round}$ denotes nearest-integer rounding and the outer maximum guarantees at least one retained mode. The sets are nested: if $\rho_1<\rho_2$, then $\Omega_{K_1}\subseteq\Omega_{K_2}$. Figure labels report the retained fraction as a sampling percentage, $100\rho\%$. For $N=64$, for example, $5\%$ sampling retains $K=\operatorname{round}(0.05\times4096)=205$ complex modes and requires $4K=820$ DMD frames in the four-mask implementation.

\subsection{Direct truncated biorthogonal reconstruction}

Reconstruction is performed directly in the two-dimensional coefficient
domain. Starting from the full left-basis coefficient matrix
$C_L(\gamma_{\mathrm e})
=\Phi_L(\gamma_{\mathrm e})O\Phi_L^T(\gamma_{\mathrm e})$
defined in ~\eqref{eq:supp_left_coeff}, we suppress its fixed
$\gamma_{\mathrm e}$ dependence for brevity, retain the first $K$
entries in the prescribed modal ordering, and set all remaining entries
to zero:

\begin{equation}
\left[C_{L}^{(K)}\right]_{i_y i_x}
=
\begin{cases}
\left[C_{L}\right]_{i_y i_x},
&
q(i_y,i_x)\in\Omega_{K},
\\[3pt]
0,
&
q(i_y,i_x)\notin\Omega_{K}
\end{cases}
\, .
\label{eq:supp_ls_ck}
\end{equation}
Here, $[C_L]_{i_y i_x}$ is the coefficient of the separable mode
$(i_y,i_x)$, and $C_L^{(K)}\in\mathbb C^{N\times N}$ is the
zero-filled coefficient matrix obtained by retaining the $K$ modes
indexed by $\Omega_K$. The superscript $(K)$ denotes the number of
retained modes. At complete sampling, $K=N^2$ and
$C_L^{(N^2)}=C_L$.

The zero-filling operation changes the number of nonzero measured
coefficients but not the matrix dimension. For $N=64$, the complete
basis contains $4096$ two-dimensional modes, while
$C_L^{(K)}$ and every reconstructed image remain $64\times64$.

\begin{figure}[htbp]
\centering
\includegraphics[width=0.98\linewidth]{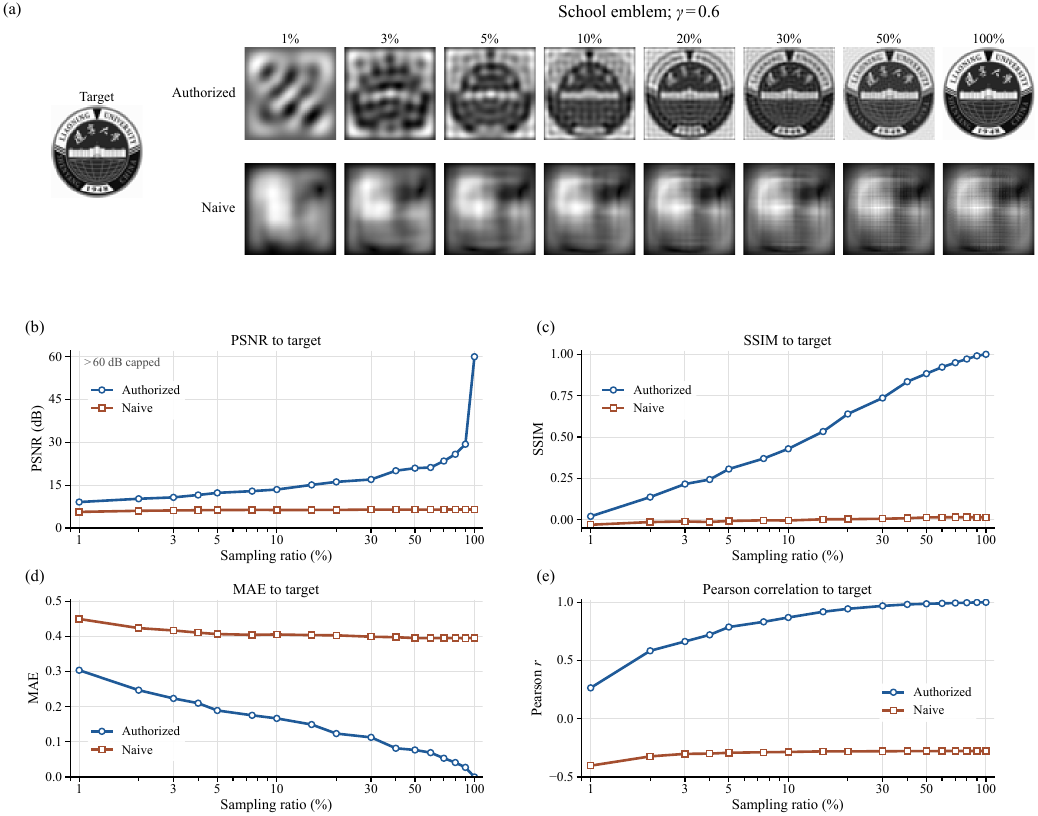}
\caption{Ordered partial-sampling simulation of authorized and
left-naive decoding for the school-emblem target, with $N=64$,
$\ell=6$, and $\gamma_{\mathrm e}=0.6$. (a) Normalized target and
authorized and left-naive reconstructions at the labeled sampling
percentages. (b)--(e) PSNR, SSIM, MAE, and Pearson correlation relative
to the normalized target. At each sampling percentage, the two
reconstruction rows use the same noiseless retained coefficient matrix
$C_L^{(K)}$. Values above $60$~dB are displayed at the plotting ceiling
defined in Sec.~III.2.}
\label{fig:s11_regular_school}
\end{figure}

For the numerical results in Figs.~\ref{fig:s11_regular_school}--\ref{fig:s14_near_ep_cameraman}, the same $C_L^{(K)}$ is supplied to the authorized and naive branches:
\begin{equation}
\begin{aligned}
O_{\mathrm{auth}}^{(K)}
&=
\Psi_R(\gamma_{\mathrm e})
C_L^{(K)}
\Psi_R^T(\gamma_{\mathrm e}),
\\
O_{\mathrm{naive}}^{(K)}
&=
\Phi_L^\dagger(\gamma_{\mathrm e})
C_L^{(K)}
\Phi_L^*(\gamma_{\mathrm e}).
\end{aligned}
\label{eq:supp_ls_decoders}
\end{equation}
The authorized branch synthesizes the retained coefficient matrix with the paired right basis, whereas the naive branch uses $\Phi_L^\dagger$ and $\Phi_L^*$, thereby acting through the same-side Gram channel rather than the biorthogonal inverse. The same authorized and naive reconstruction operators are used for both numerical and experimental data. For the experimental results in Figs.~\ref{fig:s17_exp_paired_subsampling} and \ref{fig:s18_exp_wrong_gamma_subsampling}, the authorized, naive, and parameter-mismatched reconstructions are obtained by applying the operators in ~\eqref{eq:supp_exp_decoders} to the retained measured coefficient matrix $C_L^{(K)}$. Within each column, $\Omega_K$ and $C_L^{(K)}$ are held fixed, so the row-to-row differences arise solely from the reconstruction operator, rather than from the measurements or sampling pattern.

The reconstruction consists only of deterministic ordered-mode truncation followed by direct basis synthesis. It uses no pseudoinverse, least-squares optimization, sparsity constraint, total-variation regularization, iterative solver, or learned model. The observed low-sampling behavior is therefore not a compressed-sensing guarantee; it reports how the two tested objects project onto the particular low-$\eta$ non-Hermitian modes retained by this protocol.

\subsection{Numerical validation under regular and near-EP encoding}

\begin{figure}[htbp]
\centering
\includegraphics[width=0.98\linewidth]{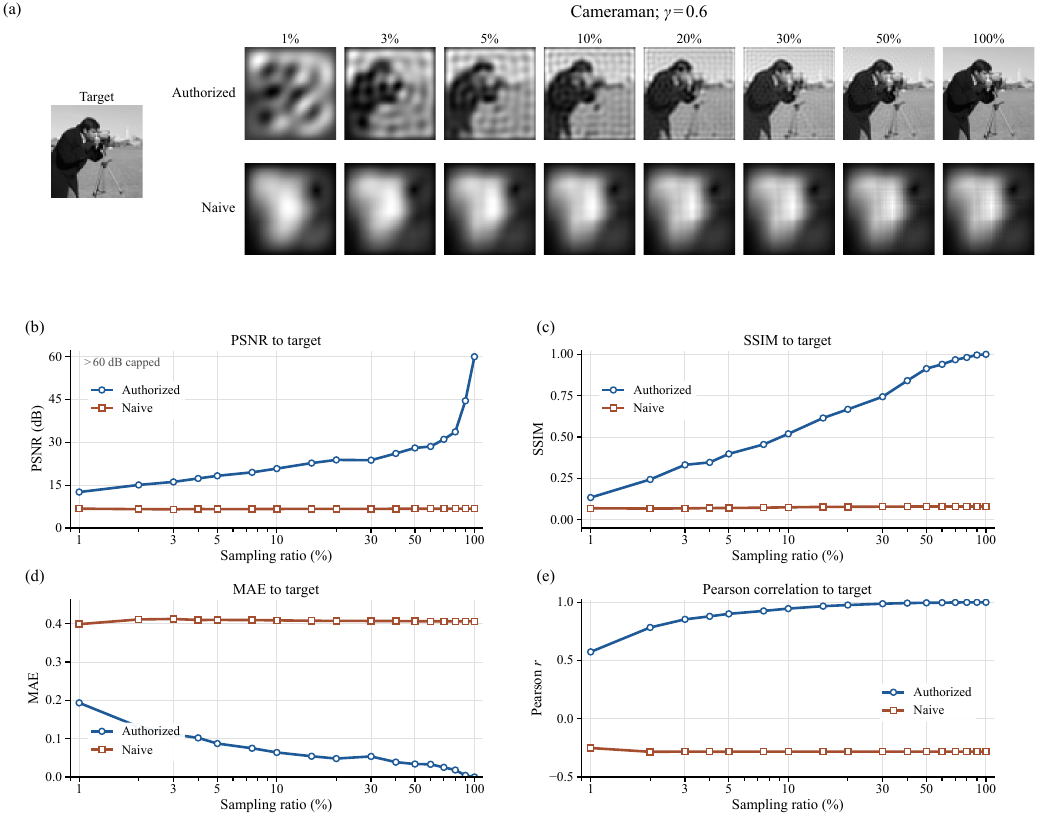}
\caption{Ordered partial-sampling simulation of authorized and
left-naive decoding for the Cameraman target, with $N=64$, $\ell=6$,
and $\gamma_{\mathrm e}=0.6$. (a) Normalized target and authorized and
left-naive reconstructions at the labeled sampling percentages.
(b)--(e) PSNR, SSIM, MAE, and Pearson correlation relative to the
normalized target. At each sampling percentage, the two reconstruction
rows use the same noiseless retained coefficient matrix $C_L^{(K)}$.
Values above $60$~dB are displayed at the plotting ceiling defined in
Sec.~III.2.}
\label{fig:s12_regular_cameraman}
\end{figure}

We first examine the regular operating coordinate $\gamma_{\mathrm e}=0.6$. Figures~\ref{fig:s11_regular_school} and \ref{fig:s12_regular_cameraman} use the same $N=64$, $\ell=6$ basis, ordered modal prefixes, and noiseless decoding rules. In each figure, panel (a) compares the target with authorized and naive reconstructions over the labeled sampling fractions. Panels (b)--(e) report PSNR, SSIM, MAE, and Pearson correlation relative to the normalized target. At each labeled sampling fraction, both rows are reconstructed from the same $C_L^{(K)}$ and differ only in the synthesis operator defined in ~\eqref{eq:supp_ls_decoders}.

Figure~\ref{fig:s11_regular_school} shows that the authorized prefix first recovers broad emblem components and then resolves the circular boundary, text band, and internal line structure. At $5\%$ sampling ($K=205$), the authorized PSNR, SSIM, and Pearson correlation are 12.31~dB, 0.306, and 0.788. At $50\%$ sampling ($K=2048$), they reach 20.95~dB, 0.883, and 0.987, while the MAE decreases from 0.189 to 0.077. The naive branch does not approach the target even when all coefficients are retained: at $100\%$ sampling, its PSNR, SSIM, and Pearson correlation are 6.46~dB, 0.015, and $-0.277$. Complete coefficient availability therefore removes truncation error but does not correct the same-side Gram metric.

Figure~\ref{fig:s12_regular_cameraman} shows the same separation for the continuous-tone Cameraman target. The authorized image is already recognizable at low sampling fractions and progressively recovers the figure, tripod, and background. Its PSNR, SSIM, and Pearson correlation increase from 18.28~dB, 0.398, and 0.901 at $5\%$ sampling to 28.04~dB, 0.913, and 0.996 at $50\%$ sampling. The authorized trend is strongly improving overall, although PSNR and MAE show a small local reversal between $20\%$ and $30\%$. Such a reversal is permitted because nested oblique projections need not be monotonic under pixel-domain metrics. By contrast, the naive branch remains broad and weakly correlated; its complete-sampling SSIM and Pearson correlation are only 0.080 and $-0.282$.

The two targets establish the same operator-level separation but different early-prefix performance. Cameraman has higher authorized correlation at very low sampling fractions, whereas the emblem requires more retained coefficients to resolve its dense circular text and internal structure. The fixed low-$\eta$ order should therefore be interpreted as a deterministic acquisition protocol, rather than as a universal best-$K$ approximation for arbitrary scenes.

\begin{figure}[htbp]
\centering
\includegraphics[width=0.98\linewidth]{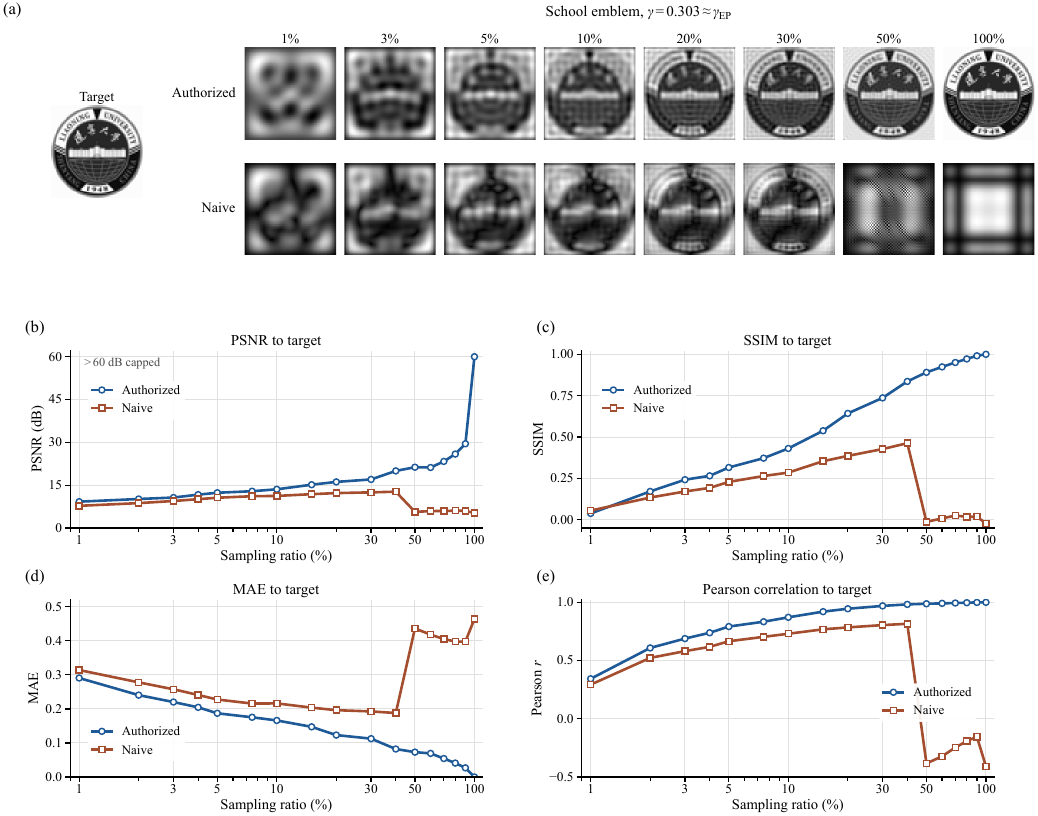}
\caption{Ordered partial-sampling simulation of authorized and
left-naive decoding for the school-emblem target near the tracked
pairwise EP, with $N=64$, $\ell=6$, and
$\gamma_{\mathrm e}=0.303$. The quoted near-EP coordinate is the rounded
value associated with the tracked finite-matrix eigenvalue pair.
(a) Normalized target and authorized and left-naive reconstructions at
the labeled sampling percentages. (b)--(e) PSNR, SSIM, MAE, and Pearson
correlation relative to the normalized target. At each sampling
percentage, the two reconstruction rows use the same noiseless retained
coefficient matrix $C_L^{(K)}$. Values above $60$~dB are displayed at
the plotting ceiling defined in Sec.~III.2.}
\label{fig:s13_near_ep_school}
\end{figure}

\begin{figure}[htbp]
\centering
\includegraphics[width=0.98\linewidth]{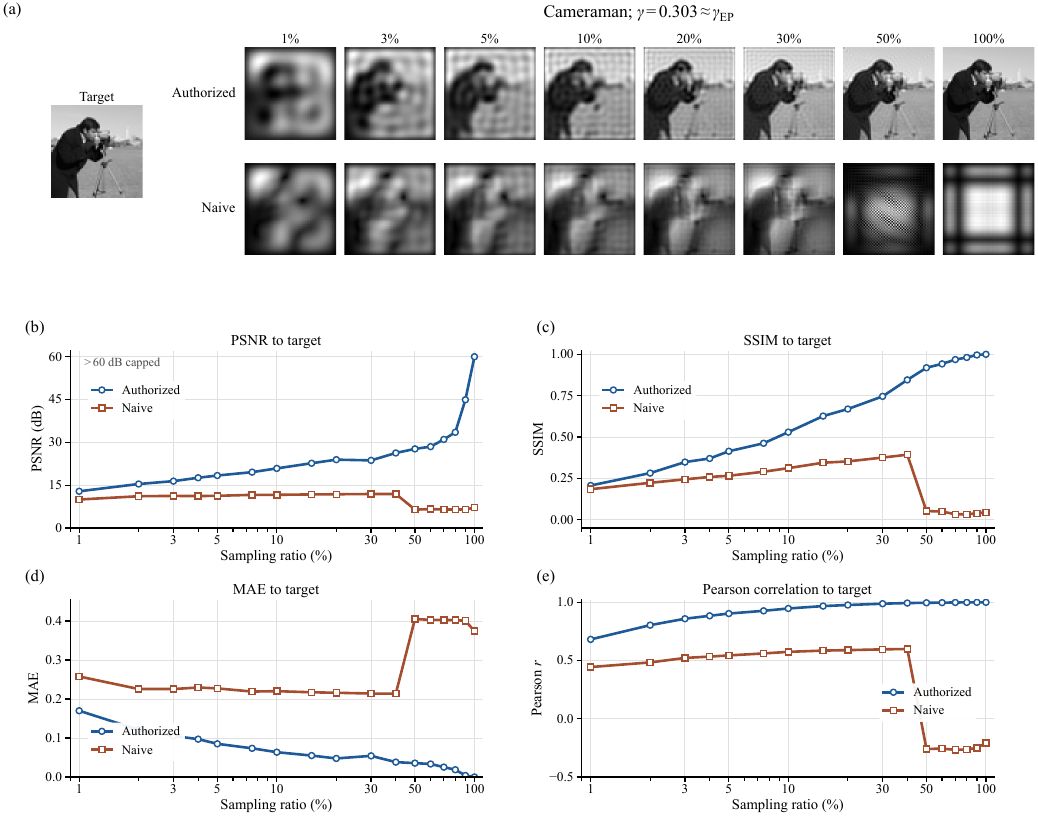}
\caption{Ordered partial-sampling simulation of authorized and
left-naive decoding for the Cameraman target near the tracked pairwise
EP, with $N=64$, $\ell=6$, and $\gamma_{\mathrm e}=0.303$. The quoted
near-EP coordinate is the rounded value associated with the tracked
finite-matrix eigenvalue pair. (a) Normalized target and authorized and
left-naive reconstructions at the labeled sampling percentages.
(b)--(e) PSNR, SSIM, MAE, and Pearson correlation relative to the
normalized target. At each sampling percentage, the two reconstruction
rows use the same noiseless retained coefficient matrix $C_L^{(K)}$.
Values above $60$~dB are displayed at the plotting ceiling defined in
Sec.~III.2.}
\label{fig:s14_near_ep_cameraman}
\end{figure}

We next repeat the protocol at the rounded near-EP coordinate $\gamma_{\mathrm e}=0.303\approx\gamma_{\mathrm{EP}}$. Figures~\ref{fig:s13_near_ep_school} and \ref{fig:s14_near_ep_cameraman} retain the targets, labeled sampling fractions, metric definitions, and noiseless coefficient sets of the regular group. This one-to-one arrangement isolates the change in finite-basis geometry.

For the school emblem, the authorized near-EP branch follows an almost identical progressive recovery and reaches the exact target at complete sampling. The naive branch initially appears to improve, reaching SSIM 0.427, Pearson correlation 0.805, and MAE 0.193 at $30\%$ sampling. Between $30\%$ and $50\%$, however, it collapses into a square-like Gram-channel pattern. At $50\%$ sampling, SSIM becomes $-0.013$, Pearson correlation becomes $-0.384$, and MAE rises to 0.436. The discontinuous deterioration shows that the newly admitted modes carry strongly amplified same-side overlap weights.

The Cameraman target exhibits the same threshold-like naive failure. Its naive SSIM, Pearson correlation, and MAE change from 0.376, 0.596, and 0.214 at $30\%$ sampling to 0.053, $-0.259$, and 0.405 at $50\%$ sampling. The detailed pre-collapse image differs from the emblem because each target weights the retained modes differently, whereas the common collapse interval and final square-lobe morphology follow the near-EP Gram channel.

These near-EP calculations do not establish a universal sampling advantage at an exceptional point. In noiseless arithmetic, paired synthesis remains algebraically consistent and reaches the target when all modes are retained. The nonmonotonic naive response demonstrates a different result: increasing the number of measurements does not guarantee improved reconstruction when the retained coefficients are synthesized through a strongly nonorthogonal, unmatched metric.

\subsection{Parameter-mismatched decoding under partial sampling}
We next test whether a larger sampling ratio can compensate for an incorrect decoding coordinate. For an object encoded at $\gamma_{\mathrm e}$ and decoded at $\gamma_{\mathrm d}$, the parameter-mismatched reconstruction is
\begin{equation}
\widetilde O^{(K)}
\left(
\gamma_{\mathrm d},
\gamma_{\mathrm e}
\right)
=
\Psi_R(\gamma_{\mathrm d})
C_L^{(K)}(\gamma_{\mathrm e})
\Psi_R^T(\gamma_{\mathrm d}),
\qquad
\gamma_{\mathrm d}\neq\gamma_{\mathrm e}.
\label{eq:supp_ls_mismatch}
\end{equation}
Here, $C_L^{(K)}(\gamma_{\mathrm e})$ is the fixed zero-filled coefficient matrix, and only the computational right-basis decoder changes. Using the cross-parameter overlap operator $M_L$ defined in ~\eqref{eq:supp_mismatch_operator}, the complete-sampling limit becomes
\begin{equation}
\widetilde O^{(N^2)}
\left(
\gamma_{\mathrm d},
\gamma_{\mathrm e}
\right)
=
M_L(\gamma_{\mathrm d},\gamma_{\mathrm e})
O
M_L^T(\gamma_{\mathrm d},\gamma_{\mathrm e}).
\label{eq:supp_ls_mismatch_limit}
\end{equation}
This limit equals $O$ only on the matched diagonal, where $\gamma_{\mathrm d}=\gamma_{\mathrm e}$ and $M_L=I_N$. Increasing the sampling fraction can therefore remove coefficient omission, but it cannot change the limiting cross-parameter channel.

\begin{figure}[htbp]
\centering
\includegraphics[width=0.98\linewidth]{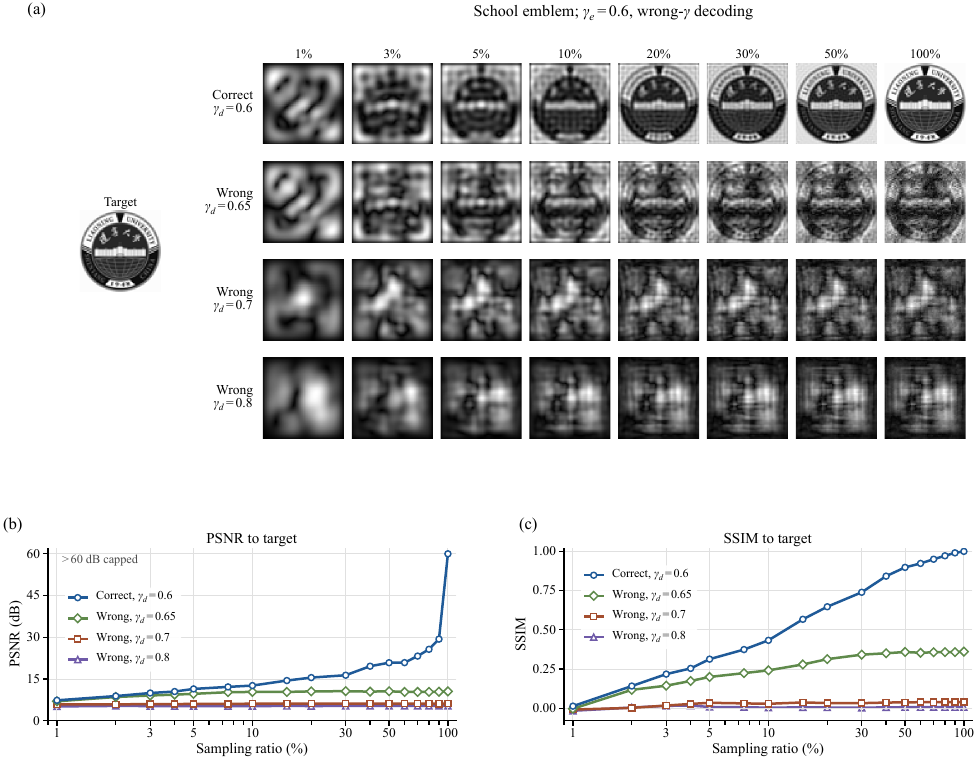}
\caption{Partial-sampling simulation of matched and
parameter-mismatched decoding for the school-emblem target, with
$N=64$, $\ell=6$, and $\gamma_{\mathrm e}=0.6$. (a) Reconstructions
for the labeled $\gamma_{\mathrm d}$ values and sampling percentages.
(b), (c) PSNR and SSIM relative to the normalized target. Within each
sampling column, all rows use the same noiseless retained coefficient
matrix $C_L^{(K)}$.}
\label{fig:s15_wrong_gamma_school}
\end{figure}

\begin{figure}[htbp]
\centering
\includegraphics[width=0.98\linewidth]{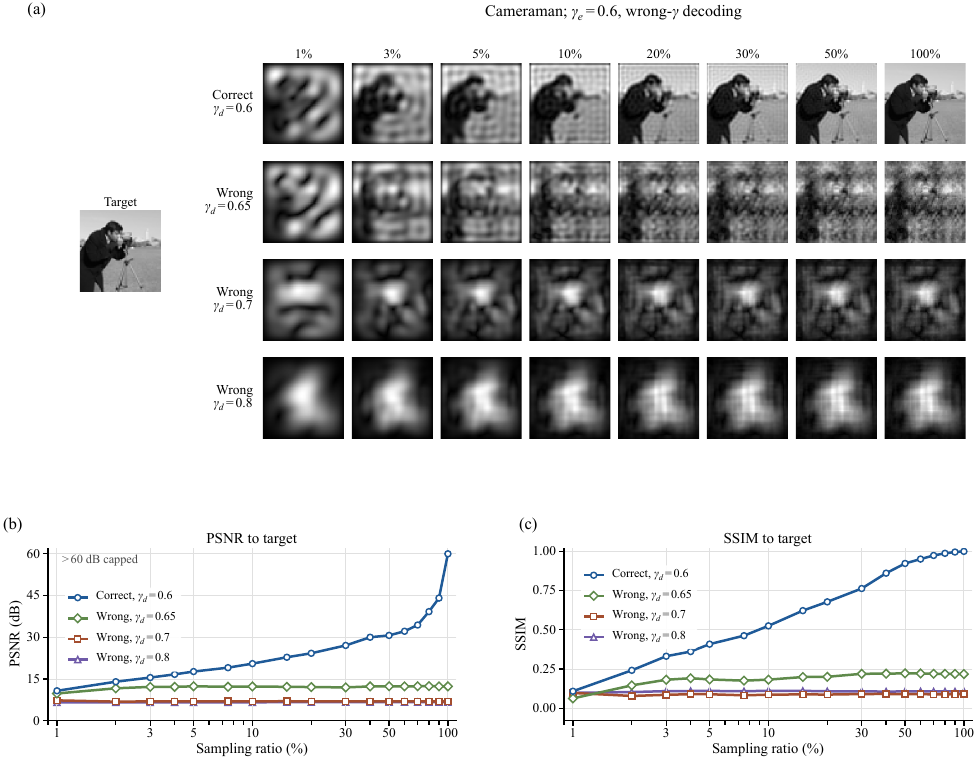}
\caption{Partial-sampling simulation of matched and
parameter-mismatched decoding for the Cameraman target, using the same
protocol parameters as in Fig.~\ref{fig:s15_wrong_gamma_school}:
$N=64$, $\ell=6$, and $\gamma_{\mathrm e}=0.6$. This cross-object
comparison evaluates the decoding response for a target with different
spatial structure. (a) Reconstructions for the labeled
$\gamma_{\mathrm d}$ values and sampling percentages. (b), (c) PSNR and
SSIM relative to the normalized target. Within each sampling column,
all rows use the same noiseless retained coefficient matrix
$C_L^{(K)}$.}
\label{fig:s16_wrong_gamma_cameraman}
\end{figure}

To determine whether the mismatch response is tied to a particular object spectrum, Figs.~\ref{fig:s15_wrong_gamma_school} and \ref{fig:s16_wrong_gamma_cameraman} compare two structurally distinct targets under the same $N=64$, $\ell=6$, and $\gamma_{\mathrm e}=0.6$ protocol. The school emblem concentrates contrast in sharp circular boundaries, lettering, and nearly binary geometric regions, whereas the Cameraman image combines localized edges with continuous gray-scale gradients and texture. Panel (a) of each figure compares the matched row with the labeled mismatched decoding coordinates over the displayed sampling fractions. Panels (b) and (c) report PSNR and SSIM over the complete calculated sampling sequence. Within each column, all rows use the same noiseless $C_L^{(K)}$, so the comparison changes only the decoding coordinate and tests which features are imposed by the cross-parameter operator and which depend on the target spectrum.

Figure~\ref{fig:s15_wrong_gamma_school} shows how the high-contrast emblem responds to the two error sources. On the matched row, the low-sampling blurred lobes develop into the circular rim, central band, lettering, and internal line structure, and the reconstruction reaches the exact noiseless full-basis limit. With the mild mismatch $\gamma_{\mathrm d}=0.65$, adding coefficients initially improves the recognizable emblem-scale layout, but the sequence then stabilizes as a grainy, boundary-fragmented image rather than converging to the target. At complete sampling, its PSNR and SSIM are only 10.56~dB and 0.361. The $\gamma_{\mathrm d}=0.7$ and $0.8$ rows are instead dominated by compact, parameter-specific lobes and cross-like structures; neither the circular boundary nor the internal emblem details are recovered. Thus, sharp target boundaries can leave recognizable fragments under a small mismatch, but they do not restore the biorthogonal identity channel.

To test whether this behavior persists after changing from a nearly binary geometric target to a continuous-tone natural scene, Fig.~\ref{fig:s16_wrong_gamma_cameraman} repeats the protocol for Cameraman without changing the basis, mode order, sampling fractions, or decoding coordinates.

On the matched row of Fig.~\ref{fig:s16_wrong_gamma_cameraman}, the global silhouette appears first, followed by the photographer, camera, tripod, and background as the sampling fraction increases; complete sampling again reaches the exact noiseless target. For $\gamma_{\mathrm d}=0.65$, the early broad components give way to a persistent fine-grained and banded distortion that does not recover the continuous gray-scale scene. Its complete-sampling PSNR is 12.36~dB, higher than the corresponding school-emblem value, whereas its SSIM is only 0.218, lower than the emblem value of 0.361. This opposite cross-target ordering shows why pixel error and structural preservation must be interpreted together: a lower average intensity error does not imply better recovery of natural-image morphology. The $\gamma_{\mathrm d}=0.7$ and $0.8$ rows remain dominated by parameter-dependent structures and never recover the joint photographer--camera--tripod geometry.

Taken together, Figs.~\ref{fig:s15_wrong_gamma_school} and
\ref{fig:s16_wrong_gamma_cameraman} establish a common operator-level
result together with a target-dependent image-level response. As the
sampling percentage increases, the matched rows progressively recover
their respective targets and reach the exact noiseless full-basis limit.
At complete sampling,
$M_L(\gamma_{\mathrm e},\gamma_{\mathrm e})=I_N$, so the matched
reconstruction reduces to the biorthogonal identity channel. For
$\gamma_{\mathrm d}\neq\gamma_{\mathrm e}$, however, the corresponding
limit remains
$M_L(\gamma_{\mathrm d},\gamma_{\mathrm e})
O M_L^T(\gamma_{\mathrm d},\gamma_{\mathrm e})$ and is therefore
constrained by a nonidentity cross-parameter operator. The
$\gamma_{\mathrm d}=0.65$ branches can improve initially as coefficient
truncation is reduced, but they subsequently approach
mismatch-limited states rather than converging to the targets. The
$\gamma_{\mathrm d}=0.7$ and $0.8$ branches remain at lower fidelity.
Because the resulting PSNR and SSIM rankings depend jointly on the
decoding coordinate and the target spectrum, their detailed variation
need not be strictly monotonic in
$|\gamma_{\mathrm d}-\gamma_{\mathrm e}|$ for every metric.

The two targets also separate the common mismatch mechanism from its target-dependent spatial manifestation. The sharp geometric boundaries of the school emblem can leave recognizable boundary fragments under a mild mismatch, whereas the continuous gray-scale gradients and textures of the Cameraman target are redistributed into fine-grained, banded, or spatially localized distortions. At low sampling percentages, the reconstructed images contain the combined effects of coefficient truncation and parameter mismatch. As the sampling percentage increases, coefficient omission progressively recedes, while the parameter-dependent transformation becomes dominant. Depending on $\gamma_d$ and the target spectrum, the mismatched reconstructions consequently evolve from broad, blurred components toward structured artifacts or parameter-specific lobes rather than recovering the true scene. The matched row is fixed by biorthogonal inversion, whereas the morphology of each mismatched row is determined by the action of the nonidentity cross-parameter operator on the specific object spectrum. Increasing the sampling percentage therefore reveals this operator more completely; it does not restore the identity channel.

\subsection{Experimental validation under partial sampling}
\begin{figure}[htbp]
\centering
\includegraphics[width=0.98\linewidth]{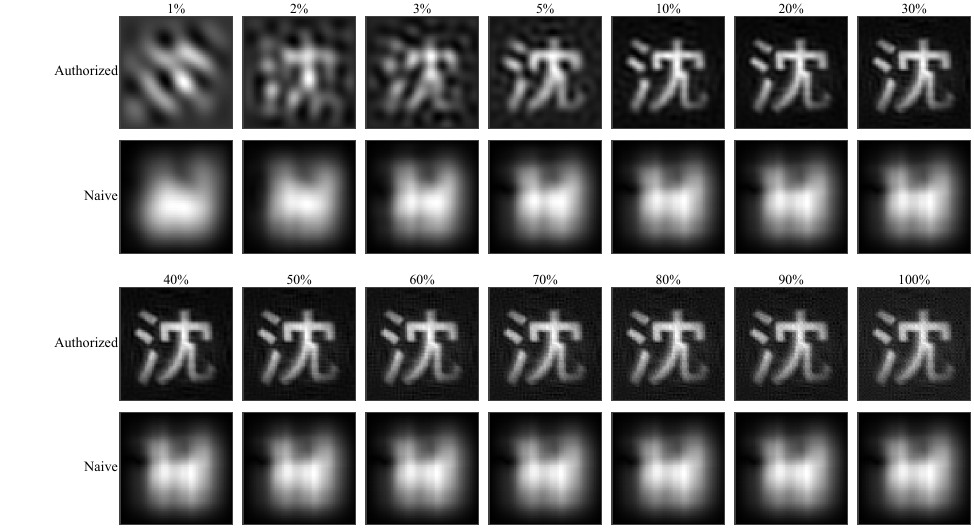}
\caption{Experimental comparison of authorized and naive reconstructions under ordered partial sampling at $\gamma_{\mathrm e}=0.6$ for the reflective Chinese-character target. The first row pair shows $1\%$--$30\%$ sampling, and the second pair shows $40\%$--$100\%$ sampling. Both rows in each column use the same measured $C_L^{(K)}$. Each retained complex coefficient is synthesized from four projected DMD patterns.}
\label{fig:s17_exp_paired_subsampling}
\end{figure}

Each experimental complex coefficient is synthesized from the four non-negative intensity measurements in ~\eqref{eq:four_mask}. Retaining $K$ modes therefore requires $4K$ projected DMD patterns. The first $K$ calibrated coefficients are returned to their prescribed entries in $C_L^{(K)}$, while all unmeasured entries are set to zero. The resulting coefficient matrix is then supplied directly to the reconstruction operators in ~\eqref{eq:supp_exp_decoders}. At a fixed sampling fraction, changing the reconstruction operator requires no additional optical measurement.

Figure~\ref{fig:s17_exp_paired_subsampling} compares the experimental authorized and naive reconstructions at $\gamma_{\mathrm e}=0.6$. The two row pairs cover the labeled sampling ranges of $1\%$--$30\%$ and $40\%$--$100\%$. Within each column, both reconstructions use the same measured $C_L^{(K)}$. The naive reconstruction is defined by
\begin{equation}
O_{\mathrm{naive}}^{(K)}
=
\Phi_L^\dagger(\gamma_{\mathrm e})
C_L^{(K)}
\Phi_L^*(\gamma_{\mathrm e}).
\end{equation}
This is the same reconstruction operator used in the numerical analysis, with measured coefficients substituted for simulated ones. The comparison therefore isolates the effect of the reconstruction operator while holding the retained experimental data fixed.

The authorized sequence initially contains only broad spatial components. The principal stroke locations emerge as the sampling fraction increases, and the Chinese character becomes recognizable near $5\%$ sampling in this data set. Higher sampling fractions improve stroke continuity, background separation, and residual texture. By contrast, the naive reconstruction retains a broad, vertically organized Gram-channel pattern and does not recover the character at complete sampling. Because both reconstructions use the same measured $C_L^{(K)}$, their persistent difference arises from the reconstruction operator rather than from coefficient availability or sampling order.

\begin{figure}[htbp]
\centering
\includegraphics[width=0.98\linewidth]{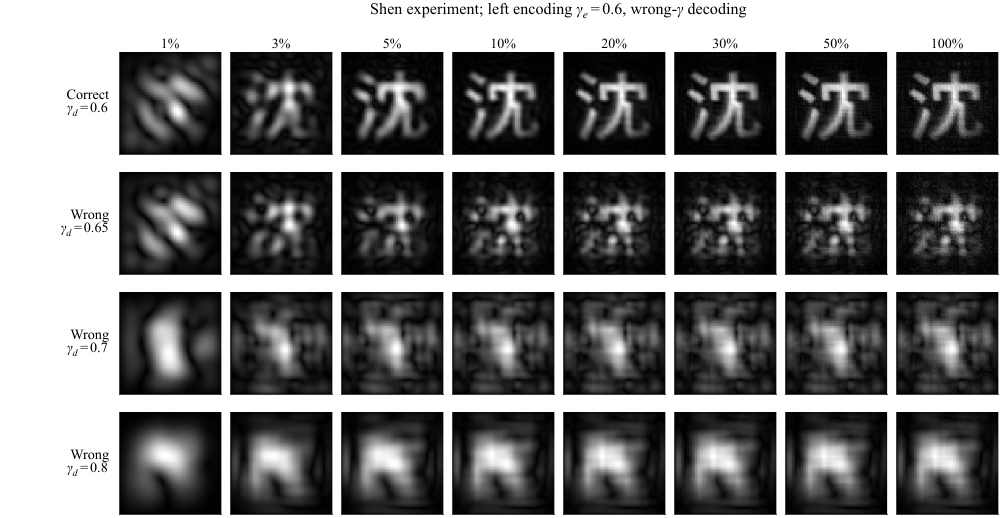}
\caption{Experimental partial-sampling reconstruction under matched and parameter-mismatched decoding for the reflective Chinese-character target. The coefficients are encoded at $\gamma_{\mathrm e}=0.6$; rows show the labeled $\gamma_{\mathrm d}$ values, and columns show the labeled sampling fractions. Every row within a column uses the same measured $C_L^{(K)}$. Each retained complex coefficient is synthesized from four projected DMD patterns.}
\label{fig:s18_exp_wrong_gamma_subsampling}
\end{figure}

Figure~\ref{fig:s18_exp_wrong_gamma_subsampling} next examines parameter-mismatched decoding of the same measured coefficients. The encoding coordinate is fixed at $\gamma_{\mathrm e}=0.6$, while the rows use the labeled decoding coordinates $\gamma_{\mathrm d}$. The columns show the labeled sampling fractions. Every row within a column is reconstructed from the same measured $C_L^{(K)}$, so the comparison isolates the effect of the computational decoding coordinate.

The matched row progressively approaches the measured full-sampling authorized reconstruction, consistent with Fig.~\ref{fig:s17_exp_paired_subsampling}. The $\gamma_{\mathrm d}=0.65$ row develops repeated and fragmented stroke-like features as additional coefficients are included. The $\gamma_{\mathrm d}=0.7$ and $0.8$ rows remain dominated by broad parameter-dependent structures even at complete sampling. Increasing the sampling fraction therefore reduces coefficient omission but cannot restore the identity channel when $\gamma_{\mathrm d}\neq\gamma_{\mathrm e}$. Instead, the additional measurements reveal the nonidentity cross-parameter operator more completely.

The partial-sampling panels are retrospective prefix reconstructions obtained from complete experimental acquisitions. They show which structures survive ordered coefficient retention and how the same retained matrix responds to different reconstruction operators. They do not constitute a real-time early-stop or dynamic-scene benchmark. Such a claim would require prospectively shortened acquisitions, repeated timing measurements, and uncertainty analysis under scene motion and detector drift.
Taken together, Figs.~\ref{fig:s17_exp_paired_subsampling} and \ref{fig:s18_exp_wrong_gamma_subsampling} experimentally separate coefficient availability from the choice of reconstruction operator. Ordered truncation determines which measured coefficients are retained. The reconstruction operator determines how those coefficients are synthesized into an image. The observed low-sampling behavior is therefore a direct truncated modal projection, rather than the output of an iterative, sparsity-based, or learned reconstruction algorithm.

\enlargethispage{2\baselineskip}

\renewcommand{\refname}{Supporting References}


\begin{thebibliography}{61}
\providecommand{\natexlab}[1]{#1}
\providecommand{\url}[1]{\texttt{#1}}
\expandafter\ifx\csname urlstyle\endcsname\relax
  \providecommand{\doi}[1]{doi: #1}\else
  \providecommand{\doi}{doi: \begingroup \urlstyle{rm}\Url}\fi

\bibitem[Bender and Boettcher(1998)]{Bender1998PRL}
Carl~M. Bender and Stefan Boettcher.
\newblock Real spectra in non-hermitian hamiltonians having {$\mathcal{PT}$}
  symmetry.
\newblock \emph{Phys. Rev. Lett.}, 80:\penalty0 5243--5246, 1998.
\newblock \doi{10.1103/PhysRevLett.80.5243}.

\bibitem[Bender et~al.(2002)Bender, Brody, and Jones]{Bender2002PRL}
Carl~M. Bender, Dorje~C. Brody, and Hugh~F. Jones.
\newblock Complex extension of quantum mechanics.
\newblock \emph{Phys. Rev. Lett.}, 89:\penalty0 270401, 2002.
\newblock \doi{10.1103/PhysRevLett.89.270401}.

\bibitem[Mostafazadeh(2002)]{Mostafazadeh2002JMP}
Ali Mostafazadeh.
\newblock Pseudo-hermiticity versus {$\mathcal{PT}$} symmetry: The necessary
  condition for the reality of the spectrum of a non-hermitian hamiltonian.
\newblock \emph{J. Math. Phys.}, 43:\penalty0 205--214, 2002.
\newblock \doi{10.1063/1.1418246}.

\bibitem[Brody(2014)]{Brody2014JPA}
Dorje~C. Brody.
\newblock Biorthogonal quantum mechanics.
\newblock \emph{J. Phys. A: Math. Theor.}, 47:\penalty0 035305, 2014.
\newblock \doi{10.1088/1751-8113/47/3/035305}.

\bibitem[Edgar et~al.(2019)Edgar, Gibson, and Padgett]{Edgar2019}
Matthew~P. Edgar, Graham~M. Gibson, and Miles~J. Padgett.
\newblock Principles and prospects for single-pixel imaging.
\newblock \emph{Nature Photonics}, 13:\penalty0 13--20, 2019.
\newblock \doi{10.1038/s41566-018-0300-7}.

\bibitem[Duarte et~al.(2008)Duarte, Davenport, Takhar, Laska, Sun, Kelly, and
  Baraniuk]{Duarte2008}
Marco~F. Duarte, Mark~A. Davenport, Dharmpal Takhar, Jason~N. Laska, Ting Sun,
  Kevin~F. Kelly, and Richard~G. Baraniuk.
\newblock Single-pixel imaging via compressive sampling.
\newblock \emph{IEEE Signal Process. Mag.}, 25\penalty0 (2):\penalty0 83--91,
  2008.
\newblock \doi{10.1109/MSP.2007.914730}.

\bibitem[Shapiro(2008)]{Shapiro2008}
Jeffrey~H. Shapiro.
\newblock Computational ghost imaging.
\newblock \emph{Phys. Rev. A}, 78:\penalty0 061802, 2008.
\newblock \doi{10.1103/PhysRevA.78.061802}.

\bibitem[Gatti et~al.(2004)Gatti, Brambilla, Bache, and Lugiato]{Gatti2004}
Alessandra Gatti, Enrico Brambilla, Morten Bache, and Luigi~A. Lugiato.
\newblock Ghost imaging with thermal light: Comparing entanglement and
  classical correlation.
\newblock \emph{Phys. Rev. Lett.}, 93:\penalty0 093602, 2004.
\newblock \doi{10.1103/PhysRevLett.93.093602}.

\bibitem[Wagadarikar et~al.(2008)Wagadarikar, John, Willett, and
  Brady]{Wagadarikar2008}
Ashwin~A. Wagadarikar, Renu John, Rebecca~M. Willett, and David~J. Brady.
\newblock Single disperser design for coded aperture snapshot spectral imaging.
\newblock \emph{Appl. Opt.}, 47\penalty0 (10):\penalty0 B44--B51, 2008.
\newblock \doi{10.1364/AO.47.000B44}.

\bibitem[Donoho(2006)]{Donoho2006}
David~L. Donoho.
\newblock Compressed sensing.
\newblock \emph{IEEE Trans. Inf. Theory}, 52:\penalty0 1289--1306, 2006.
\newblock \doi{10.1109/TIT.2006.871582}.

\bibitem[Cand{\`e}s et~al.(2006)Cand{\`e}s, Romberg, and Tao]{Candes2006}
Emmanuel~J. Cand{\`e}s, Justin Romberg, and Terence Tao.
\newblock Robust uncertainty principles: Exact signal reconstruction from
  highly incomplete frequency information.
\newblock \emph{IEEE Trans. Inf. Theory}, 52:\penalty0 489--509, 2006.
\newblock \doi{10.1109/TIT.2005.862083}.

\bibitem[Miao et~al.(1999)Miao, Charalambous, Kirz, and Sayre]{Miao1999}
Jianwei Miao, Pambos Charalambous, Janos Kirz, and David Sayre.
\newblock Extending the methodology of {X}-ray crystallography to allow imaging
  of micrometre-sized non-crystalline specimens.
\newblock \emph{Nature}, 400\penalty0 (6742):\penalty0 342--344, 1999.
\newblock \doi{10.1038/22498}.

\bibitem[Rodenburg and Faulkner(2004)]{Rodenburg2004}
J.~M. Rodenburg and H.~M.~L. Faulkner.
\newblock A phase retrieval algorithm for shifting illumination.
\newblock \emph{Appl. Phys. Lett.}, 85\penalty0 (20):\penalty0 4795--4797,
  2004.
\newblock \doi{10.1063/1.1823034}.

\bibitem[Zheng et~al.(2013)Zheng, Horstmeyer, and Yang]{Zheng2013}
Guoan Zheng, Roarke Horstmeyer, and Changhuei Yang.
\newblock Wide-field, high-resolution {Fourier} ptychographic microscopy.
\newblock \emph{Nat. Photonics}, 7\penalty0 (9):\penalty0 739--745, 2013.
\newblock \doi{10.1038/nphoton.2013.187}.

\bibitem[Popoff et~al.(2010)Popoff, Lerosey, Carminati, Fink, Boccara, and
  Gigan]{Popoff2010}
S.~M. Popoff, G.~Lerosey, R.~Carminati, M.~Fink, A.~C. Boccara, and S.~Gigan.
\newblock Measuring the transmission matrix in optics: An approach to the study
  and control of light propagation in disordered media.
\newblock \emph{Phys. Rev. Lett.}, 104:\penalty0 100601, 2010.
\newblock \doi{10.1103/PhysRevLett.104.100601}.

\bibitem[Bertolotti et~al.(2012)Bertolotti, van Putten, Blum, Lagendijk, Vos,
  and Mosk]{Bertolotti2012}
Jacopo Bertolotti, Elbert~G. van Putten, Christian Blum, Ad~Lagendijk,
  Willem~L. Vos, and Allard~P. Mosk.
\newblock Non-invasive imaging through opaque scattering layers.
\newblock \emph{Nature}, 491:\penalty0 232--234, 2012.
\newblock \doi{10.1038/nature11578}.

\bibitem[Rotter and Gigan(2017)]{Rotter2017}
Stefan Rotter and Sylvain Gigan.
\newblock Light fields in complex media: Mesoscopic scattering meets wave
  control.
\newblock \emph{Reviews of Modern Physics}, 89:\penalty0 015005, 2017.
\newblock \doi{10.1103/RevModPhys.89.015005}.

\bibitem[Llull et~al.(2013)Llull, Liao, Yuan, Yang, Kittle, Carin, Sapiro, and
  Brady]{Llull2013OptExpress}
Patrick Llull, Xuejun Liao, Xin Yuan, Jianbo Yang, David Kittle, Lawrence
  Carin, Guillermo Sapiro, and David~J. Brady.
\newblock Coded aperture compressive temporal imaging.
\newblock \emph{Opt. Express}, 21\penalty0 (9):\penalty0 10526--10545, 2013.
\newblock \doi{10.1364/OE.21.010526}.

\bibitem[Yuan et~al.(2021)Yuan, Brady, and Katsaggelos]{Yuan2021SPM}
Xin Yuan, David~J. Brady, and Aggelos~K. Katsaggelos.
\newblock Snapshot compressive imaging: Theory, algorithms, and applications.
\newblock \emph{IEEE Signal Process. Mag.}, 38\penalty0 (2):\penalty0 65--88,
  2021.
\newblock \doi{10.1109/MSP.2020.3023869}.

\bibitem[Velten et~al.(2012)Velten, Willwacher, Gupta, Veeraraghavan, Bawendi,
  and Raskar]{Velten2012}
Andreas Velten, Thomas Willwacher, Otkrist Gupta, Ashok Veeraraghavan,
  Moungi~G. Bawendi, and Ramesh Raskar.
\newblock Recovering three-dimensional shape around a corner using ultrafast
  time-of-flight imaging.
\newblock \emph{Nat. Commun.}, 3:\penalty0 745, 2012.
\newblock \doi{10.1038/ncomms1747}.

\bibitem[O'Toole et~al.(2018)O'Toole, Lindell, and Wetzstein]{OToole2018}
Matthew O'Toole, David~B. Lindell, and Gordon Wetzstein.
\newblock Confocal non-line-of-sight imaging based on the light-cone transform.
\newblock \emph{Nature}, 555\penalty0 (7696):\penalty0 338--341, 2018.
\newblock \doi{10.1038/nature25489}.

\bibitem[Faccio et~al.(2020)Faccio, Velten, and Wetzstein]{Faccio2020}
Daniele Faccio, Andreas Velten, and Gordon Wetzstein.
\newblock Non-line-of-sight imaging.
\newblock \emph{Nature Reviews Physics}, 2:\penalty0 318--327, 2020.
\newblock \doi{10.1038/s42254-020-0174-8}.

\bibitem[Sun et~al.(2013)Sun, Edgar, Bowman, Vittert, Welsh, Bowman, and Padgett]{Sun2013Science}
Baoqing Sun, Matthew~P. Edgar, Richard Bowman, Leroy~E. Vittert, Stuart Welsh, Adrian Bowman, and Miles~J. Padgett.
\newblock 3D computational imaging with single-pixel detectors.
\newblock \emph{Science}, 340:\penalty0 844--847, 2013.
\newblock \doi{10.1126/science.1234454}.

\bibitem[Gehm et~al.(2007)Gehm, John, Brady, Willett, and Schulz]{Gehm2007OptExpress}
Michael~E. Gehm, Renu John, David~J. Brady, Rebecca~M. Willett, and Timothy~J. Schulz.
\newblock Single-shot compressive spectral imaging with a dual-disperser architecture.
\newblock \emph{Opt. Express}, 15\penalty0 (21):\penalty0 14013--14027, 2007.
\newblock \doi{10.1364/OE.15.014013}.

\bibitem[Clemente et~al.(2013)Clemente, Dur{\'a}n, Tajahuerce, Andr{\'e}s, Climent, and Lancis]{Clemente2013OL}
Pere Clemente, Vicente Dur{\'a}n, Enrique Tajahuerce, Pedro Andr{\'e}s, Vicent Climent, and Jes{\'u}s Lancis.
\newblock Compressive holography with a single-pixel detector.
\newblock \emph{Opt. Lett.}, 38:\penalty0 2524--2527, 2013.
\newblock \doi{10.1364/OL.38.002524}.

\bibitem[Teague(1983)]{Teague1983JOSA}
Michael~Reed Teague.
\newblock Deterministic phase retrieval: A Green's function solution.
\newblock \emph{J. Opt. Soc. Am.}, 73:\penalty0 1434--1441, 1983.
\newblock \doi{10.1364/JOSA.73.001434}.

\bibitem[Park et~al.(2018)Park, Depeursinge, and Popescu]{Park2018QPI}
YongKeun Park, Christian Depeursinge, and Gabriel Popescu.
\newblock Quantitative phase imaging in biomedicine.
\newblock \emph{Nat. Photonics}, 12:\penalty0 578--589, 2018.
\newblock \doi{10.1038/s41566-018-0253-x}.

\bibitem[Li et~al.(2021)Li, Bian, Zheng, Maiden, Liu, Li, Suo, Dai, and Zhang]{Li2021OL}
Meng Li, Liheng Bian, Guoan Zheng, Andrew Maiden, Yang Liu, Yiming Li, Jinli Suo, Qionghai Dai, and Jun Zhang.
\newblock Single-pixel ptychography.
\newblock \emph{Opt. Lett.}, 46:\penalty0 1624--1627, 2021.
\newblock \doi{10.1364/OL.417039}.

\bibitem[Cand{\`e}s and Wakin(2008)]{CandesWakin2008IEEESPM}
Emmanuel~J. Cand{\`e}s and Michael~B. Wakin.
\newblock An introduction to compressive sampling.
\newblock \emph{IEEE Signal Process. Mag.}, 25:\penalty0 21--30, 2008.
\newblock \doi{10.1109/MSP.2007.914731}.

\bibitem[R{\'e}fr{\'e}gier and Javidi(1995)]{Refregier1995OptLett}
Philippe R{\'e}fr{\'e}gier and Bahram Javidi.
\newblock Optical image encryption based on input plane and fourier plane
  random phase encoding.
\newblock \emph{Opt. Lett.}, 20:\penalty0 767--769, 1995.
\newblock \doi{10.1364/OL.20.000767}.

\bibitem[Situ and Zhang(2004)]{Situ2004OptLett}
Guohai Situ and Jingjuan Zhang.
\newblock Double random-phase encoding in the fresnel domain.
\newblock \emph{Opt. Lett.}, 29:\penalty0 1584--1586, 2004.
\newblock \doi{10.1364/OL.29.001584}.

\bibitem[Matoba et~al.(2009)Matoba, Nomura, P{\'e}rez-Cabr{\'e}, Mill{\'a}n,
  and Javidi]{Matoba2009ProcIEEE}
Osamu Matoba, Takanori Nomura, Elisabet P{\'e}rez-Cabr{\'e}, Mar{\'i}a~S.
  Mill{\'a}n, and Bahram Javidi.
\newblock Optical techniques for information security.
\newblock \emph{Proc. IEEE}, 97:\penalty0 1128--1148, 2009.
\newblock \doi{10.1109/JPROC.2009.2018367}.

\bibitem[Frauel et~al.(2007)Frauel, Castro, Naughton, and Javidi]{Frauel2007OptExpress}
Yann Frauel, Albertina Castro, Thomas~J. Naughton, and Bahram Javidi.
\newblock Resistance of the double random phase encryption against various attacks.
\newblock \emph{Opt. Express}, 15\penalty0 (16):\penalty0 10253--10265, 2007.
\newblock \doi{10.1364/OE.15.010253}.

\bibitem[Popoff et~al.(2010)Popoff, Lerosey, Fink, Boccara, and Gigan]{Popoff2010NatCommunImage}
S{\'e}bastien~M. Popoff, Geoffroy Lerosey, Mathias Fink, A.~Claude Boccara, and Sylvain Gigan.
\newblock Image transmission through an opaque material.
\newblock \emph{Nat. Commun.}, 1:\penalty0 81, 2010.
\newblock \doi{10.1038/ncomms1078}.

\bibitem[Yu et~al.(2024)Yu, Li, Zhao, Huang, Lin, Yao, Li, Zhao, Wu, Li,
  Genevet, Song, and Lai]{Yu2024NatCommun}
Zhipeng Yu, Huanhao Li, Wannian Zhao, Po-Sheng Huang, Yu-Tsung Lin, Jing Yao,
  Wenzhao Li, Qi~Zhao, Pin~Chieh Wu, Bo~Li, Patrice Genevet, Qinghua Song, and
  Puxiang Lai.
\newblock High-security learning-based optical encryption assisted by
  disordered metasurface.
\newblock \emph{Nat. Commun.}, 15:\penalty0 2607, 2024.
\newblock \doi{10.1038/s41467-024-46946-w}.

\bibitem[Jiao et~al.(2020)Jiao, Feng, Gao, Lei, and Yuan]{Jiao2020OptExpress}
Shuming Jiao, Jun Feng, Yang Gao, Ting Lei, and Xiaocong Yuan.
\newblock Visual cryptography in single-pixel imaging.
\newblock \emph{Opt. Express}, 28\penalty0 (5):\penalty0 7301--7313, 2020.
\newblock \doi{10.1364/OE.383240}.

\bibitem[Peng et~al.(2006)Peng, Wei, and Zhang]{Peng2006OptLett}
Xiang Peng, Hengzheng Wei, and Peng Zhang.
\newblock Chosen-plaintext attack on lensless double-random phase encoding in
  the fresnel domain.
\newblock \emph{Opt. Lett.}, 31\penalty0 (22):\penalty0 3261--3263, 2006.
\newblock \doi{10.1364/OL.31.003261}.

\bibitem[Guo et~al.(2009)Guo, Salamo, Duchesne, Morandotti, Volatier-Ravat,
  Aimez, Siviloglou, and Christodoulides]{Guo2009PRL}
A.~Guo, G.~J. Salamo, D.~Duchesne, R.~Morandotti, M.~Volatier-Ravat, V.~Aimez,
  G.~A. Siviloglou, and D.~N. Christodoulides.
\newblock Observation of {$\mathcal{PT}$}-symmetry breaking in complex optical
  potentials.
\newblock \emph{Phys. Rev. Lett.}, 103:\penalty0 093902, 2009.
\newblock \doi{10.1103/PhysRevLett.103.093902}.

\bibitem[Rotter(2009)]{Rotter2009JPhysA}
Ingrid Rotter.
\newblock A non-Hermitian Hamilton operator and the physics of open quantum systems.
\newblock \emph{J. Phys. A: Math. Theor.}, 42:\penalty0 153001, 2009.
\newblock \doi{10.1088/1751-8113/42/15/153001}.

\bibitem[Lin et~al.(2011)Lin, Ramezani, Eichelkraut, Kottos, Cao, and
  Christodoulides]{Lin2011PRL}
Zin Lin, Hamidreza Ramezani, Toni Eichelkraut, Tsampikos Kottos, Hui Cao, and
  Demetrios~N. Christodoulides.
\newblock Unidirectional invisibility induced by {$\mathcal{PT}$}-symmetric
  periodic structures.
\newblock \emph{Phys. Rev. Lett.}, 106:\penalty0 213901, 2011.
\newblock \doi{10.1103/PhysRevLett.106.213901}.

\bibitem[R{\"u}ter et~al.(2010)R{\"u}ter, Makris, El-Ganainy, Christodoulides, Segev, and Kip]{Ruter2010NatPhys}
Christian~E. R{\"u}ter, Konstantinos~G. Makris, Ramy El-Ganainy, Demetrios~N. Christodoulides, Mordechai Segev, and Detlef Kip.
\newblock Observation of parity--time symmetry in optics.
\newblock \emph{Nat. Phys.}, 6:\penalty0 192--195, 2010.
\newblock \doi{10.1038/nphys1515}.

\bibitem[Heiss(2012)]{Heiss2012JPA}
W.~D. Heiss.
\newblock The physics of exceptional points.
\newblock \emph{Journal of Physics A: Mathematical and Theoretical},
  45\penalty0 (44):\penalty0 444016, 2012.
\newblock \doi{10.1088/1751-8113/45/44/444016}.

\bibitem[El-Ganainy et~al.(2018)El-Ganainy, Makris, Khajavikhan, Musslimani,
  Rotter, and Christodoulides]{ElGanainy2018NatPhys}
Ramy El-Ganainy, Konstantinos~G. Makris, Mercedeh Khajavikhan, Ziad~H.
  Musslimani, Stefan Rotter, and Demetrios~N. Christodoulides.
\newblock Non-hermitian physics and {$\mathcal{PT}$} symmetry.
\newblock \emph{Nat. Phys.}, 14:\penalty0 11--19, 2018.
\newblock \doi{10.1038/nphys4323}.

\bibitem[Feng et~al.(2017)Feng, El-Ganainy, and Ge]{Feng2017NatPhot}
Liang Feng, Ramy El-Ganainy, and Li~Ge.
\newblock Non-hermitian photonics based on parity-time symmetry.
\newblock \emph{Nat. Photonics}, 11:\penalty0 752--762, 2017.
\newblock \doi{10.1038/s41566-017-0031-1}.

\bibitem[Miri and Al{\`u}(2019)]{Miri2019Science}
Mohammad-Ali Miri and Andrea Al{\`u}.
\newblock Exceptional points in optics and photonics.
\newblock \emph{Science}, 363:\penalty0 eaar7709, 2019.
\newblock \doi{10.1126/science.aar7709}.

\bibitem[{\"O}zdemir et~al.(2019){\"O}zdemir, Rotter, Nori, and
  Yang]{Ozdemir2019NatMater}
{\c{S}}ahin~Kaya {\"O}zdemir, Stefan Rotter, Franco Nori, and Lan Yang.
\newblock Parity-time symmetry and exceptional points in photonics.
\newblock \emph{Nat. Mater.}, 18:\penalty0 783--798, 2019.
\newblock \doi{10.1038/s41563-019-0304-9}.

\bibitem[Garrison and Wright(1988)]{Garrison1988PLA}
J.~C. Garrison and E.~M. Wright.
\newblock Complex geometrical phases for dissipative systems.
\newblock \emph{Phys. Lett. A}, 128:\penalty0 177--181, 1988.
\newblock \doi{10.1016/0375-9601(88)90905-X}.

\bibitem[Ahmed(2001)]{Ahmed2001PLA}
Zafar Ahmed.
\newblock Pseudo-hermiticity of hamiltonians under imaginary shift of the
  coordinate: Real spectrum of complex potentials.
\newblock \emph{Phys. Lett. A}, 290\penalty0 (1--2):\penalty0 19--22, 2001.
\newblock \doi{10.1016/S0375-9601(01)00622-3}.

\bibitem[Fornberg(1988)]{Fornberg1988MathComp}
Bengt Fornberg.
\newblock Generation of finite difference formulas on arbitrarily spaced grids.
\newblock \emph{Mathematics of Computation}, 51\penalty0 (184):\penalty0
  699--706, 1988.
\newblock \doi{10.1090/S0025-5718-1988-0935077-0}.

\bibitem[Van Loan(2000)]{VanLoan2000JCAM}
Charles~F. Van Loan.
\newblock The ubiquitous Kronecker product.
\newblock \emph{J. Comput. Appl. Math.}, 123\penalty0 (1--2):\penalty0 85--100, 2000.
\newblock \doi{10.1016/S0377-0427(00)00393-9}.

\bibitem[Trefethen and Bau(2022)]{TrefethenBau2022}
Lloyd~N. Trefethen and David Bau, III.
\newblock \emph{Numerical Linear Algebra}.
\newblock Society for Industrial and Applied Mathematics, Philadelphia,
  twenty-fifth anniversary edition edition, 2022.
\newblock ISBN 978-1-61197-715-8.
\newblock \doi{10.1137/1.9781611977165}.

\bibitem[Zhang et~al.(2015)Zhang, Ma, and Zhong]{Zhang2015NatCommun}
Zibang Zhang, Xiao Ma, and Jingang Zhong.
\newblock Single-pixel imaging by means of fourier spectrum acquisition.
\newblock \emph{Nat. Commun.}, 6:\penalty0 6225, 2015.
\newblock \doi{10.1038/ncomms7225}.

\bibitem[Sun et~al.(2017)Sun, Meng, Edgar, Padgett, and Radwell]{Sun2017SciRep}
Ming-Jie Sun, Ling-Tong Meng, Matthew~P. Edgar, Miles~J. Padgett, and Neal
  Radwell.
\newblock A {Russian Dolls} ordering of the {Hadamard} basis for compressive
  single-pixel imaging.
\newblock \emph{Scientific Reports}, 7:\penalty0 3464, 2017.
\newblock \doi{10.1038/s41598-017-03725-6}.

\bibitem[Yu et~al.(2020)Yu, Stantchev, Yang, and
  Pickwell-MacPherson]{Yu2020SciRep}
Xiao Yu, Rayko~Ivanov Stantchev, Fan Yang, and Emma Pickwell-MacPherson.
\newblock Super sub-nyquist single-pixel imaging by total variation ascending
  ordering of the {Hadamard} basis.
\newblock \emph{Scientific Reports}, 10:\penalty0 9338, 2020.
\newblock \doi{10.1038/s41598-020-66371-5}.

\bibitem[Wang et~al.(2004)Wang, Bovik, Sheikh, and Simoncelli]{Wang2004TIP}
Zhou Wang, Alan~C. Bovik, Hamid~R. Sheikh, and Eero~P. Simoncelli.
\newblock Image quality assessment: From error visibility to structural
  similarity.
\newblock \emph{IEEE Trans. Image Process.}, 13\penalty0 (4):\penalty0
  600--612, 2004.
\newblock \doi{10.1109/TIP.2003.819861}.

\bibitem[Wiersig(2014)]{Wiersig2014PRL}
Jan Wiersig.
\newblock Enhancing the sensitivity of frequency and energy splitting detection by using exceptional points: Application to microcavity sensors for single-particle detection.
\newblock \emph{Phys. Rev. Lett.}, 112:\penalty0 203901, 2014.
\newblock \doi{10.1103/PhysRevLett.112.203901}.

\bibitem[Chen et~al.(2017)Chen, {\"O}zdemir, Zhao, Wiersig, and Yang]{Chen2017Nature}
Weijian Chen, {\c{S}}ahin~Kaya {\"O}zdemir, Guangming Zhao, Jan Wiersig, and Lan Yang.
\newblock Exceptional points enhance sensing in an optical microcavity.
\newblock \emph{Nature}, 548:\penalty0 192--196, 2017.
\newblock \doi{10.1038/nature23281}.

\bibitem[Wang et~al.(2020)Wang, Lai, Yuan, Suh, and Vahala]{Wang2020NatCommun}
Heming Wang, Yu-Hung Lai, Zhiquan Yuan, Myoung-Gyun Suh, and Kerry Vahala.
\newblock Petermann-factor sensitivity limit near an exceptional point in a Brillouin ring laser gyroscope.
\newblock \emph{Nat. Commun.}, 11:\penalty0 1610, 2020.
\newblock \doi{10.1038/s41467-020-15341-6}.

\bibitem[Ferri et~al.(2010)Ferri, Magatti, Lugiato, and Gatti]{Ferri2010PRL}
F.~Ferri, D.~Magatti, L.~A. Lugiato, and A.~Gatti.
\newblock Differential ghost imaging.
\newblock \emph{Physical Review Letters}, 104:\penalty0 253603, 2010.
\newblock \doi{10.1103/PhysRevLett.104.253603}.

\bibitem[Pastuszczak et~al.(2016)Pastuszczak, Szczygie{\l}, Miko{\l}ajczyk, and
  Koty{\'n}ski]{Pastuszczak2016AO}
Anna Pastuszczak, Bart{\l}omiej Szczygie{\l}, Micha{\l} Miko{\l}ajczyk, and
  Rafa{\l} Koty{\'n}ski.
\newblock Efficient adaptation of complex-valued noiselet sensing matrices for
  compressed single-pixel imaging.
\newblock \emph{Applied Optics}, 55\penalty0 (19):\penalty0 5141--5148, 2016.
\newblock \doi{10.1364/AO.55.005141}.

\bibitem[Vellekoop and Mosk(2007)]{Vellekoop2007OptLett}
I.~M. Vellekoop and A.~P. Mosk.
\newblock Focusing coherent light through opaque strongly scattering media.
\newblock \emph{Opt. Lett.}, 32\penalty0 (16):\penalty0 2309--2311, 2007.
\newblock \doi{10.1364/OL.32.002309}.

\end{thebibliography}

\begin{thebibliography}{18}
\providecommand{\natexlab}[1]{#1}
\providecommand{\url}[1]{\texttt{#1}}
\expandafter\ifx\csname urlstyle\endcsname\relax
  \providecommand{\doi}[1]{doi: #1}\else
  \providecommand{\doi}{doi: \begingroup \urlstyle{rm}\Url}\fi

\bibitem[Garrison and Wright(1988)]{SI-Garrison1988PLA}
J.~C. Garrison and E.~M. Wright.
\newblock Complex geometrical phases for dissipative systems.
\newblock \emph{Phys. Lett. A}, 128:\penalty0 177--181, 1988.
\newblock \doi{10.1016/0375-9601(88)90905-X}.

\bibitem[Mostafazadeh(2002)]{SI-Mostafazadeh2002JMP}
Ali Mostafazadeh.
\newblock Pseudo-hermiticity versus {$\mathcal{PT}$} symmetry: The necessary
  condition for the reality of the spectrum of a non-hermitian hamiltonian.
\newblock \emph{J. Math. Phys.}, 43:\penalty0 205--214, 2002.
\newblock \doi{10.1063/1.1418246}.

\bibitem[Brody(2014)]{SI-Brody2014JPA}
Dorje~C. Brody.
\newblock Biorthogonal quantum mechanics.
\newblock \emph{J. Phys. A: Math. Theor.}, 47:\penalty0 035305, 2014.
\newblock \doi{10.1088/1751-8113/47/3/035305}.

\bibitem[Ahmed(2001)]{SI-Ahmed2001PLA}
Zafar Ahmed.
\newblock Pseudo-hermiticity of hamiltonians under imaginary shift of the
  coordinate: Real spectrum of complex potentials.
\newblock \emph{Phys. Lett. A}, 290\penalty0 (1--2):\penalty0 19--22, 2001.
\newblock \doi{10.1016/S0375-9601(01)00622-3}.

\bibitem[Fornberg(1988)]{SI-Fornberg1988MathComp}
Bengt Fornberg.
\newblock Generation of finite difference formulas on arbitrarily spaced grids.
\newblock \emph{Mathematics of Computation}, 51\penalty0 (184):\penalty0
  699--706, 1988.
\newblock \doi{10.1090/S0025-5718-1988-0935077-0}.

\bibitem[Trefethen and Bau(2022)]{SI-TrefethenBau2022}
Lloyd~N. Trefethen and David Bau, III.
\newblock \emph{Numerical Linear Algebra}.
\newblock Society for Industrial and Applied Mathematics, Philadelphia,
  twenty-fifth anniversary edition edition, 2022.
\newblock ISBN 978-1-61197-715-8.
\newblock \doi{10.1137/1.9781611977165}.

\bibitem[Bender and Boettcher(1998)]{SI-Bender1998PRL}
Carl~M. Bender and Stefan Boettcher.
\newblock Real spectra in non-hermitian hamiltonians having {$\mathcal{PT}$}
  symmetry.
\newblock \emph{Phys. Rev. Lett.}, 80:\penalty0 5243--5246, 1998.
\newblock \doi{10.1103/PhysRevLett.80.5243}.

\bibitem[El-Ganainy et~al.(2018)El-Ganainy, Makris, Khajavikhan, Musslimani,
  Rotter, and Christodoulides]{SI-ElGanainy2018NatPhys}
Ramy El-Ganainy, Konstantinos~G. Makris, Mercedeh Khajavikhan, Ziad~H.
  Musslimani, Stefan Rotter, and Demetrios~N. Christodoulides.
\newblock Non-hermitian physics and {$\mathcal{PT}$} symmetry.
\newblock \emph{Nat. Phys.}, 14:\penalty0 11--19, 2018.
\newblock \doi{10.1038/nphys4323}.

\bibitem[Feng et~al.(2017)Feng, El-Ganainy, and Ge]{SI-Feng2017NatPhot}
Liang Feng, Ramy El-Ganainy, and Li~Ge.
\newblock Non-hermitian photonics based on parity-time symmetry.
\newblock \emph{Nat. Photonics}, 11:\penalty0 752--762, 2017.
\newblock \doi{10.1038/s41566-017-0031-1}.

\bibitem[Heiss(2012)]{SI-Heiss2012JPA}
W.~D. Heiss.
\newblock The physics of exceptional points.
\newblock \emph{Journal of Physics A: Mathematical and Theoretical},
  45\penalty0 (44):\penalty0 444016, 2012.
\newblock \doi{10.1088/1751-8113/45/44/444016}.

\bibitem[Miri and Al{\`u}(2019)]{SI-Miri2019Science}
Mohammad-Ali Miri and Andrea Al{\`u}.
\newblock Exceptional points in optics and photonics.
\newblock \emph{Science}, 363:\penalty0 eaar7709, 2019.
\newblock \doi{10.1126/science.aar7709}.

\bibitem[{\"O}zdemir et~al.(2019){\"O}zdemir, Rotter, Nori, and
  Yang]{SI-Ozdemir2019NatMater}
{\c{S}}ahin~Kaya {\"O}zdemir, Stefan Rotter, Franco Nori, and Lan Yang.
\newblock Parity-time symmetry and exceptional points in photonics.
\newblock \emph{Nat. Mater.}, 18:\penalty0 783--798, 2019.
\newblock \doi{10.1038/s41563-019-0304-9}.

\bibitem[Wang et~al.(2004)Wang, Bovik, Sheikh, and Simoncelli]{SI-Wang2004TIP}
Zhou Wang, Alan~C. Bovik, Hamid~R. Sheikh, and Eero~P. Simoncelli.
\newblock Image quality assessment: From error visibility to structural
  similarity.
\newblock \emph{IEEE Trans. Image Process.}, 13\penalty0 (4):\penalty0
  600--612, 2004.
\newblock \doi{10.1109/TIP.2003.819861}.

\bibitem[Duarte et~al.(2008)Duarte, Davenport, Takhar, Laska, Sun, Kelly, and
  Baraniuk]{SI-Duarte2008}
Marco~F. Duarte, Mark~A. Davenport, Dharmpal Takhar, Jason~N. Laska, Ting Sun,
  Kevin~F. Kelly, and Richard~G. Baraniuk.
\newblock Single-pixel imaging via compressive sampling.
\newblock \emph{IEEE Signal Process. Mag.}, 25\penalty0 (2):\penalty0 83--91,
  2008.
\newblock \doi{10.1109/MSP.2007.914730}.

\bibitem[Edgar et~al.(2019)Edgar, Gibson, and Padgett]{SI-Edgar2019}
Matthew~P. Edgar, Graham~M. Gibson, and Miles~J. Padgett.
\newblock Principles and prospects for single-pixel imaging.
\newblock \emph{Nature Photonics}, 13:\penalty0 13--20, 2019.
\newblock \doi{10.1038/s41566-018-0300-7}.

\bibitem[Ferri et~al.(2010)Ferri, Magatti, Lugiato, and Gatti]{SI-Ferri2010PRL}
F.~Ferri, D.~Magatti, L.~A. Lugiato, and A.~Gatti.
\newblock Differential ghost imaging.
\newblock \emph{Physical Review Letters}, 104:\penalty0 253603, 2010.
\newblock \doi{10.1103/PhysRevLett.104.253603}.

\bibitem[Pastuszczak et~al.(2016)Pastuszczak, Szczygie{\l}, Miko{\l}ajczyk, and
  Koty{\'n}ski]{SI-Pastuszczak2016AO}
Anna Pastuszczak, Bart{\l}omiej Szczygie{\l}, Micha{\l} Miko{\l}ajczyk, and
  Rafa{\l} Koty{\'n}ski.
\newblock Efficient adaptation of complex-valued noiselet sensing matrices for
  compressed single-pixel imaging.
\newblock \emph{Applied Optics}, 55\penalty0 (19):\penalty0 5141--5148, 2016.
\newblock \doi{10.1364/AO.55.005141}.

\bibitem[Zhang et~al.(2015)Zhang, Ma, and Zhong]{SI-Zhang2015NatCommun}
Zibang Zhang, Xiao Ma, and Jingang Zhong.
\newblock Single-pixel imaging by means of fourier spectrum acquisition.
\newblock \emph{Nat. Commun.}, 6:\penalty0 6225, 2015.
\newblock \doi{10.1038/ncomms7225}.

\end{thebibliography}
\end{document}